\documentclass[11pt, notitlepage]{article} % {article}
\usepackage{a4}
\usepackage{epstopdf}
\usepackage{xcolor}

\definecolor{TUMblue}{RGB}{0, 101, 189}
\definecolor{TUMlightblue}{RGB}{100,160,200}
\definecolor{TUMgreen}{RGB}{162,173,0}
\definecolor{TUMorange}{RGB}{227,114,034}
\definecolor{TUMivory}{RGB}{218,215,203}

\usepackage{natbib}

\usepackage{hyperref}
\hypersetup{
	colorlinks=true,
	linkcolor=TUMblue,
	citecolor=TUMblue,
	filecolor=TUMblue,
	urlcolor=TUMblue
}

\usepackage{etoolbox}
\makeatletter

\pretocmd{\NAT@citex}{%
	\let\NAT@hyper@\NAT@hyper@citex
	\def\NAT@postnote{#2}%
	\setcounter{NAT@total@cites}{0}%
	\setcounter{NAT@count@cites}{0}%
	\forcsvlist{\stepcounter{NAT@total@cites}\@gobble}{#3}}{}{}
\newcounter{NAT@total@cites}
\newcounter{NAT@count@cites}
\def\NAT@postnote{}

\def\NAT@hyper@citex#1{%
	\stepcounter{NAT@count@cites}%
	\hyper@natlinkstart{\@citeb\@extra@b@citeb}#1%
	\ifnumequal{\value{NAT@count@cites}}{\value{NAT@total@cites}}
	{\ifNAT@swa\else\if*\NAT@postnote*\else%
		\NAT@cmt\NAT@postnote\global\def\NAT@postnote{}\fi\fi}{}%
	\ifNAT@swa\else\if\relax\NAT@date\relax
	\else\NAT@@close\global\let\NAT@nm\@empty\fi\fi% avoid compact citations
	\hyper@natlinkend}
\renewcommand\hyper@natlinkbreak[2]{#1}

\patchcmd{\NAT@citex}
{\ifNAT@swa\else\if*#2*\else\NAT@cmt#2\fi
	\if\relax\NAT@date\relax\else\NAT@@close\fi\fi}{}{}{}
\patchcmd{\NAT@citex}
{\if\relax\NAT@date\relax\NAT@def@citea\else\NAT@def@citea@close\fi}
{\if\relax\NAT@date\relax\NAT@def@citea\else\NAT@def@citea@space\fi}{}{}

\makeatother

\usepackage{courier}
\usepackage{amssymb, graphicx}
\usepackage{subfigure}
\usepackage{amsmath}
\usepackage{amsthm}
\usepackage{verbatim}
\usepackage{dsfont}
\usepackage{geometry}
\usepackage{pdflscape}
\usepackage{multirow}
\usepackage{aliascnt}
\usepackage{verbatim}
\usepackage{graphicx}
\usepackage{float}
\usepackage{tikz}
\usetikzlibrary{calc,shapes,arrows,decorations.pathmorphing,graphs,positioning,backgrounds,arrows.meta}
\usepackage{latexsym}
\usepackage{mathtools}
\usepackage{mathrsfs}
\usepackage{bbm}
\usepackage{shadethm}
\usepackage{enumerate}
\usepackage{colortbl}
\usepackage{framed}
\colorlet{shadecolor}{gray!25}
\usepackage{booktabs}
\usepackage{longtable}
\usepackage{multirow}
\usepackage{rotating}
\usepackage{chngpage}
\usepackage{appendix}
\usepackage{pdfpages}

\usepackage{algorithm}
\usepackage[noend]{algpseudocode}
\algnewcommand{\IIf}[1]{\State\algorithmicif\ #1\ \algorithmicthen}
\algnewcommand{\EndIIf}{\unskip\ \algorithmicend\ \algorithmicif}
\makeatletter
\def\BState{\State\hskip-\ALG@thistlm}
\makeatother

\newcommand{\mynewtheorem}[2]{
	\newaliascnt{#1}{dummy}
	\newtheorem{#1}[#1]{#2}
	\aliascntresetthe{#1}
	\expandafter\def\csname #1autorefname\endcsname{#2}
}

\bibpunct[\textcolor{TUMblue}{, }]{\textcolor{TUMblue}{(}}{\textcolor{TUMblue}{)}}{\textcolor{TUMblue}{;}}{\textcolor{TUMblue}{a}}{\textcolor{TUMblue}{}}{\textcolor{TUMblue}{,}}
\makeatletter
\renewcommand\eqref[1]{%
	\textup{\color{TUMblue}\tagform@{\ref{#1}}}%
}

\newcommand{\myrel}[2]{\genfrac{}{}{0pt}{1}{#1}{#2}}

\theoremstyle{definition}
\mynewtheorem{thm}{Theorem}
\mynewtheorem{defi}{Definition}%[section]
\mynewtheorem{lem}{Lemma}%[section]
\mynewtheorem{cor}{Corollary}%[section]
\mynewtheorem{prop}{Proposition}%[section]
\mynewtheorem{exa}{Example}%[section]
\mynewtheorem{alg}{Algorithm}%[section]
\mynewtheorem{rem}{Remark}%[section]
\mynewtheorem{bsp}{Example}

\usepackage{etoolbox}
\makeatletter
\patchcmd{\hyper@makecurrent}{%
	\ifx\Hy@param\Hy@chapterstring
	\let\Hy@param\Hy@chapapp
	\fi
}{%
	\iftoggle{inappendix}{%true-branch
		\@checkappendixparam{chapter}%
		\@checkappendixparam{section}%
		\@checkappendixparam{subsection}%
		\@checkappendixparam{subsubsection}%
		\@checkappendixparam{paragraph}%
		\@checkappendixparam{subparagraph}%
	}{}%
}{}{\errmessage{failed to patch}}

\newcommand*{\@checkappendixparam}[1]{%
	\def\@checkappendixparamtmp{#1}%
	\ifx\Hy@param\@checkappendixparamtmp
	\let\Hy@param\Hy@appendixstring
	\fi
}
\makeatletter

\newtoggle{inappendix}
\togglefalse{inappendix}

\apptocmd{\appendix}{\toggletrue{inappendix}}{}{\errmessage{failed to patch}}
\apptocmd{\subappendices}{\toggletrue{inappendix}}{}{\errmessage{failed to patch}}
\usepackage{sectsty}
\allsectionsfont{\sffamily}

\usepackage{titlesec}
\titleformat{\chapter}[display]
{\normalfont\sffamily\LARGE\bfseries\centering}
{\chaptertitlename\ \thechapter}{20pt}{\LARGE}

\usepackage{caption}
\usepackage{footnote}

\newcommand{\sC}{\mathbbm{C}} % sC for survival copula
\newcommand{\scd}{\mathbbm{c}} % sc for survival copula density

\begin{document}
	
	{	\renewcommand*{\thefootnote}{\fnsymbol{footnote}}
		\title{\textbf{\sffamily Dependence modeling for recurrent event times\\ subject to right-censoring with D-vine copulas}}
		
		\date{\small \today}
		\newcounter{savecntr1}% Save footnote counter
		\newcounter{restorecntr1}% Restore footnote counter
		\newcounter{savecntr2}% Save footnote counter
		\newcounter{restorecntr2}% Restore footnote counter
		
		\author{Nicole Barthel\setcounter{savecntr1}{\value{footnote}}\thanks{Department of Mathematics, Technische Universit{\"a}t M{\"u}nchen, Boltzmannstra{\ss}e 3, 85748 Garching, Germany (email: \href{mailto:nicole.barthel@tum.de}{nicole.barthel@tum.de} (corresponding author), \href{mailto:cczado@ma.tum.de}{cczado@ma.tum.de})}, Candida Geerdens\setcounter{savecntr2}{\value{footnote}}\thanks{Center for Statistics, I-BioStat, Universiteit Hasselt, Agoralaan 1, B-3590 Diepenbeek, Belgium (email: \href{mailto:candida.geerdens@uhasselt.be}{candida.geerdens@uhasselt.be}, \href{mailto:paul.janssen@uhasselt.be}{paul.janssen@uhasselt.be})}, Claudia Czado\setcounter{restorecntr1}{\value{footnote}}%
			\setcounter{footnote}{\value{savecntr1}}\footnotemark% Print footnotemark
			\setcounter{footnote}{\value{restorecntr1}} \ and Paul Janssen\setcounter{restorecntr2}{\value{footnote}}%
		\setcounter{footnote}{\value{savecntr2}}\footnotemark% Print footnotemark
	\setcounter{footnote}{\value{restorecntr2}}}
		
		\maketitle
	}

%\vspace*{2cm}
\begin{abstract}
   In many time-to-event studies, the event of interest is recurrent. Here, the data for each sample unit corresponds to a series of gap times between the subsequent events. Given a limited follow-up period, the last gap time might be right-censored. In contrast to classical analysis, gap times and censoring times cannot be assumed independent, i.e.\ the sequential nature of the data induces dependent censoring. Also, the recurrences typically vary between sample units leading to unbalanced data. To model the association pattern between gap times, so far only parametric margins combined with the restrictive class of Archimedean copulas have been considered. Here, taking the specific data features into account, we extend existing work in several directions: we allow for nonparametric margins and consider the flexible class of D-vine copulas. A global and sequential (one- and two-stage) likelihood approach are suggested. We discuss the computational efficiency of each estimation strategy. 
   Extensive simulations show good finite sample performance of the proposed methodology. It is used to analyze the association in recurrent asthma attacks in children. The analysis reveals that a D-vine copula detects relevant insights, on how dependence changes in strength and type over time.\\

%	\vspace*{1cm}
	\vspace*{.5cm}
\end{abstract}

%\newpage
%\tableofcontents

%\newpage
\section{Introduction}\label{sec:Introduction}
In survival analysis interest is in the time to a predefined event. In a number of e.g.\ biomedical, sociological or engineering studies, this event is recurrent for each sample unit. For example, one may investigate the time to an asthma attack in children. Since a child can experience multiple subsequent asthma attacks, a series of event times is observed for each child. %Moreover, 
Due to limited follow-up, the time to the last recurrence may not be recorded, but it may be right-censored. %instead a lower time is observed, i.e.\ the event times are subject to right-censoring. 
While the sample units, called clusters (e.g.\ a child), are independent, the event times %or gap times (periods between subsequent events)
within a cluster are typically dependent.
%\\\\

%Popular survival models that account for within-cluster association are the marginal model and the (shared) frailty model. A marginal model usually corresponds to the classical Cox proportional hazards model \citep{wei1989regression}, where one accounts for the within-cluster association in the calculation of the variance of model parameter estimates, e.g.\ via the grouped jackknife method. A (shared) frailty model is a hazards model supplemented with a multiplicative cluster-specific random effect, named the frailty \citep{duchateau2008frailty}. Thus, it is a conditional version of the classical Cox proportional hazards model, where the within-cluster association is generated via the frailty. Details on these models and their application to recurrent event time data are given in \cite{cook2007statistical}. Clearly, the marginal model and the (shared) frailty model account for the within-cluster association in an indirect way.
%\\\\
Popular survival models that account for within-cluster association are the marginal model \citep{wei1989regression} and the (shared) frailty model %, a conditional version of the classical Cox proportional hazards model, where the within-cluster association is generated via a multiplicative cluster-specific random effect 
\citep{duchateau2008frailty}. %Details on these models and applications to recurrent event time data are given in \cite{cook2007statistical}. 
While these models account for the within-cluster association %in clustered event time data 
in an indirect way, copulas can be used for direct dependence modeling. A copula model
describes the joint survival function of the event times or gap times (periods between subsequent events) via their survival margins and a function, called the copula, that fully captures the within-cluster association \citep{Sklar59}. Thus, copulas are an attractive tool when interest is in the dependence itself. %See e.g.\ \cite{Nelsen2006} for a thorough introduction to copulas.
%\\\\

Typically, copulas are applied to clusters of equal size, a feature that recurrent event time data often lack. For example, one child could have two asthma attacks, while another child experiences three or more asthma attacks. \cite{meyer2015bayesian} and \cite{prenen2017extending} study copula based inference for unbalanced right-censored clustered event time data focusing on the class of 
Archimedean copulas. Unfortunately, the latter only allow for a restrictive dependence structure: %it is assumed that 
all time pairs in a cluster exhibit the same type and strength of association. For recurrent event times, however, the type and strength of dependence may evolve over time. %, making the use of Archimedean copulas problematic. 
In this paper, we advocate D-vine copulas as a flexible alternative to Archimedean copulas
\citep{aas2009pair,czado2010pair,kurowicka2010dependence}. D-vines are built from freely chosen bivariate (conditional) copulas such that complex association patterns with various types and strengths of dependence can be modeled. In particular, the serial dependence inherent for recurrent events is naturally captured. Further, their construction principle allows to easily handle the unbalanced data setting.
%\\\\

We focus on the analysis of the gap times and so an extra challenge arises: not only are the gap times in a cluster associated, due to the recurrent nature of the data they are also subject to induced dependent right-censoring (\autoref{Sec:SettingNotation}). In their analysis, \cite{meyer2015bayesian} assume parametric survival functions in a likelihood based global one-stage estimation strategy. To increase model flexibility we also consider nonparametrically estimated survival margins together with global two-stage estimation. For both modeling approaches, alternative sequential estimation techniques are presented. They facilitate the global optimization procedures for high-dimensional data.
%\\\\

In summary: %for recurrent event time data subject to right-censoring, or equivalently, 
for gap time data subject to induced dependent right-censoring (i) we show that D-vines provide a natural way to unravel a possibly complex association structure; (ii) we propose estimation procedures that allow %unspecified (
nonparametric
%)
survival margins and (iii) we compare parametric and nonparametric as well as global and sequential estimation strategies.
%\\\\

The paper is organized as follows. A motivating example on recurrent asthma attacks in children is introduced in \autoref{Sec:AsthmaMotivation}. The general data setting and notation are given in \autoref{Sec:SettingNotation}. In \autoref{Sec:DepModeling}, Archimedean copulas and D-vine copulas are introduced as the two copula classes considered for dependence modeling. Four estimation strategies are presented in \autoref{Sec:Methodology}: parametric versus nonparametric combined with global versus sequential.
%\autoref{Sec:1stage_global} and \autoref{Sec:1stage_seq} are concerned with parametric one-stage estimation considering both global likelihood optimization and sequentially conducted likelihood maximization. A modified nonparametric estimator that can handle the induced dependent right-censoring within the marginal estimation is introduced in \autoref{Sec:2StageMargins}. Then, semiparametric two-stage estimation approaches are developed. Global and sequential likelihood optimization is discussed in \autoref{Sec:2stage_global} and \autoref{Sec:2stage_seq}.
A simulation study in three dimensions is used to point out diverse aspects and challenges of modeling gap time data.
%\textcolor{red}{Some practical guidelines are formulated.}
A simulation study in four dimensions in \autoref{Sec:SimStudy} further %, a simulation study in four dimensions is used to further evaluate the finite sample performance of the proposed methods. 
demonstrates the flexibility of D-vines compared to Archimedean copulas in modeling complex dependence structures. %Based on the developed methods, 
The asthma data are investigated in \autoref{Sec:DataApplication}. In \autoref{Sec:Discussion}, concluding remarks are given. This paper comes with extensive online supplementary material.

\section{Motivating example: the asthma data}\label{Sec:AsthmaMotivation}
We consider a study on $232$ children with a high risk of developing asthma. Asthma is a chronic lung disease that inflames and narrows the airways. It causes consecutive episodes of wheezing, chest tightness and/or shortness of breath, commonly referred to as asthma attacks. The children enter the study at the age of $6$ months, at which they are randomized into a placebo group ($113$ children) or a treatment group ($119$ children). They are followed up for about $18$ months. The data have been analyzed by \cite{duchateau2003} and \cite{meyer2015bayesian}, employing a frailty model resp. a copula model. % In this paper, interest is in the dependence structure of subsequent asthma attacks. Hence, a copula approach is taken.

Only few children have more than four asthma attacks (see Table 17 in the supplementary material), making accurate estimation of the marginal survival functions and of the association from the fifth gap time on rather difficult. As in \cite{meyer2015bayesian}, we focus on the association between the first four gap times, i.e.\ we use the data of attack $1$ up to attack $4$ even if there is information on subsequent attacks. By doing so, each child experiences at least one asthma attack and $97$ children have at least four attacks. For $25$ of these children the last asthma attack is right-censored ($8$ in the treatment group and $17$ in the control group). By only considering the first four asthma attacks, the overall censoring rate is $22.13\%$.  Table 18 of the supplementary material gives a concise overview of the considered data.

\section{General setting and notation}\label{Sec:SettingNotation}

Suppose a study includes $n$ independent individuals that are followed-up for a recurrent event. For individual $i$ ($i=1,\ldots,n$) let $d_i$ denote the total number of consecutive events. Thus, individual $i$ corresponds to % can be considered as 
a cluster of size $d_i$. Let $T_{i,j}$ be the true $j$-th event time for cluster $i$, where $T_{i,j} > 0$ and $T_{i,1} < \ldots < T_{i,d_{i}}$ ($i=1,\ldots,n$ and $j=1,\ldots,d_i$). Due to a limited study period, the follow-up time of cluster $i$ is subject to right-censoring by $C_{i}$. %($i=1,\ldots,n$). 
The censoring times are assumed to be non-informative and independent of the event times. The intervals between two subsequent events are referred to as gap times $G_{i,j}$ and are defined by
\begin{align*}
G_{i,1} = T_{i,1} \quad \text{and} \quad G_{i,j} = T_{i,j} - T_{i,j-1} \quad \text{for} \quad i = 1,\ldots,n \quad \text{and} \quad j = 2,\ldots, d_i.
\end{align*}
It follows that gap time $G_{i,1}$ is subject to right-censoring by $C_i$, while the subsequent gap times $G_{i,j}$ ($j = 2, \ldots, d_i$) are subject to right-censoring by
%\begin{align*}
$C_i-T_{i,j-1} = C_i - \sum_{\ell=1}^{j-1} G_{i,\ell}$,
%\end{align*}
which naturally depends on previous gap times. Since the dependence between gap times and censoring times is a direct consequence of the recurrent nature of the underlying data and thus induced by the data structure itself, we say that gap times are subject to induced dependent right-censoring. Note that %, due to the subsequent nature of recurrent event time data, 
only the last gap time $G_{i,d_i}$ of cluster $i$ can be right-censored. Hence, for cluster $i$ of size $d_i$ the observed data are given by
\begin{align*}
Y_{i,j} = G_{i,j} \quad \text{and} \quad Y_{i,d_i} = \min\big(G_{i,d_i}, C_i -  \sum_{\ell=1}^{d_i-1} G_{i,\ell}\big)
\end{align*}
together with the right-censoring indicator $\delta_{i,d_i} = I\left(Y_{i,d_i} = G_{i,d_i}\right)$ ($i=1,\ldots,n$ and $j=1,\ldots,d_i-1$). For observed times $Y_{i,j}$ with $j<d_i$, we set $\delta_{i,j} = 1$.
%\begin{align}\label{eq:deltatilde}
%\widetilde{\delta}_{i,j} =
%\begin{cases}
%1, \quad \ j < d_i,\\
%\delta_{i}, \quad j = d_i.
%\end{cases}
%\end{align}
Typically, not all $n$ individuals experience the same number of events, i.e.\ the cluster size $d_i$ varies among clusters, resulting in an unbalanced cluster setting. Let the maximum cluster size be $d = \max\lbrace d_i | i = 1, \ldots, n\rbrace$. Denote by $n_j$ ($j=1,\ldots,d$) the number of clusters with size $j$ such that $n=n_1+n_2+\ldots+n_{d-1}+n_d$. Throughout we assume that data are ordered by decreasing cluster size. See Table 1 in the supplementary material for details on the required data format.

\section{Dependence modeling using copulas}\label{Sec:DepModeling}
Interest is in the dependence of the gap times $\left(G_1,\ldots,G_{d_i}\right)$ for $i=1,\ldots,n$. We consider the random vector $\left(G_1,\ldots,G_{d}\right)$ with survival margins and joint survival function given by
\begin{align*}
S_j(g) = P(G_j > g), \quad j = 1, \ldots, d, \quad \text{ and } \quad S(g_1,\ldots,g_{d}) = P(G_1 > g_1,\ldots,G_{d} > g_{d}).
\end{align*}
For dependence modeling, copulas are a popular and useful tool to apply. A $d$-dimensional copula $\sC$ is a distribution function on $[0,1]^{d}$ with uniform marginal distribution functions. According to \cite{Sklar59} the copula $\sC$ provides a connection between the survival margins $S_j$ of $G_j$ ($j=1,\ldots,d$) and thereby models the joint survival function $S$ of $\left(G_1,\ldots,G_{d}\right)$:
\begin{align}
S(g_1,\ldots,g_{d}) = \sC\{S_1(g_1),\ldots,S_{d}(g_{d})\}.
\label{eq:SklarCDF}
\end{align}
The copula $\sC$ fully captures the dependence structure of $\left(G_1,\ldots,G_{d}\right)$. The decomposition \eqref{eq:SklarCDF} is unique, when $\left(G_1,\ldots,G_{d}\right)$ is absolutely continuous, which we will assume throughout the paper. The joint density $f$ of $\left(G_1,\ldots,G_{d}\right)$ is then given by
\begin{align*}
f(g_1,\ldots,g_{d}) = \scd\{S_1(g_1),\ldots,S_{d}(g_{d})\}f_1\left(g_1\right)\cdots f_d\left(g_d\right),
%\label{eq:SklarPDF}
\end{align*}
where $f_j$ denotes the marginal density function of $G_j$ ($j=1,\ldots,d$) and $\scd$ is the copula density. With $U_{j} \coloneqq S_j\left(G_{j}\right) (j=1,\ldots,d)$
the joint distribution function of $\left(U_1,\ldots,U_{d}\right)$ is the copula $\sC$. %describing $\left(G_1,\ldots,G_d\right)$. 
Two classes of copulas, Archimedean copulas and D-vine copulas, are studied and compared.

\subsection{Archimedean copulas}\label{Sec:ArchCops}
A copula $\sC$ is called an Archimedean copula if it admits the representation
\begin{align}
\sC(u_1,\ldots,u_d) = \phi\{\phi^{-1}(u_1) + \ldots + \phi^{-1}(u_d)\},
\label{Eq:ArchimedeanCop}
\end{align}
where $\phi:[0,\infty[ \rightarrow [0,1]$ is a continuous strictly decreasing function with $\phi(0)=1, \phi(\infty)=0$ and satisfies the complete monotonicity condition, i.e.\ the derivatives of $\phi$ must alternate in sign. Common Archimedean copulas are Clayton, Gumbel and Frank (see Table 2 in the supplementary material for details). %For $d=2$, details are listed in \autoref{table:Copulas}. 
From %For $d>2$, 
\eqref{Eq:ArchimedeanCop} it follows that an Archimedean copula is fully determined by the choice of $\phi$. As a result, a restrictive dependence structure is implied, e.g.\ all marginal copulas show exactly the same type and strength of association. Note that Archimedean copulas %like Clayton, Gumbel and Frank 
having the same global strength of association as expressed by Kendall's $\tau$, do have a quite divers local dependence structure: a Clayton copula is lower tail-dependent, a Gumbel copula is upper tail-dependent and a Frank copula shows no tail-behavior. For a detailed study %of Archimedean copulas 
see e.g.\ \cite{Nelsen2006}, \cite{Embrechts03} or \cite{Joe1997}.

%\begin{table}[t]
%\small
%%\begin{centering}
%\caption{Popular bivariate Archimedean copulas with the range of their dependence parameter $\theta$ and the formula of $\phi$. \textcolor{red}{Include Kendall's $\tau$?}}
%\label{table:Copulas}
%\begin{tabular}{ p{1cm} c p{3.5cm} p{4.75cm} c} 			\midrule
%Family  &  $\theta \in $ & \centering $\phi(s)$
%& \centering$\sC(u_{1},u_{2})$ & Kendall's $\tau$\\  [1ex] \hline \midrule
%Clayton   & $(0,\infty)$    & \centering$(1+\theta s)^{-1/ \theta}$
%& \centering$ (u_{1}^{-\theta}+u_{2}^{-\theta}-1)^{-\frac{1}{\theta}}$ & {\centering $\tau = \frac{\theta}{\theta+2}$}\\
%Gumbel    & $[1,\infty) $ &  \centering $e^{-s^{1/\theta}}$
%& \centering$e^{\Big\{ - [(-\ln u_{1})^{\theta} + (- \ln u_{2})^{\theta} ]^{\frac{1}{\theta}} \Big\}}$ & {$\tau = 1-\frac{1}{\theta}$} \\ [1.25ex]
%Frank   & $(-\infty,\infty) \setminus \{0\}$ &  \centering $-\frac{1}{\theta} \ln(1-(1-e^{-\theta})e^{-s})$
%& \centering $- \frac{1}{\theta}  \ln \Big\{  1+ \frac{(e^{-\theta u_{1}}-1)    (e^{-\theta u_{2}}-1)}{   e^{-\theta }-1} \Big\}   $ & {\centering$\tau =
%1-\frac{4}{\theta}+4\frac{D_1(\theta)}{\theta} $} \\
%& &
%&  & {\centering{\scriptsize with} $D_1(\theta) =
%\int_0^{\theta} \frac{t/\theta}{e^t-1}dt$} \\
%\midrule			\midrule
%\end{tabular}
%%\end{centering}
%\\
%%{\footnotesize where $D_{1}(\theta) = \frac{1}{\theta}  \int_{0}^{\theta} \frac{t}{e^{t}-1} dt $ is the Debye function.}
%\end{table}

\subsection{D-vine copulas}\label{DVineCops}
A flexible alternative to Archimedean copulas is given by regular (R) vine copulas, also referred to as pair-copula constructions (PCC). An R-vine copula is based on a decomposition of the copula density $\scd$ into a cascade of $d(d-1)/2$ bivariate (un)conditional copula densities, which can be chosen arbitrarily from a large catalogue of bivariate copula families \citep{joe1996multivariate,bedford2002vines}. Possible candidates are Clayton, Gumbel and Frank.

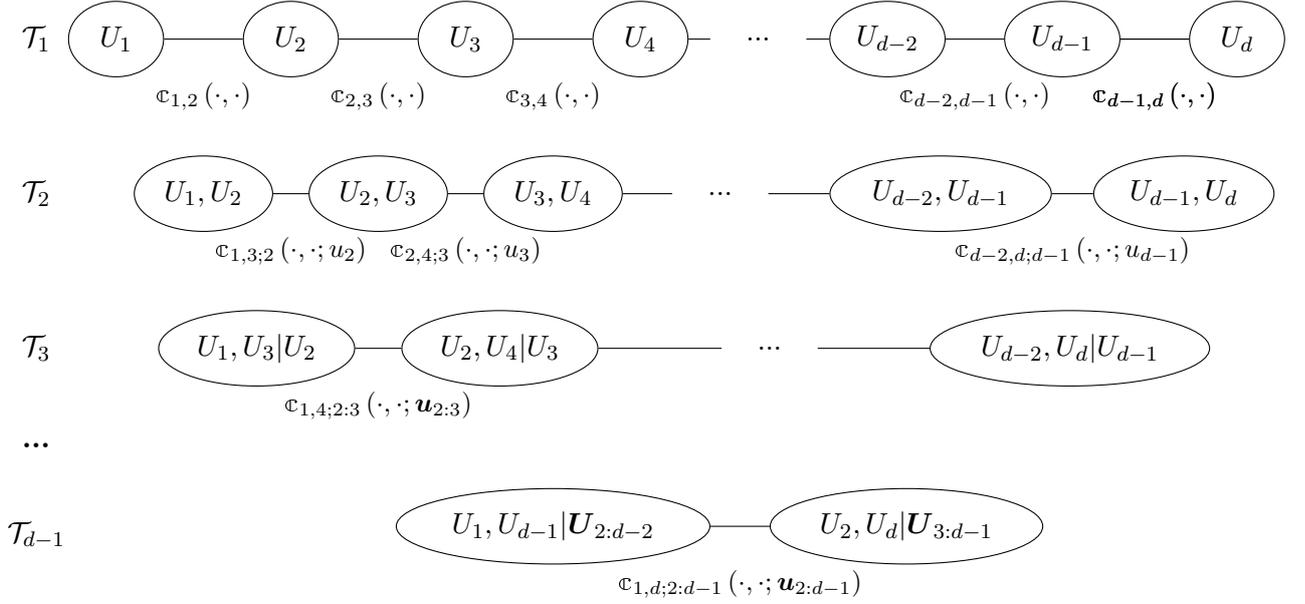
\begin{figure}
	\tikzstyle{VineNode} = [ellipse, fill = white, draw = black, text = black, align = center, minimum height = 1cm, minimum width = 1.25cm]
	\tikzstyle{TreeLabels} = [draw = none, fill = none, text = black, font = \bfseries]
	\tikzstyle{DummyNode}  = [draw = none, fill = none, text = black]
	\newcommand{\yshift}{-.9cm}
	\newcommand{\yshiftLabel}{-0.25cm}
	\newcommand{\labelsize}{\small}
	\newcommand{\fontsizeEdge}{\small}
	\centering
	\hspace*{-1cm}
	\begin{tikzpicture}	[every node/.style = VineNode, node distance =1.15cm]
	
	\node (1){$U_1$}
	node[DummyNode] (D1-2) [right of = 1]{}
	node  (2)   [right of = D1-2]{$U_2$}
	node[DummyNode] (D2-3) [right of = 2]{}
	node  (3)   [right of = D2-3]{$U_3$}
	node[DummyNode] (D3-4) [right of = 3]{}
	node  (4)   [right of = D3-4]{$U_4$}
	node[DummyNode] (D4-5) [right of = 4, xshift = -.75cm]{}
	node[DummyNode]  (cdots)   [right of = D4-5]{...}
	node[DummyNode] (Dcdots-d-1) [right of = cdots, xshift = -.6cm]{}
	node  (d-2)   [right of = Dcdots-d-1]{$U_{d-2}$}
	node[DummyNode] (Dd-2-d-1) [right of = d-2]{}
	node  (d-1)   [right of = Dd-2-d-1]{$U_{d-1}$}
	node[DummyNode] (Dd-1-d) [right of = d-1]{}
	node  (d)   [right of = Dd-1-d]{$U_d$}
	
	node (1-2)  [below of = D1-2, yshift = \yshift]{$U_1, U_2$}
	node[DummyNode] (D1-3_2) [below of = 2, yshift = \yshift]{}
	node  (2-3)   [right of = D1-3_2]{$U_2, U_3$}
	node[DummyNode] (D2-4_3) [right of = 2-3]{}
	node  (3-4)   [below of = D3-4, yshift = \yshift]{$U_3, U_4$}
	node[DummyNode] (D3-5_4) [right of = 3-4, xshift = -1cm]{}
	node  (d-1-d)   [below of = Dd-1-d, yshift = \yshift, xshift = .45cm]{$U_{d-1}, U_d$}
	node  (d-2-d-1)   [below of = Dd-2-d-1, yshift = \yshift, xshift = -.45cm]{$U_{d-2}, U_{d-1}$}
	node[DummyNode]  (cdots2)   [left of = d-2-d-1, xshift = -1.75cm]{...}
	node[DummyNode] (Dcdots-d-2) [right of = cdots2, xshift = -1cm]{}
	
	node (1-3_2) [below of = D1-3_2, yshift = \yshift, xshift = -.45cm]{$U_1, U_3\vert U_2$}
	node[DummyNode] (D1-4_2-3) [below of = 2, yshift = \yshift]{}
	node (2-4_3) [below of = D2-4_3, yshift = \yshift, xshift = .45cm]{$U_2, U_4\vert U_3$}
	
	node (d-2-d) [below of = d-1-d, yshift = \yshift, xshift = -1.5cm]{$U_{d-2}, U_d\vert U_{d-1}$}
	node[DummyNode]  (cdots3)   [left of = d-2-d, xshift = -2.8cm]{...}
	
	node (1-d-1) [below of = D3-4, yshift =5.9*\yshift]{$U_1, U_{d-1}\vert \boldsymbol{U}_{2:d-2}$}
	node (2-d) [right of = 1-d-1, xshift = 3.5cm]{$U_2, U_{d}\vert \boldsymbol{U}_{3:d-1}$}
	;
	\draw (1) to node[draw=none, fill = none, font = \fontsizeEdge,
	below, yshift = \yshiftLabel] {$\scd_{1,2}\left(\cdot,\cdot\right)$} (2);
	\draw (2) to node[draw=none, fill = none, font = \fontsizeEdge,
	below, yshift = \yshiftLabel] {$\scd_{2,3}\left(\cdot,\cdot\right)$} (3);
	\draw (3) to node[draw=none, fill = none, font = \fontsizeEdge,
	below, yshift = \yshiftLabel] {$\scd_{3,4}\left(\cdot,\cdot\right)$} (4);
	\draw (d-2) to node[draw=none, fill = none, font = \fontsizeEdge, below, yshift = \yshiftLabel] {$\scd_{d-2,d-1}\left(\cdot,\cdot\right)$} (d-1);
	\draw (d-1) to node[draw=none, fill = none, font = \fontsizeEdge, below, yshift = \yshiftLabel] {$\scd_{d-1,d}\left(\cdot,\cdot\right)$} (d);
	\draw (4) to node[draw=none, fill = none, font = \fontsizeEdge, below, yshift = \yshiftLabel] {} (cdots);
	\draw (cdots) to node[draw=none, fill = none, font = \fontsizeEdge, below, yshift = \yshiftLabel] {} (d-2);
	\draw (3-4) to node[draw=none, fill = none, font = \fontsizeEdge, below, yshift = \yshiftLabel] {} (cdots2);
	\draw (cdots2) to node[draw=none, fill = none, font = \fontsizeEdge, below, yshift = \yshiftLabel] {} (d-2-d-1);
	\draw (2-4_3) to node[draw=none, fill = none, font = \fontsizeEdge, below, yshift = \yshiftLabel] {} (cdots3);
	\draw (cdots3) to node[draw=none, fill = none, font = \fontsizeEdge, below, yshift = \yshiftLabel] {} (d-2-d);
	\draw (d-1) to node[draw=none, fill = none, font = \fontsizeEdge, below, yshift = \yshiftLabel] {$\scd_{d-1,d}\left(\cdot,\cdot\right)$} (d);
	\draw (1-2) to node[draw=none, fill = none, font = \fontsizeEdge,
	below, yshift = \yshiftLabel] {$\scd_{1,3;2}\left(\cdot,\cdot;u_2\right)$} (2-3);
	\draw (2-3) to node[draw=none, fill = none, font = \fontsizeEdge,
	below, yshift = \yshiftLabel] {$\scd_{2,4;3}\left(\cdot,\cdot;u_3\right)$} (3-4);
	\draw (d-2-d-1) to node[draw=none, fill = none, font = \fontsizeEdge,
	below, yshift = \yshiftLabel] {$\scd_{d-2,d;d-1}\left(\cdot,\cdot;u_{d-1}\right)$} (d-1-d);
	\draw (1-3_2) to node[draw=none, fill = none, font = \fontsizeEdge,
	below, yshift = \yshiftLabel] {$\scd_{1,4;2:3}\left(\cdot,\cdot;\boldsymbol{u}_{2:3}\right)$} (2-4_3);
	\draw (1-d-1) to node[draw=none, fill = none, font = \fontsizeEdge,
	below, yshift = \yshiftLabel] {$\scd_{1,d;2:d-1}\left(\cdot,\cdot;\boldsymbol{u}_{2:d-1}\right)$} (2-d);
	
	\node[TreeLabels] (CT1)  [left of = 1, xshift = +.1cm] {$\mathcal{T}_1$}
	node[TreeLabels] (CT2)  [below of = CT1, yshift = \yshift] {$\mathcal{T}_2$}
	node[TreeLabels] (CT3)  [below of = CT2, yshift = \yshift] {$\mathcal{T}_3$}
	node[TreeLabels] (CTdots)  [below of = CT3, yshift = .15*\yshift] {...}
	node[TreeLabels] (CTd-1)  [below of = CTdots, yshift = .05*\yshift] {$\mathcal{T}_{d-1}$}
	;			
	\end{tikzpicture}
	\caption{D-vine tree structure with order $1-2-\ldots-d$.}
	\label{fig:D-vineGeneral}
\end{figure}

In this paper, we focus on a special class of R-vine copulas named D-vine copulas.
Due to their construction, D-vine copulas overcome the restrictive dependence pattern of Archimedean copulas and, as will be explained, easily capture the inherent sequential nature of recurrent event time data.	
\autoref{fig:D-vineGeneral} illustrates the construction of a D-vine copula with ordering $1-2-\ldots-d$, referred to as ordered D-vine. In tree $\mathcal{T}_1$, the nodes correspond to the random variables $U_j = S_j\left(G_j\right)$ ($j = 1, \ldots, d$), while the edges refer to the bivariate copula density $\scd_{k,k+1}\left(\cdot, \cdot\right)$ ($k=1,\ldots, d - 1$) related to the bivariate distribution of $\left(U_k, U_{k+1}\right)$. In tree $\mathcal{T}_{\ell}$ ($\ell = 2, \ldots, d - 1$), we define for $k = 1,\ldots, d - \ell$ the vector $\boldsymbol{u}_{k+1:k+\ell-1} \coloneqq \left(u_{k+1}, \ldots, u_{k+\ell-1}\right)$ and denote by $\scd_{k,k+\ell;k+1:k+\ell-1}\left(\cdot, \cdot; \boldsymbol{u}_{k+1:k+\ell-1}\right)$ the bivariate conditional copula density linked to the conditional distribution of $\left(U_k, U_{k+\ell}\right)$ given $\boldsymbol{U}_{k+1:k+\ell-1} = \boldsymbol{u}_{k+1:k+\ell-1}$. Thus, for a D-vine copula conditioning is on intermediate variables.
As derived in detail in \cite{czado2010pair} and illustrated for $d=4$ in \autoref{Exa:DVineDensities}, the copula density $\scd_{1:d}$ of $(U_1,\ldots,U_d)$ can be expressed as a $d$-dimensional ordered D-vine copula density as follows:
\begin{align}
\small
& \scd_{1:d}(u_{1},\ldots,u_{d})\label{eq:d_DVineDensity}\\
& \phantom{\partial} = \prod_{\ell = 1}^{d-1} \prod_{k=1}^{d - \ell} \scd_{k,k+\ell;k+1:k+\ell-1}\{\sC_{k|k+1:k+\ell-1}(u_{k}|\boldsymbol{u}_{k+1:k+\ell-1}), \sC_{k+\ell|k+1:k+\ell-1}(u_{k+\ell}|\boldsymbol{u}_{k+1:k+\ell-1})\},\notag
\end{align}
where $\sC_{k|k+1:k+\ell-1}\left(\cdot|\boldsymbol{u}_{k+1:k+\ell-1}\right)$, resp. $\sC_{k+\ell|k+1:k+\ell-1}\left(\cdot|\boldsymbol{u}_{k+1:k+\ell-1}\right)$, denotes the univariate conditional distribution of $U_k$ given $\boldsymbol{U}_{k+1:k+\ell-1} = \boldsymbol{u}_{k+1:k+\ell-1}$, resp. of $U_{k+\ell}$ given $\boldsymbol{U}_{k+1:k+\ell-1} = \boldsymbol{u}_{k+1:k+\ell-1}$. As common within the vine copula framework, we assume in \eqref{eq:d_DVineDensity} that\\ $\scd_{k,k+\ell;k+1:k+\ell-1}\left(\cdot, \cdot; \boldsymbol{u}_{k+1:k+\ell-1}\right) \equiv \scd_{k,k+\ell;k+1:k+\ell-1}\left(\cdot, \cdot\right)$, i.e.\ the conditional pair-copulas\\ $\scd_{k,k+\ell;k+1:k+\ell-1}$ in trees $\mathcal{T}_\ell$ ($\ell = 2, \ldots, d-1$) do not depend on the conditioning vector $\boldsymbol{u}_{k+1:k+\ell-1}$. Their arguments  $\sC_{k|k+1:k+\ell-1}\left(u_k|\boldsymbol{u}_{k+1:k+\ell-1}\right)$ and $\sC_{k+\ell|k+1,k+\ell-1}\left(u_{k+\ell}|\boldsymbol{u}_{k+1:k+\ell-1}\right)$ indeed do depend on $\boldsymbol{u}_{k+1:k+\ell-1}$. For details on this so-called simplifying assumption, see e.g.\ \cite{haff2010simplified}, \cite{kurz2017testing} or \cite{stoeber2013simplified}.

\begin{exa}\label{Exa:DVineDensities}
	For a 3-dimensional copula density $\scd_{1:3}$, resp.\ 4-dimensional copula density $\scd_{1:4}$, the construction corresponding to an ordered D-vine copula is given by
	\begin{align}
	&\scd_{1:3}\left(u_1,u_2,u_3\right) =  \scd_{1,2}\left(u_1,u_2\right)\scd_{2,3}\left(u_2,u_3\right)\scd_{1,3;2}\{\sC_{1|2}\left(u_1|u_2\right),\sC_{3|2}\left(u_3|u_2\right)\},\label{eq:3dExampleDensity}\\
	%\end{align}
	%For a 4-dimensional copula density $\scd_{1:4}$ it holds %the construction corresponding to an ordered D-vine copula is given by
	%\begin{align}
	\text{resp.\ \ } &\scd_{1:4}\left(u_1,u_2,u_3,u_4\right) =  \scd_{1,2}\left(u_1,u_2\right)\scd_{2,3}\left(u_2,u_3\right)\scd_{3,4}\left(u_3,u_4\right)\scd_{1,3;2}\{\sC_{1|2}\left(u_1|u_2\right),\sC_{3|2}\left(u_3|u_2\right)\}\notag\\
	& \times \scd_{2,4;3}\{\sC_{2|3}\left(u_2|u_3\right),\sC_{4|3}\left(u_4|u_3\right)\}\scd_{1,4;2,3}\{\sC_{1|2:3}\left(u_1|\boldsymbol{u}_{2:3}\right),\sC_{4|2:3}\left(u_4|\boldsymbol{u}_{2:3}\right)\}\notag\\
	= & \ \scd_{1:3}\left(u_1,u_2,u_3\right)\label{eq:4dExampleDensity}\\
	& \times \scd_{3,4}\left(u_3,u_4\right)\scd_{2,4;3}\{\sC_{2|3}\left(u_2|u_3\right),\sC_{4|3}\left(u_4|u_3\right)\}\scd_{1,4;2,3}\{\sC_{1|2:3}\left(u_1|\boldsymbol{u}_{2:3}\right),\sC_{4|2:3}\left(u_4|\boldsymbol{u}_{2:3}\right)\},\notag
	\end{align}
	where the second equality for $\scd_{1:4}$ is based on \eqref{eq:3dExampleDensity}. As can be seen from \eqref{eq:4dExampleDensity} the construction of an ordered D-vine implies that lower dimensional D-vine copula densities are embedded within higher dimensional D-vine copula densities. In particular, this holds true for all marginal copula densities $\scd_{1:\tilde{d}}$ of random vectors $\left(U_1,\ldots,U_{\tilde{d}}\right)$ with $\tilde{d}=2,\ldots,d-1$.
\end{exa}

%In general, we have  are nested within \eqref{eq:d_DVineDensity}, i.e.\
%\begin{align*}
%\small
%\textcolor{red}{\scd_{1:d}}&(u_{1},\ldots,u_{d})\\%\label{eq:nested_DVineDensity}\\
%\phantom{\partial} = \ & \scd_{1:d_i}(u_{1},\ldots,u_{d_i})\\
%& \times \prod_{\ell = 1}^{d-1}\prod_{k=\max\left(1,d_i-\ell+1\right)}^{d - \ell} \scd_{k,k+\ell;k+1:k+\ell-1}\big(\sC_{k|k+1:k+\ell-1}(u_{k}|\boldsymbol{u}_{k+1:k+\ell-1}),\\
%& \phantom{\times \prod_{\ell = 1}^{d-1}\prod_{k=\max\left(1,d_i-\ell+1\right)}^{d - \ell} \scd_{k,k+\ell;k+1:k+\ell-1}\big(\ } \sC_{k+\ell|k+1:k+\ell-1}(u_{k+\ell}|\boldsymbol{u}_{k+1:k+\ell-1})\big),\notag
%\end{align*}

The arguments of the pair-copulas from tree $\mathcal{T}_2$ on are univariate conditional distributions,
%of the form $\sC_{k|k+1:k+\ell-1}\left(u_k|\boldsymbol{u}_{k+1:k+\ell-1}\right)$ and $\sC_{k+\ell|k+1,k+\ell-1}\left(u_{k+\ell}|\boldsymbol{u}_{k+1:k+\ell-1}\right)$
which are derived from the underlying copula $\sC_{1:d}$. We now show that these conditional distributions can be rewritten in terms of (derivatives) of pair-copula components; an attractive feature when performing likelihood inference. For this, so-called h-functions are instrumental. With $k = 1, \ldots, d-\ell$ and $\ell = 1, \dots, d-1$ they are defined by 
\begin{align}
&h_{k|k+\ell;k+1:k+\ell-1}\left(u|v\right)  \coloneqq \frac{\partial}{\partial v} \sC_{k,k+\ell;k+1:k+\ell-1}\left(u, v\right)\label{eq:hfunction1}\\
\text{and }\quad &h_{k+\ell|k;k+1:k+\ell-1}\left(v|u\right)  \coloneqq \frac{\partial}{\partial u}  \sC_{k,k+\ell;k+1:k+\ell-1}\left(u, v\right).\label{eq:hfunction2}
\end{align}
\noindent
Using these h-functions, \cite{Joe1997} shows that the following recursion formulas hold true:
\begin{align}
&  \sC_{k|k+1:k+\ell}\left(u_k|\boldsymbol{u}_{k+1:k+\ell}\right)\label{eq:Recursion1}\\
& \phantom{\partial} = h_{k|k+\ell;k+1:k+\ell-1}\{ \sC_{k|k+1:k+\ell-1}\left(u_k|\boldsymbol{u}_{k+1:k+\ell-1}\right)| \sC_{k+\ell|k+1:k+\ell-1}\left(u_{k+\ell}|\boldsymbol{u}_{k+1:k+\ell-1}\right)\},\notag\\
&  \sC_{k+\ell|k:k+\ell-1}\left(u_{k+\ell}|\boldsymbol{u}_{k:k+\ell-1}\right)\label{eq:Recursion2}\\
& \phantom{\partial} = h_{k+\ell|k;k+1:k+\ell-1}\{ \sC_{k+\ell|k+1:k+\ell-1}\left(u_{k+\ell}|\boldsymbol{u}_{k+1:k+\ell-1}\right)| \sC_{k|k+1:k+\ell-1}\left(u_{k}|\boldsymbol{u}_{k+1:k+\ell-1}\right)\}\notag.
\end{align}
The conditional distributions on the left in \eqref{eq:Recursion1} and \eqref{eq:Recursion2} are arguments of pair-copulas in tree $\mathcal{T}_{\ell+1}$. The h-functions on the right are linked to the copula $\sC_{k,k+\ell;k+1:k+\ell-1}$, which corresponds to tree $\mathcal{T}_{\ell}$. Further, the arguments of the h-functions are univariate conditional distributions themselves and can again be written as h-functions now linked to pair-copulas in $\mathcal{T}_{\ell-1}$. Thus, the arguments of the pair-copulas in $\mathcal{T}_{\ell}$  $(\ell=2,\ldots,d-1)$ can be evaluated using the pair-copulas in trees $\mathcal{T}_{\ell-1}$ up to $\mathcal{T}_1$. An illustration for $d=4$ is given in \autoref{Exa:DVineDensities} (continued).

For the remainder of the paper, we consider only one-parametric bivariate pair-copula families %(\autoref{table:Copulas}) 
and denote by $\theta_{k,k+\ell;k+1:k+\ell-1}$ the copula parameter corresponding to the bivariate copula density $\scd_{k,k+\ell;k+1:k+\ell-1}$. The collection of parameters of an ordered $d$-dimensional D-vine is then given by
%\begin{align*}
$\boldsymbol{\theta}_{1:d} = \lbrace \boldsymbol{\theta}_{k,k+\ell;k+1:k+\ell-1}|k = 1, \ldots, d-\ell, \ell = 1, \dots, d-1 \rbrace$.
%\end{align*}
Thus, the dependence structure among $d$ variables is described by $d(d-1)/2$ copula parameters. Unless unclear, we do not explicitly include the parameters in the notation of a D-vine copula.

\setcounter{exa}{0}
\begin{exa}[\textbf{continued}]
	In the ordered D-vine copula density corresponding to $\scd_{1:4}$ the univariate conditional distribution $\sC_{1|2:3}\left(u_1|\boldsymbol{u}_{2:3}\right)$ is the first argument of the pair-copula density $\scd_{1,4;2:3}$ that appears in tree $\mathcal{T}_3$ of the corresponding D-vine tree structure. 	
	Following \eqref{eq:Recursion1} with $k=1$ and $\ell=2$, we obtain
	%\begin{align*}
	$	\sC_{1|2:3}\left(u_1|\boldsymbol{u}_{2:3}\right) =
	h_{1|3;2}\{\sC_{1|2}\left(u_1|u_2\right)|\sC_{3|2}\left(u_3|u_2\right)\}$.
	%\end{align*}	
	Further, using \eqref{eq:Recursion1} with $k=1$ and $\ell=1$, resp. using \eqref{eq:Recursion2} with $k=2$ and $\ell=1$, it follows that
	%\begin{align*}
	$\sC_{1|2}\left(u_1|u_2\right) = h_{1|2}\left(u_1|u_2\right), \text{ resp. }
	\sC_{3|2}\left(u_3|u_2\right) = h_{3|2}\left(u_3|u_2\right)$.
	%\end{align*}
	Combining the two recursion steps, leads to
	%\begin{align*}
	$\sC_{1|2:3}\left(u_1|\boldsymbol{u}_{2:3}\right) =
	h_{1|3;2}\{h_{1|2}\left(u_1|u_2\right)|h_{3|2}\left(u_3|u_2\right)\}$.
	%\end{align*}
	Thus, while being derived from the underlying copula $\sC_{1:4}$ the function $\sC_{1|2:3}\left(u_1|\boldsymbol{u}_{2:3}\right)$ only depends on the parameters $\theta_{1,2}$ in $h_{1|2}\left(u_1|u_2\right)$, $\theta_{2,3}$ in $h_{3|2}\left(u_3|u_2\right)$ and $\theta_{1,3;2}$ in $h_{1|3;2}(\cdot|\cdot)$ of the corresponding ordered D-vine, i.e.\ pair-copula parameters enter an ordered D-vine in a sequential way.
\end{exa}

\begin{rem}\label{Remark:Clayton}
	A $d$-dimensional Clayton copula with parameter $\theta$ can be expressed as a D-vine copula (called a Clayton vine) in which all pair-copulas are Clayton with, in tree $\mathcal{T}_\ell$ ($\ell=1,\ldots,d-1$), all parameters %the same and 
	given by $\theta/\{\left(\ell-1\right)\theta + 1\}$ \citep{stoeber2013simplified}.
\end{rem}

\section{Methodology}\label{Sec:Methodology}
In this section, we develop several procedures to estimate, for gap times subject to induced dependent right-censoring, the parameters of Archimedean and D-vine copulas. %\textcolor{red}{Through discussion of the diverse estimation approaches and illustrating simulations} the specific aspects and challenges of recurrent event time data will be disclosed. 
We distinguish two %main 
approaches: one-stage parametric and two-stage semiparametric estimation. For D-vine based models, we consider a global and a sequential strategy within each approach. %Simulation based evidence supports the good finite sample performance of the proposed strategies and illustrates that D-vine copulas as a well-suited class to capture the naturally inherent serial dependence.

Recall that cluster $i$ ($i=1,\ldots,n$) contains $d_i \leq d$ observations. A natural approach to describe the dependence structure of unbalanced data is to choose the copula $\sC_{1:d}$ for the maximum available cluster size $d$ and to take the induced $d_i$-dimensional marginal copula $\sC_{1:d_i}$ for clusters of size $2\leq d_i < d$. Denote by $\scd_{1:d}$, resp. $\scd_{1:d_i}$, the copula density and by $\boldsymbol{\theta}_{1:d}$, resp. $\boldsymbol{\theta}_{1:d_i}$, the corresponding parameter vector. Recall that in case of $\sC_{1:d}$ being an Archimedean copula, $\boldsymbol{\theta}_{1:d}$ is one-dimensional and $\boldsymbol{\theta}_{1:d_i} = \boldsymbol{\theta}_{1:d}$ for all $2 \leq d_i < d$. For a D-vine, we have $\boldsymbol{\theta}_{1:d_i} \subset  \boldsymbol{\theta}_{1:d}$ where $\boldsymbol{\theta}_{1:d_i}$ contains $d_i(d_i-1)/2$ elements ($2\leq d_i < d$) (see %\eqref{eq:4dExampleDensity} in
\autoref{Exa:DVineDensities}).

\subsection{One-stage parametric estimation approach}\label{Sec:1stage}
A parametric form with parameters $\boldsymbol{\alpha} \coloneqq \left(\boldsymbol{\alpha}_1,\ldots,\boldsymbol{\alpha}_d\right)$ for the survival margins and parameters $\boldsymbol{\theta}_{1:d}$ for the copula is taken. As mentioned before the marginal data are subject to induced dependence between gap times and censoring times, such that standard univariate likelihood inference for each survival margin is no longer applicable. However, since the induced dependent right-censoring is a direct consequence of the association between subsequent gap times, joint estimation of the margins and the dependence structure resolves the issue.
%Based on \cite{Sklar59} and assuming $\left(G_1,\ldots,G_d\right)$ to be absolutely continuous we have
%\begin{align*}
%f\left(g_1,\ldots,g_d\right) = \scd\left(S_1\left(g_1;\boldsymbol{\alpha}_1\right),\ldots,S_d\left(g_d;\boldsymbol{\alpha}_d\right);\boldsymbol{\theta}\right)\cdot f_1\left(g_1,\boldsymbol{\alpha}_1\right)\cdot \ldots \cdot f_d\left(g_d,\boldsymbol{\alpha}_d\right),
%\end{align*}
%where $\scd$ is the copula density associated with the joint survival function of $\left(G_1,\ldots,G_d\right)$. Further,  $\boldsymbol{\alpha} \coloneqq \left(\boldsymbol{\alpha}_1,\ldots,\boldsymbol{\alpha}_d\right)$ collects all marginal parameters and $\boldsymbol{\theta}$ is the vector of copula parameters.

\subsubsection{Global likelihood inference}\label{Sec:1stage_global}
For cluster $i$ ($i=1,\ldots,n$) of size $d_i$ the observed data are
\begin{align*}
\left(y_{i,1},\ldots,y_{i,d_i-1},y_{i,d_i}\right) \coloneqq \{g_{i,1},\ldots,g_{i,d_i-1},\min\big(g_{i,d_i},c_i-\sum_{\ell=1}^{d_i-1}g_{i,\ell}\big)\}
\end{align*}
with censoring indicator $\delta_{i,d_i} = I\left(y_{i,d_i} = g_{i,d_i}\right)$. The loglikelihood contribution of cluster $i$ is defined by
\begin{align}
\ell_{i,d_i}^{\text{1stage}}&(y_{i,1},\ldots,y_{i,d_i},\delta_{i,d_i})\label{eq:loglik}\\
= & \quad \delta_{i,d_i} \log \left[ \scd_{1:d_i}\{S_1\left(y_{i,1};\boldsymbol{\alpha}_1\right),\ldots,S_{d_i}\left(y_{i,d_i};\boldsymbol{\alpha}_{d_i}\right);\boldsymbol{\theta}_{1:d_i}\}\cdot f_1\left(y_{i,1};\boldsymbol{\alpha}_1\right)\cdot \ldots \cdot f_{d_i}\left(y_{i,d_i};\boldsymbol{\alpha}_{d_i}\right) \right] \nonumber \\
& + (1-\delta_{i,d_i}) \log \left[ \left(-1\right)^{d_i-1} \frac{\partial^{d_i-1}}{\partial y_{i,1} \cdots \partial y_{i,{d_i-1}}}\sC_{1:d_i}\{S_1\left(y_{i,1};\boldsymbol{\alpha}_1\right),\ldots,S_{d_i}\left(y_{i,d_i};\boldsymbol{\alpha}_{d_i}\right),\boldsymbol{\theta}_{1:d_i}\} \right].\notag
\end{align}
The first term in \eqref{eq:loglik} covers the case of $y_{i,d_i}$ being a true gap time, i.e.\ the last event was observed, the second term in \eqref{eq:loglik} corresponds to the case of $y_{i,d_i}$ being a censored gap time. For D-vines an explicit expression of the loglikelihood contributions in terms of pair-copula components is given in Section 3 of the supplementary material.

For one-stage global parametric estimation, the loglikelihood for induced dependent right-censored gap time data is then given by
\begin{align}
\ell^{\text{1stage}}(\boldsymbol{\alpha},\boldsymbol{\theta}_{1:d}) = \displaystyle \sum_{i=1}^n \ell_{i,d_i}^{\text{1stage}}(y_{i,1},\ldots,y_{i,d_i},\delta_{i,d_i}),
\label{eq:1stageGlobal}
\end{align}
which is to be optimized with respect to the marginal parameters $\boldsymbol{\alpha}$ and the copula parameters $\boldsymbol{\theta}_{1:d}$. In case of an Archimedean copula all clusters contribute to the estimation of %the single parameter 
$\boldsymbol{\theta}_{1:d}$. For a D-vine, with $\ell = 1,\dots, d-1, k = \max\left(1,j-\ell+1\right), \ldots, d-\ell$ and $j = 1,\ldots,d$ estimation of the %pair-copula 
parameters $ \boldsymbol{{\theta}}_{k,k+\ell;k+1:k+\ell-1}$  is based only on clusters $i$ of size $d_i>j$.

\subsubsection{Sequential estimation approach}\label{Sec:1stage_seq}
The global one-stage parametric estimation approach is valid for both Archimedean and D-vine copulas. However, for data of maximum cluster size $d$ this requires, for D-vines, the joint estimation of $d(d-1)/2$ copula parameters together with the parameters of the $d$ survival margins. Given this high computational demand we aim for a more parsimonious estimation strategy by proceeding sequentially.

We use the fact that for each cluster size $2 \leq d_i < d$ ($i=1,\ldots,n$), the copula density $\scd_{1:d_i}$ is embedded within the copula density $\scd_{1:d}$ (see Example \ref{Exa:DVineDensities}). Proceeding sequentially means that the number of considered gap times increases stepwise from $1$ to $d$. In each step $j$ ($j=1,\ldots,d$) estimates obtained from previous steps are fixed such that only the marginal parameters of the $j$-th gap time and the pair-copulas incorporating the $j$-th gap time are to be estimated. The details are given in \autoref{1stageSequential}. Looking at \autoref{fig:D-vineGeneral}, estimation proceeds from left to right. For a model having e.g.\ two-parametric marginal models (like Weibull), the $(d(d-1)/2+2d)$-dimensional optimization problem is split into $d$ optimization problems, where in step $j$ $(j=1,\ldots,d)$ a $(j+1)$-dimensional optimization problem needs to be solved. A similar sequential procedure is used in \cite{barthel2017vine} for multivariate right-censored event time data in a balanced data setting and shows good finite sample performance.

\begin{algorithm}[h]
	{\fontsize{12pt}{18}\selectfont	
		\caption{Sequential left-right one-stage estimation.}\label{1stageSequential}
		\textbf{Input:} gap time data $(y_{i,1}, y_{i,2},\ldots,y_{i,d_i}, \delta_{i,1},\delta_{i,2}, \ldots, \delta_{i,d_i})$, $i=1,\ldots,n$, subject to induced dependent right-censoring ordered by decreasing cluster size.\\\noindent
		\textbf{Output:} parameter estimates $\boldsymbol{\hat{\alpha}} = (\boldsymbol{\hat{\alpha}}_1,\boldsymbol{\hat{\alpha}}_2,\ldots,\boldsymbol{\hat{\alpha}}_d)$ and $\boldsymbol{\hat{\theta}}_{1:d}$ with $d = \max\lbrace d_i| i=1,\ldots,n\rbrace$.
		\begin{algorithmic}[1]
			%\Procedure{}{}
			\State Set $d = \max\lbrace d_i| i=1,\ldots,n\rbrace$.
			\State Set $N=n_d+\ldots+n_1$.
			\State Maximize  $\sum_{i=1}^{N}\ell_{i,1}^{\text{1stage}}\left(y_{i,1},\delta_{i,1}\right)$ with respect to $\boldsymbol{\alpha}_1$. Denote the maximizer by $\boldsymbol{\hat{\alpha}}_1$.
			%\State Set $\boldsymbol{\alpha}_1 = \boldsymbol{\hat{\alpha}}_1$.
			\State Set $N=n_d+\ldots+n_2$.
			\State Fix $\boldsymbol{\alpha}_1$ at $\boldsymbol{\hat{\alpha}}_1$.
			\State Maximize $\sum_{i=1}^{N}\ell_{i,2}^{\text{1stage}}\left(y_{i,1},y_{i,2},\delta_{i,2}\right)$ with respect to $\boldsymbol{\alpha}_2$ and $\theta_{1,2}$. Denote the maximizers by $\boldsymbol{\hat{\alpha}}_2$ and $\hat{\theta}_{1,2}$.
			%\State Set $\boldsymbol{\alpha}_2 = \boldsymbol{\hat{\alpha}}_2$ and $\theta_{1,2} = \hat{\theta}_{1,2}$.			
			\For {$j=3,\ldots,d$}
			\IIf {$j < d$} Set $N = n_d+\ldots+n_j$. \EndIIf
			\IIf {$j = d$} Set $N = n_d$. \EndIIf
			\State  Fix $\boldsymbol{\alpha}_1,\ldots,\boldsymbol{\alpha}_{j-1}$  at
			$\boldsymbol{\hat{\alpha}}_1,\ldots,\boldsymbol{\hat{\alpha}}_{j-1}$ and $\boldsymbol{\theta}_{1:j-1}$ at $\boldsymbol{\hat{\theta}}_{1:j-1}$.
			\State	Maximize
			$\sum_{i=1}^{N}\ell_{i,j}^{\text{1stage}}\left(y_{i,1},\ldots,y_{i,j},\delta_{i,j}\right)$ with respect to $\boldsymbol{\alpha}_{j}$ and $\boldsymbol{\theta}_{1:j}\backslash\boldsymbol{\theta}_{1:j-1}$. The \hspace*{.5cm} estimates obtained in steps 1 to $j$ are $\boldsymbol{\hat{\alpha}}_1, \ldots, \boldsymbol{\hat{\alpha}}_j$, $\boldsymbol{\hat{\theta}}_{1:j}$.
			%\State Set $\boldsymbol{\alpha}_j = \boldsymbol{\hat{\alpha}}_j$ and $\boldsymbol{\theta}_{1:j}\backslash\boldsymbol{\theta}_{1:j-1}=\widehat{\boldsymbol{\theta}_{1:j}\backslash\boldsymbol{\theta}_{1:j-1}}$.
			\EndFor
		\end{algorithmic}
	}	
\end{algorithm}
%\newpage
\subsubsection{Illustrating simulations}
To investigate the finite sample performance of the suggested one-stage parametric approaches, a wide range of scenarios inspired by the asthma data is considered. The procedure for sampling induced right-censored unbalanced recurrent event time data is explained in Section 4 of the supplementary material. In each scenario, the results are based on 250 data sets. We consider samples of 250 and 500 clusters, each with a maximum size of 3. The gap times and the censoring times are assumed to follow a Weibull distribution, i.e.\
$S\left(g\right) = \exp\left(-\lambda g^\rho\right).$ The considered scale ($\lambda$) and shape ($\rho$) parameters were chosen such that data shows about 15\% or 30\% censoring. A third scenario is chosen such that it yields 30\% censoring but with censored observations mainly located at late time points (heavy tail - HT) (for details see Table 3 and Figure 1 in the supplementary material).
%Consequently, the tail of the Nelsen-Aalen or Kaplan-Meier survival estimate of the total times is heavily censored leading to an early leveling off away from zero (heavy tail - HT). 
It is assumed that gap 1 differs from gap 2 and gap 3, i.e.\ the latter are expected to be shorter, reflecting a weakening of the lungs after a first asthma attack.

The dependence between the gap times is modeled via a copula. %To provide evidence for the general good performance of the proposed estimation approach, 
First, we look at a simple three-dimensional (3d) Archimedean copula, where one single parameter controls the dependence between all gap times. We focus on an intermediate dependence strength as expressed by a Kendall's $\tau$ of $0.5$ and investigate the scenario of a Gumbel copula (upper tail-dependent) and a Clayton copula (lower tail-dependent). For D-vine copulas, we need to specify three values for Kendall's $\tau$ corresponding to the parameters $\boldsymbol{\theta}_{1:3} = \left(\theta_{1,2}, \theta_{2,3}, \theta_{1,3;2}\right)$. Using the one-to-one relationship between a Clayton copula and a Clayton vine (\autoref{Remark:Clayton}), we obtain for
$\tau = 0.5$ in the Clayton copula values of $\tau_{1,2} = \tau_{2,3} = 0.5$ and $\tau_{1,3;2} = 0.25$ in the corresponding Clayton vine. We consider scenarios where both pair-copulas in tree $\mathcal{T}_1$ are Clayton or Gumbel. The pair-copula in tree $\mathcal{T}_2$ is assumed to be Frank. Note that the considered Archimedean copulas and D-vine copulas describe different dependence structures. %\autoref{table:CopulaSimSettings_3d} summarizes all underlying copula models using C for Clayton, G for Gumbel and F for Frank.    		

The results are obtained under a correct specification of the marginal and the copula format. Since focus lies on dependence modeling, the results in \autoref{Table:3d_ClaytonGumbel_1stage_tau} are reported in terms of Kendall's $\tau$ values. Corresponding results for the copula parameters as well as for the marginal parameters are given in Table 5 to Table 7 of the supplementary material. On average and taking the standard deviation into account, all parameters are estimated close to their target value. %As expected,
Estimation improves with increasing sample size, but deteriorates with increasing censoring rate and -- under fixed censoring rate ($30\%$ and $30\%$ HT) -- if censored observations are mainly located at late time points. %the censoring density puts more weight early in time. %Note that 
Based on empirical mean and empirical standard deviation the results for the Clayton based copulas in the top panels of \autoref{Table:3d_ClaytonGumbel_1stage_tau} are somewhat less accurate than those for the Gumbel based copulas in the bottom panels. This is a direct consequence of the lower tail-property of a Clayton copula, which makes it %the latter 
more sensitive to right-censoring when modeling a survival function. For the D-vines, results of %the 
global and %the 
sequential optimization are quite similar, indicating that the latter is a valid alternative for the computationally more demanding global approach.

\begin{table}[H]
	\centering
	\small
	\caption{Simulation results using a \textbf{one-stage parametric estimation approach} for three-dimensional data. A \textbf{Clayton (3dC) copula} (top panel right) and a \textbf{Gumbel (3dG) copula} (bottom panel right) each with Kendall's $\tau = 0.5$ are considered. A \textbf{D-vine copula including Clayton copulas} (top panel left), \textbf{resp. Gumbel copulas} (bottom panel left), with $\tau_{1,2} = \tau_{2,3} = 0.5$ in $\mathcal{T}_1$ and a Frank (F) copula with $\tau_{1,3;2} = 0.25$ in $\mathcal{T}_2$ is considered. For the D-vine copulas \textbf{global and sequential likelihood estimation} is reported. The \textbf{empirical mean (empirical standard deviation) for the Kendall's $\tau$ estimates} are presented based on 250 replications and samples of size 250 and 500 affected by either 15\%, 30\% or heavy tail 30\% right-censoring.}
	\label{Table:3d_ClaytonGumbel_1stage_tau}
	\begin{tabular}{cccccccc}
		\midrule
		& & & & \multicolumn{3}{c}{D-vine copula model} & Archimedean copula\\
		&  &  &  & C; \ $\tau_{1,2}: 0.50$ & C; \ $\tau_{2,3}: 0.50$ & F; \ $\tau_{1,3;2}: 0.25$ & 3dC; \ $\tau: 0.50$ \\
		\midrule\midrule
		\multirow{12}{*} {\begin{sideways} parametric one-stage \end{sideways}} & \multirow{6}{*} {\begin{sideways} global \end{sideways}} & \multirow{2}{*} {15\%} & $250$ & 0.502 (0.035) & 0.505 (0.041) & 0.250 (0.052) & 0.503 (0.033) \\
		&  &  & $500$ & 0.501 (0.027) & 0.503 (0.029) & 0.251 (0.036) & 0.501 (0.024) \\
		\cmidrule{3-8}
		&  & \multirow{2}{*} {30\%} & $250$ & 0.501 (0.049) & 0.504 (0.058) & 0.250 (0.068) & 0.505 (0.041) \\
		&  &  & $500$ & 0.501 (0.033) & 0.505 (0.041) & 0.251 (0.046) & 0.501 (0.029) \\
		\cmidrule{3-8}
		&  & \multirow{2}{*} {30\% HT} & $250$ & 0.503 (0.059) & 0.502 (0.083) & 0.247 (0.080) & 0.504 (0.051) \\
		&  &  & $500$ & 0.503 (0.041) & 0.501 (0.057) & 0.249 (0.053) & 0.500 (0.038) \\
		\cmidrule{2-8}
		& \multirow{6}{*} {\begin{sideways} sequential \end{sideways}} & \multirow{2}{*} {15\%} & $250$ & 0.501 (0.036) & 0.505 (0.042) & 0.250 (0.052) & \\
		&  &  & $500$ & 0.501 (0.028) & 0.503 (0.030) & 0.251 (0.036) &\\
		\cmidrule{3-7}
		&  & \multirow{2}{*} {30\%} & $250$ & 0.500 (0.049) & 0.504 (0.059) & 0.250 (0.068) &\\
		&  &  & $500$ & 0.501 (0.033) & 0.505 (0.041) & 0.251 (0.046) &\\
		\cmidrule{3-7}
		&  & \multirow{2}{*} {30\% HT} & $250$ & 0.503 (0.060) & 0.502 (0.081) & 0.247 (0.080) &\\
		&  &  & $500$ & 0.503 (0.041) & 0.501 (0.057) & 0.249 (0.053) &\\
		\midrule\midrule
		&  &  &  & G; \ $\tau_{1,2}: 0.50$ & G; \ $\tau_{2,3}: 0.50$ & F; \ $\tau_{1,3;2}: 0.25$ & 3dG; \ $\tau: 0.5$ \\
		\midrule\midrule
		\multirow{12}{*} {\begin{sideways} parametric one-stage \end{sideways}} & \multirow{6}{*} {\begin{sideways} global \end{sideways}} & \multirow{2}{*} {15\%} & $250$ & 0.498 (0.033) & 0.501 (0.036) & 0.251 (0.050) & 0.500 (0.030)\\
		&  &  & $500$ & 0.499 (0.027) & 0.501 (0.027) & 0.250 (0.035) & 0.501 (0.021) \\
		\cmidrule{3-8}
		&  & \multirow{2}{*} {30\%} & $250$ & 0.501 (0.039) & 0.503 (0.044) & 0.249 (0.066) & 0.504 (0.035) \\
		&  &  & $500$ & 0.502 (0.030) & 0.504 (0.033) & 0.251 (0.046) & 0.502 (0.025)\\
		\cmidrule{3-8}
		&  & \multirow{2}{*} {30\% HT} & $250$ & 0.507 (0.042) & 0.507 (0.046) & 0.245 (0.077) & 0.504 (0.040) \\
		&  &  & $500$ & 0.506 (0.031) & 0.506 (0.035) & 0.248 (0.048) & 0.503 (0.029) \\
		\cmidrule{2-8}
		& \multirow{6}{*} {\begin{sideways} sequential \end{sideways}} & \multirow{2}{*} {15\%} & $250$ & 0.498 (0.034) & 0.501 (0.035) & 0.250 (0.050) &  \\
		&  &  & $500$ & 0.499 (0.027) & 0.501 (0.027) & 0.250 (0.035) &\\
		\cmidrule{3-7}
		&  & \multirow{2}{*} {30\%} & $250$ & 0.501 (0.039) & 0.503 (0.044) & 0.249 (0.066) &\\
		&  &  & $500$ & 0.500 (0.030) & 0.503 (0.033) & 0.252 (0.046) &\\
		\cmidrule{3-7}
		&  & \multirow{2}{*} {30\% HT} & $250$ & 0.499 (0.045) & 0.502 (0.047) & 0.247 (0.078) &\\
		&  &  & $500$ & 0.501 (0.033) & 0.502 (0.036) & 0.249 (0.049) &\\
		\midrule\midrule
	\end{tabular}
\end{table}
%\newpage
\subsection{Two-stage semiparametric estimation approach}
In spite of the good performance of the one-stage parametric approaches, %(marginal)
model flexibility is increased when using % a 
two-stage semiparametric estimation. % approach. 
In stage 1, the survival margins ($S_j$) are estimated nonparametrically ($\hat{S}_j$). In stage 2, the pseudo-data $\hat{u}_{i,j}=\hat{S}_j(y_{i,j})$ are used to estimate the copula parameters via likelihood optimization. This %two-stage 
approach goes back to \cite{Shih95}. They considered %the case of 
bivariate survival data subject to independent right-censoring. Extensions %of their idea 
to clustered survival data of dimension more than two are in \cite{Chen2010129}, \cite{Geerdens2014} and \cite{barthel2017vine}. They all use Kaplan-Meier or Nelson-Aalen estimators to obtain %the 
marginal estimates. For gap time data subject to induced dependent right-censoring these standard %nonparametric 
estimators are no longer consistent \citep{cook2007statistical, meyer2015bayesian} and nonparametric alternatives %  estimators 
are needed.

\subsubsection{Marginal modeling}\label{Sec:2StageMargins}
In case of induced dependent right-censoring  \cite{UnaAlvarezMachado2008nonparaCensGap} proposed a consistent nonparametric estimator for the survival margins. As an estimate for the joint distribution $F$ of $\left(G_1,\ldots,G_d\right)$ they
define
\begin{align*}
\widehat{F}(g_1, \ldots, g_{d}) = \displaystyle \sum_{i=1}^n W^{\text{KM}}_i I(y_{i,1} \leq g_1, \ldots, y_{i,d} \leq g_{d}),
\end{align*}
where $W^{\text{KM}}_i$ is the jump of the Kaplan-Meier estimate obtained from the observations
$\left(\tilde{y}_i, \delta_{i,d_i}\right)$ with $\tilde{y}_i$ the total follow-up time for cluster $i$, i.e.\
%\begin{align*}
$\tilde{y}_i = \min\left(c_i, t_{i,d_i}\right) = \sum_{j=1}^{d_i}y_{i,j}$ $(i = 1, \ldots, n)$.
%\end{align*}
%The Kaplan-Meier weights are given by
%\begin{align*}
%W_i = \frac{\delta_i}{n - R_i + 1} \prod_{\ell = 1}^{i-1} \left(1- \frac{\delta_{\ell}}{n - R_{\ell} + 1}\right) \text{ with } R_i = \textup{Rank}\left(\tilde{Y_i}\right).
%\end{align*}
An estimate for the $j$th marginal survival function is then given by
\begin{align*}
\widehat{S}^{\text{KM}}_j(g) = \displaystyle 1 - \sum_{i = 1}^{n} W^{\text{KM}}_i I(y_{i,j} \leq g), \quad j=1,\ldots,d.
\end{align*}
Note that the Kaplan-Meier estimator drops to zero, whenever the largest observed total time is a true event. After applying the probability integral transform this results in a zero value for the corresponding copula data value. To avoid numerical difficulties in the likelihood maximization, we propose to modify the \cite{UnaAlvarezMachado2008nonparaCensGap} estimator as follows: instead of using the Kaplan-Meier estimate for the survival function of the observed total times, we apply the Nelson-Aalen estimator to obtain a nonparametric estimate for the cumulative hazard function $\Lambda(t)$ of the total times. The corresponding survival jumps $W_i^{\text{NA}}$ ($i=1,\ldots,n$) are then obtained via the exponential transformation $\exp(-\Lambda(t))$. Following this approach, no zero values can occur. Given the unbalanced data setting, the pseudo copula data then are:%Recall that, due to the unbalanced cluster setting, cluster $i$ contains no data for $j=d_i + 1,\ldots,d$. Consequently, the pseudo copula data are:
\begin{align}\label{eq:PseudoCopData}
\widehat{u}_{i,j} = \widehat{S}^{\text{NA}}_j(y_{i,j}) = \displaystyle 1 \ \ - \hspace*{-.5cm} \sum_{\ell \in \lbrace i|1 \leq i \leq n,  d_i \geq j \rbrace} \hspace*{-.5cm} W^{\text{NA}}_\ell I(y_{\ell,j} \leq y_{i,j}), \quad i=1,\ldots,n \text{ and } j=1,\ldots,d_i.
\end{align}
%Finally, note that for non-censored data, $W_i = n^{-1}$ ($i = 1, \ldots, n$) and $\widehat{F}$ coincides with the joint empirical distribution function.

\subsubsection{Global likelihood inference}\label{Sec:2stage_global}
Based on the pseudo copula data in \eqref{eq:PseudoCopData}, the copula parameters $\boldsymbol{\theta}_{1:d}$ are estimated using maximum likelihood optimization. As for one-stage parametric estimation, the presence of right-censoring needs to be taken into account. % within the copula likelihood expression.
For cluster $i$ of size $d_i$ ($i=1,\ldots,n$) the loglikelihood contribution then equals
\begin{align}\label{eq:2stage_contribution}
&\ell_{i,d_{i}}^{\text{2stage}}\left(\widehat{u}_{i,1},\ldots,\widehat{u}_{i,d_i}, \delta_{i,d_i}\right)\\
&= \delta_{i,d_i} \log \left[\scd_{1:d_i}\{\widehat{u}_{i,1},\ldots,\widehat{u}_{i,d_i};\boldsymbol{\theta}_{1:d_i}\} \right] + \left(1-\delta_{i,d_i}\right) \log \left[\frac{\partial^{d_i-1}}{\partial \widehat{u}_{i,1}\cdots \partial \widehat{u}_{i,d_i-1}}\sC_{1:d_i}\{\widehat{u}_{i,1},\ldots,\widehat{u}_{i,d_i};\boldsymbol{\theta}_{1:d_i}\}\right].\notag
\end{align}
The loglikelihood function for induced dependent right-censored gap time data in a two-stage estimation approach, which needs to be optimized with respect to $\boldsymbol{\theta}_{1:d}$, is then given by
\begin{align}
\ell^\text{2stage}\left(\boldsymbol{\theta}_{1:d}\right)  = \sum_{i=1}^{n} \ell_{i,d_i}^{\text{2stage}}\left(\widehat{u}_{i,1},\ldots,\widehat{u}_{i,d_i}, \delta_{i,d_i}\right).
%\label{eq:likli2stage}
\end{align}
%which needs to be optimized with respect to the copula parameters $\boldsymbol{\theta}_{1:d}$.

\newpage
\subsubsection{Sequential estimation approach}\label{Sec:2stage_seq}

For D-vines, also in case of two-stage estimation, a flexible sequential procedure for likelihood maximization is feasible. It relies on the recursive nature of the arguments of the pair-copulas, which can be written as h-functions corresponding to pair-copulas of lower tree levels (\autoref{DVineCops}). In \autoref{fig:D-vineGeneral}, estimation proceeds from top to bottom. First, all pair-copula parameters in $\mathcal{T}_1$ are estimated separately. Based on the fitted pair-copulas, the arguments needed in $\mathcal{T}_2$ are calculated by application of the corresponding h-functions. Using the so-obtained pseudo data all pair-copula parameters in $\mathcal{T}_2$ can be estimated separately, etc. The procedure has been developed for %and applied 
complete data \citep{aas2009pair,dissmann2013selecting}. In case of right-censoring, an extra challenge arises: from tree $\mathcal{T}_2$ on estimation is no longer based on the observed copula data, but on pseudo data, namely univariate conditional distribution functions, which are evaluated at the observed copula data. For these pseudo observations censoring indicators need to be defined. Recall that within a cluster only the last gap time can be right-censored. Given the construction of an ordered D-vine, the value on the copula scale corresponding to the last gap-time can only occur as conditioned variable in the univariate conditional functions. Further, the latter are monotonously increasing in their conditioned argument. Hence, the pseudo observations inherit the censoring status of their observed conditioned variable. Detailed steps %of this sequential proceeding for right-censored recurrent data 
are given in \autoref{2stageSequential}. By doing so, the $d$-dimensional optimization %estimation problem 
is split into $d(d-1)/2$ bivariate ones and the estimation of a high-dimensional D-vine becomes tractable and computationally easier. For complete and balanced data \cite{haff2013parameter}, \cite{schepsmeier2014derivatives} and \cite{stober2013estimating} give asymptotic properties of this approach. %Using D-vine copula models,
\cite{killiches2017d} model unbalanced recurrent data without censoring.

\begin{algorithm}	
	{\fontsize{12pt}{18}\selectfont
		\caption{Sequential top-down two-stage estimation.}\label{2stageSequential}
		\textbf{Input:} gap time data $(y_{i,1}, y_{i,2},\ldots,y_{i,d_i}, \delta_{i,1},\delta_{i,2}, \ldots, \delta_{i,d_i})$, $i=1,\ldots,n$, subject to induced dependent right-censoring ordered by decreasing cluster size.\\
		\noindent
		\textbf{Output:} parameter estimates $\boldsymbol{\hat{\theta}}_{1:d}$ with $d = \max\lbrace d_i|i= 1,\ldots,n\rbrace$.
		\begin{algorithmic}[1]
			\State Set $d = \max\lbrace d_i|i= 1,\ldots,n\rbrace$.
			\For {$j=1,\ldots,d$}
			\IIf {$j < d$} Set $N = n_d+\ldots+n_j$. \EndIIf
			\IIf {$j = d$} Set $N = n_d$. \EndIIf
			\State With $(y_{i,j},\delta_{i,j})$, $i=1,\ldots,N$, estimate $\hat{S}_j$ nonparametrically (\autoref{Sec:2StageMargins}).
			\State Obtain pseudo-copula data $(\hat{u}_{i,j},\delta_{i,j})$, $i=1,\ldots,N$, by $\hat{u}_{i,j} = \hat{S}_j(y_{i,j})$.
			\EndFor
			\For {$k=1,\ldots,d-1$}
			\IIf {$k < d-1$} Set $N = n_d+\ldots+n_{k+1}$. \EndIIf
			\IIf {$k = d-1$} Set $N = n_d$. \EndIIf
			\State Select a copula family for $\scd_{k,k+1}$ and with $(\hat{u}_{i,k},\hat{u}_{i,k+1},\delta_{i,k+1})$,  $i=1,\ldots,N$, maximize 
			%\begin{align*}
			$\hspace*{.5cm} \sum_{i=1}^{N}\ell_{i,2}^{\text{2stage}}\left(\hat{u}_{i,k},\hat{u}_{i,k+1},\delta_{i,k+1}\right)$
			%\end{align*}
			with respect to $\theta_{k,k+1}$.
			\State Using the fitted copula $\sC_{k,k+1}(\cdot,\cdot; \hat{\theta}_{k,k+1})$ apply the h-functions \eqref{eq:hfunction1} and \eqref{eq:hfunction2} to calculate  $\hspace*{.65cm} \sC_{k|k+1}(\hat{u}_{i,k}|\hat{u}_{i,k+1})$ and $\sC_{k+1|k}(\hat{u}_{i,k+1}|\hat{u}_{i,k})$, $i=1,\ldots,N$.
			\EndFor	
			\For {$\ell=2,\ldots,d-1$}
			\For {$k=1,\ldots,d-\ell$}
			\IIf {$k < d-\ell$} Set $N = n_d+\ldots+n_{k+\ell}$. \EndIIf
			\IIf {$k = d-\ell$} Set $N = n_d$. \EndIIf
			\State For $i=1,\ldots,N$, set $u_{i} = \sC_{k|k+1:k+\ell-1}(\hat{u}_{i,k}|\boldsymbol{\hat{u}}_{i,k+1:k+\ell-1})$\newline \hspace*{4.15cm} and $v_{i} = \sC_{k+\ell|k+1:k+\ell-1}(\hat{u}_{i,k+\ell}|\boldsymbol{\hat{u}}_{i,k+1:k+\ell-1})$.\newline
			\hspace*{1.2cm }Set censoring indicator $\delta_i$ corresponding to $v_{i}$ to\newline \hspace*{4.95cm}
			%\begin{align*}
			$\delta_i = I(d_i > k+\ell) + I(d_i = k+\ell)\delta_{i,k+\ell}.$
			%\end{align*}
			\State Select a copula family for $\scd_{k,k+\ell;k+1:k+\ell-1}$ and with $(u_{i},v_{i}, \delta_{i})$, $i=1,\ldots,N$,\newline  $\hspace*{1.2cm}\text{maximize}$
			%\begin{align*}
			$\sum_{i=1}^{N}\ell_{i,2}^{\text{2stage}}\left(u_{i},v_{i},\delta_{i}\right)$
			%\end{align*}
			with respect to $\theta_{k,k+\ell;k+1:k+\ell-1}$.
			\State Using the fitted copula $\sC_{k,k+\ell;k+1:k+\ell-1}(\cdot,\cdot;\hat{\theta}_{k,k+\ell;k+1:k+\ell-1})$ apply the h-functions \hspace*{1.15cm} \eqref{eq:hfunction1} and \eqref{eq:hfunction2} to calculate $\sC_{k|k+1:k+\ell}(\hat{u}_{i,k}|\boldsymbol{\hat{u}}_{i,k+1:k+\ell})=h_{k|k+\ell;k+1:k+\ell-1}(u_i|v_i)$ and  \hspace*{1.25cm} $\sC_{k+\ell|k:k+\ell-1}(\hat{u}_{i,k+\ell}|\boldsymbol{\hat{u}}_{i,k:k+\ell-1})=h_{k+\ell|k;k+1:k+\ell-1}(v_i|u_i)$, $i=1,\ldots,N$.
			\EndFor	
			\EndFor					
		\end{algorithmic}
	}	
\end{algorithm}

\subsubsection{Illustrating simulations}
To investigate the finite sample performance of %both 
the global and %the 
sequential two-stage semiparametric estimation approach, the same simulation settings as for %the 
one-stage parametric estimation 
%approaches 
are used. 
%Note that 
Now, also %In addition %to sample sizes 250 and 500, 
a sample size of 1000 is considered.

The obtained results for Kendall's $\tau$ are in \autoref{Table:3d_ClaytonGumbel_2stage_tau}, while those for the copula parameters are in Table 8 of the supplementary material. Results are calculated under the assumption of a correct copula format. Compared to one-stage parametric estimation, some additional uncertainty is induced by nonparametric marginal modeling. For $15\%$ and $30\%$ censoring, the Kendall's $\tau$ and parameter estimates are (on average) close to their target values. However, for $30\%$ censoring with a heavy tail, estimation is off, i.e.\ the empirical mean estimates are too high and the empirical standard deviations are larger. Increasing the sample size slightly improves estimation. Clearly, in a two-stage estimation approach not only the amount of censoring but also the censoring position plays a role. In case of many large censored total times, the Nelsen-Aalen estimator for the survival function of the total times (usually) levels off away from zero. As such, the estimated marginal survival functions do not drop sufficiently low to zero, which in turn affects the copula data and hence distorts estimation. Note that this issue did not appear in the one-stage estimation approaches (\autoref{Sec:1stage}). Consequently, we recommend to use the latter whenever the tail of the Nelsen-Aalen estimate for the survival function of the total times is heavily affected by censoring (leveling off away from zero). The censoring effect is more manifest for a Clayton copula and a D-vine with Clayton parts.% (top panels in \autoref{Table:3d_ClaytonGumbel_2stage_tau}). %Given that a Clayton copula is lower tail-dependent, the latter is to be expected. The simulation results also indicate that, for D-vine copulas, %the sequential strategy is a good alternative for the computationally more challenging global approach.

\begin{table}[H]
	\centering
	\small
	\caption{Simulation results using a \textbf{two-stage semiparametric estimation approach} for three-dimensional data. A \textbf{Clayton (3dC) copula} (top panel right) and a \textbf{Gumbel (3dG) copula} (bottom panel right) with Kendall's $\tau = 0.5$ is considered. A \textbf{D-vine copula including Clayton copulas} (top panel left), \textbf{resp. Gumbel copulas} (bottom panel left), with $\tau_{1,2} = \tau_{2,3} = 0.5$ in $\mathcal{T}_1$ and a Frank (F) copula with $\tau_{1,3;2} = 0.25$ in $\mathcal{T}_2$ is considered. For the D-vine copulas \textbf{global and sequential likelihood estimation} is reported. The \textbf{empirical mean (empirical standard deviation) for the Kendall's $\tau$ estimates} are presented based on 250 replications and samples of size 250, 500 and 1000 affected by either 15\%, 30\% or heavy tail 30\% right-censoring.}
	\label{Table:3d_ClaytonGumbel_2stage_tau}
	\begin{tabular}{cccccccc}
		\midrule
		\small
		& & & & \multicolumn{3}{c}{D-vine copula model} & Archimedean copula \\
		&  &  &  & C; \ $\tau_{1,2}: 0.50$ & C; \ $\tau_{2,3}: 0.50$ & F; \ $\tau_{1,3;2}: 0.25$ & 3dC; \ $\tau: 0.5$ \\
		\midrule\midrule
		\multirow{18}{*} {\begin{sideways} semiparametric two-stage \end{sideways}} & \multirow{9}{*} {\begin{sideways} global \end{sideways}} & \multirow{3}{*} {15\%} & $250$ & 0.495 (0.042) & 0.497 (0.046) & 0.253 (0.053) & 0.496 (0.040)\\
		&  &  & $500$ & 0.496 (0.034) & 0.498 (0.035) & 0.253 (0.037) &0.497 (0.028)\\
		&  &  & $1000$ & 0.498 (0.023) & 0.498 (0.023) & 0.251 (0.025) & 0.499 (0.019)\\
		\cmidrule{3-8}
		&  & \multirow{3}{*} {30\%} & $250$ & 0.504 (0.071) & 0.505 (0.074) & 0.249 (0.070) & 0.504 (0.061) \\
		&  &  & $500$ & 0.501 (0.044) & 0.504 (0.049) & 0.253 (0.047) & 0.498 (0.042)\\
		&  &  & $1000$ & 0.503 (0.035) & 0.502 (0.037) & 0.251 (0.035) & 0.498 (0.031)\\
		\cmidrule{3-8}
		&  & \multirow{3}{*} {30\% HT} & $250$ & 0.558 (0.110) & 0.556 (0.101) & 0.245 (0.088) & 0.546 (0.094) \\
		&  &  & $500$ & 0.558 (0.089) & 0.549 (0.092) & 0.246 (0.061) & 0.543 (0.081)\\
		&  &  & $1000$ & 0.551 (0.069) & 0.536 (0.072) & 0.242 (0.042) &0.538 (0.066)\\
		\cmidrule{2-8}
		& \multirow{9}{*} {\begin{sideways} sequential  \end{sideways}} & \multirow{3}{*} {15\%} & $250$ & 0.495 (0.042) & 0.497 (0.045) & 0.252 (0.053) &\\
		&  &  & $500$ & 0.497 (0.034) & 0.498 (0.035) & 0.253 (0.037) &\\
		&  &  & $1000$ & 0.498 (0.023) & 0.498 (0.023) & 0.251 (0.025) &\\
		\cmidrule{3-7}
		&  & \multirow{3}{*} {30\%} & $250$ & 0.504 (0.070) & 0.511 (0.071) & 0.246 (0.069) &\\
		&  &  & $500$ & 0.502 (0.043) & 0.510 (0.048) & 0.251 (0.046) &\\
		&  &  & $1000$ & 0.503 (0.035) & 0.507 (0.035) & 0.250 (0.035) &\\
		\cmidrule{3-7}
		&  & \multirow{3}{*} {30\% HT} & $250$ & 0.558 (0.107) & 0.564 (0.089) & 0.242 (0.086) &\\
		&  &  & $500$ & 0.557 (0.087) & 0.556 (0.083) & 0.243 (0.060) &\\
		&  &  & $1000$ & 0.550 (0.067) & 0.544 (0.064) & 0.240 (0.041) &\\
		\midrule\midrule
		&  &  &  & G; \ $\tau_{1,2}: 0.50$ & G; \ $\tau_{2,3}: 0.50$ & F; \ $\tau_{1,3;2}: 0.25$ & 3dG; \ $\tau: 0.5$ \\
		\midrule\midrule
		\multirow{18}{*} {\begin{sideways} semiparametric two-stage \end{sideways}} & \multirow{9}{*} {\begin{sideways} global \end{sideways}} & \multirow{3}{*} {15\%} & $250$ & 0.501 (0.038) & 0.506 (0.039) & 0.251 (0.051) & 0.504 (0.033)\\
		&  &  & $500$ & 0.503 (0.028) & 0.505 (0.028) & 0.251 (0.036) & 0.503 (0.022)\\
		&  &  & $1000$ & 0.501 (0.019) & 0.502 (0.020) & 0.250 (0.025) & 0.504 (0.015)\\
		\cmidrule{3-8}
		&  & \multirow{3}{*} {30\%} & $250$ & 0.503 (0.049) & 0.511 (0.049) & 0.249 (0.067) & 0.511 (0.042) \\
		&  &  & $500$ & 0.507 (0.033) & 0.510 (0.037) & 0.254 (0.047) & 0.508 (0.028)\\
		&  &  & $1000$ & 0.505 (0.024) & 0.505 (0.025) & 0.249 (0.037) & 0.506 (0.020)\\
		\cmidrule{3-8}
		&  & \multirow{3}{*} {30\% HT} & $250$ & 0.535 (0.065) & 0.531 (0.058) & 0.251 (0.087) & 0.534 (0.063) \\
		&  &  & $500$ & 0.536 (0.053) & 0.527 (0.052) & 0.253 (0.061) & 0.531 (0.048)\\
		&  &  & $1000$ & 0.533 (0.039) & 0.523 (0.035) & 0.249 (0.044) & 0.526 (0.039)\\
		\cmidrule{2-8}
		& \multirow{9}{*} {\begin{sideways} sequential  \end{sideways}} & \multirow{3}{*} {15\%} & $250$ & 0.501 (0.038) & 0.505 (0.039) & 0.250 (0.051) &\\
		&  &  & $500$ & 0.503 (0.028) & 0.504 (0.028) & 0.251 (0.036) &\\
		&  &  & $1000$ & 0.501 (0.020) & 0.501 (0.021) & 0.250 (0.025) &\\
		\cmidrule{3-7}
		&  & \multirow{3}{*} {30\%} & $250$ & 0.504 (0.049) & 0.515 (0.050) & 0.247 (0.067) &\\
		&  &  & $500$ & 0.507 (0.033) & 0.513 (0.037) & 0.253 (0.046) &\\
		&  &  & $1000$ & 0.505 (0.024) & 0.507 (0.025) & 0.249 (0.036) &\\
		\cmidrule{3-7}
		&  & \multirow{3}{*} {30\% HT} & $250$ & 0.535 (0.065) & 0.531 (0.058) & 0.251 (0.087) &\\
		&  &  & $500$ & 0.537 (0.053) & 0.533 (0.050) & 0.251 (0.059) &\\
		&  &  & $1000$ & 0.534 (0.039) & 0.528 (0.033) & 0.247 (0.043) &\\
		\midrule\midrule
	\end{tabular}
\end{table}

\subsection{Overview and guidelines for the different estimation strategies}
Based on the findings of the illustrating simulations with all four estimation strategies \autoref{fig:FlowchartEstMethods} serves as a guideline to decide for the best suitable model approach given specific data characteristics. It also gives an overview of the four proposed estimation techniques.

\tikzstyle{ClassicalNode} = [rectangle, fill = gray!7, draw = gray, text = black, align = center]
\tikzstyle{TreeLabels} = [draw = none, fill = none, text = black, font = \bf]
\tikzstyle{DummyNode}  = [draw = none, fill = none, text = white]
\tikzset{
	block/.style = {rectangle, draw = gray, text width=14.0cm,
		line width=.5pt,minimum height=1.25cm},
	block_half/.style = {block, text width=5cm},
}
\newcommand{\yshift}{-.5cm}
\newcommand{\yshiftlabel}{-0.1cm}
\newcommand{\labelsize}{\small}
%\begin{landscape}
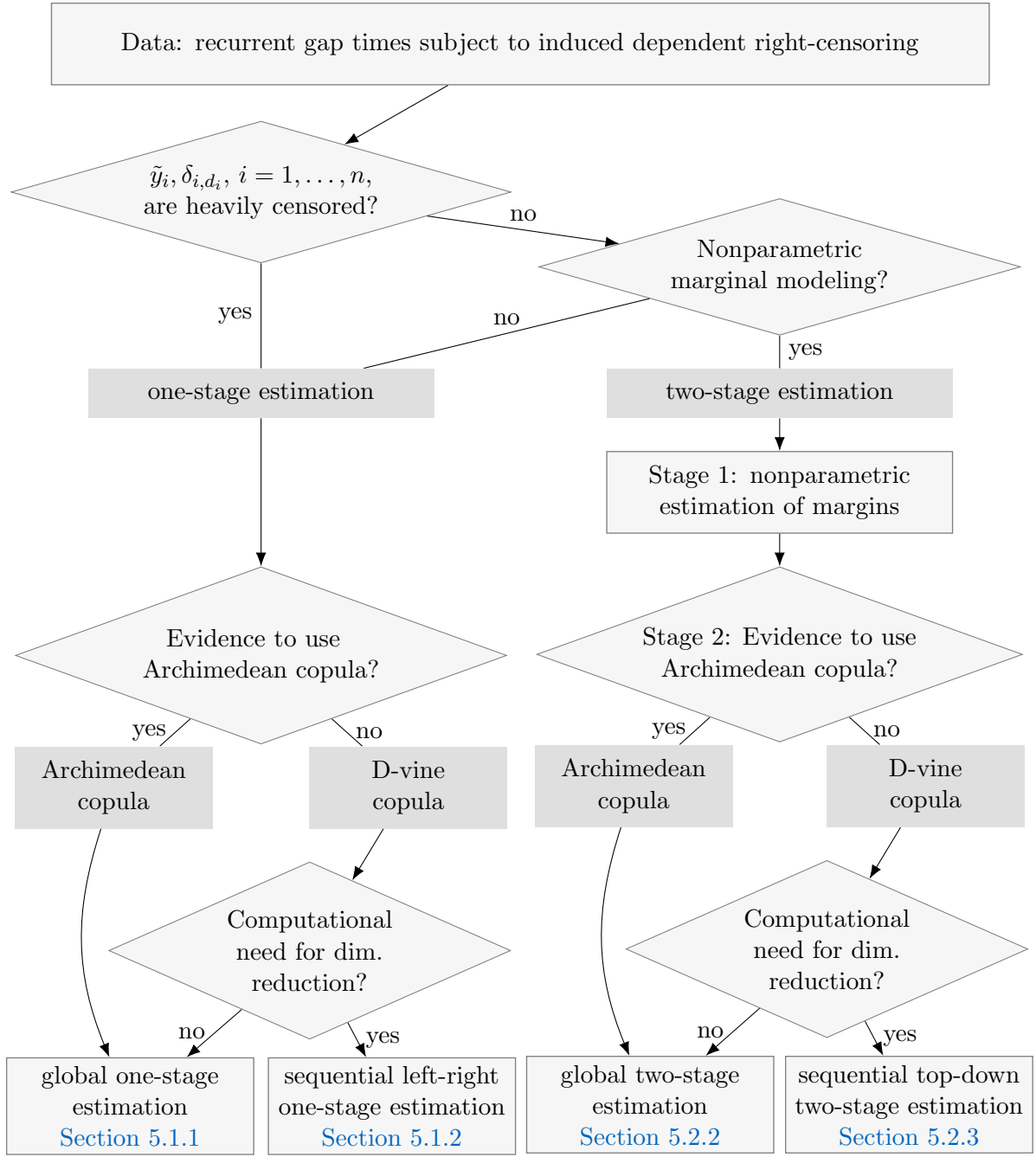
\begin{figure}[H]
	\centering
	\begin{tikzpicture}	[every node/.style = ClassicalNode, node distance =1cm, font = \normalsize]
	\node [block, anchor=north east] (Data) {Data: recurrent gap times subject to induced dependent right-censoring};
	\node[DummyNode, below = of Data, yshift = \yshift] (Dummy){};
	
	\node[diamond, aspect = 3.5, white, left = of Dummy, xshift = +1cm] (Dummy1Stage_0){\textcolor{white}{$\tilde{y}_{i},\delta_{i,d_i}$, $i=1,\ldots,n$,}\\ \textcolor{white}{are heavily censored?}};
	\node[diamond, aspect = 3.5, white, right = of Dummy, xshift = -1cm] (Dummy2Stage_1){\textcolor{white}{$\tilde{y}_{i},\delta_{i,d_i}$, $i=1,\ldots,n$,}\\ \textcolor{white}{are heavily censored?}};
	
	\node [diamond, white, aspect = 3.5, below = of Dummy1Stage_0, yshift = \yshift + 2.5cm] (Dummy1Stage_1) {\textcolor{white}{Nonparametric}\\ \textcolor{white}{marginal modeling?}};
	\node [diamond, aspect = 3.5, below = of Dummy2Stage_1, yshift = \yshift + 2.5cm] (nonparametric) {Nonparametric\\ marginal modeling?};
	
	\node [diamond, aspect = 3.5, left = of Dummy, xshift = +1cm] (censoring) { $\tilde{y}_{i},\delta_{i,d_i}$, $i=1,\ldots,n$,\\ are heavily censored?};
	
	\node [block_half, minimum height = .75cm, draw = none, fill = gray!25, below = of Dummy1Stage_1, yshift = \yshift + 1cm] (1stage) {one-stage estimation};
	\node [block_half, minimum height = .75cm, draw = none, fill = gray!25,below = of nonparametric, yshift = \yshift + 1cm] (2stage) {two-stage estimation};
	
	\node [block_half, white, below = of 1stage, yshift = \yshift + 1cm] (Dummy1Stage_2) {\textcolor{white}{Stage 1: nonparametric estimation of margins}};
	\node [block_half, below = of 2stage, yshift = \yshift + 1cm] (2stage1) {Stage 1: nonparametric estimation of margins};
	\node [diamond, aspect = 2.75, below = of 2stage1, yshift = \yshift + 1cm] (Arch2) {Stage 2: Evidence to use\\ Archimedean copula?};
	\node [diamond, aspect = 2.75, below = of Dummy1Stage_2, yshift = \yshift + 1cm] (Arch1) {\phantom{Sta} Evidence to use \phantom{ge 1:}\\ Archimedean copula?};
	\node [block_half, draw = none, fill = gray!25, text width = 2.75cm, below left = of Arch1, yshift = \yshift+.5cm, xshift = +1.85cm] (1_Arch) {Archimedean\\
		copula};
	\node [block_half, draw = none, fill = gray!25, text width = 2.75cm, below right = of Arch1, yshift = \yshift+.5cm, xshift = -1.85cm] (1_vine) {D-vine\\ copula};
	\node [block_half, draw = none, fill = gray!25, text width = 2.75cm, below left = of Arch2, yshift = \yshift+.5cm, xshift = +1.85cm] (2_Arch) {Archimedean\\ copula};
	\node [block_half, draw = none, fill = gray!25, text width = 2.75cm, below right = of Arch2, yshift = \yshift+.5cm, xshift = -1.85cm] (2_vine) {D-vine\\ copula};
	\node [diamond, aspect = 2.25, below = of 1_vine, yshift = \yshift + 1cm, xshift = -1.5cm] (1_dim) {Computational\\ need for dim.\\ reduction?};
	\node [diamond, aspect = 2.25, below = of 2_vine, yshift = \yshift + 1cm, xshift = -1.5cm] (2_dim) {Computational\\ need for dim.\\ reduction?};
	\node [block_half, text width = 3.5cm, below = of 1_Arch, yshift = \yshift-2cm, xshift = +.25cm] (1_global) {global one-stage estimation\\ \autoref{Sec:1stage_global}};
	\node [block_half, text width = 3.5cm, below = of 1_vine, yshift = \yshift-2cm, xshift = -.25cm] (1_seq) {sequential left-right one-stage estimation\\ \autoref{Sec:1stage_seq}};	
	\node [block_half, text width = 3.5cm, below = of 2_Arch, yshift = \yshift-2cm, xshift = +.25cm] (2_global) {global two-stage estimation\\ \autoref{Sec:2stage_global}};
	\node [block_half, text width = 3.5cm, below = of 2_vine, yshift = \yshift-2cm, xshift = -.25cm] (2_seq) {sequential top-down two-stage estimation\\ \autoref{Sec:2stage_seq}};
	
	\draw[-{Latex[scale=1.25]}] (Data) to node[draw=none, fill = none] {} (censoring);
	\draw[-] (censoring) to node[draw=none, fill = none, left] {yes} (1stage);
	\draw[-{Latex[scale=1.25]}] (censoring) to node[draw=none, fill = none, above] {no} (nonparametric);
	\draw[-] (nonparametric) to node[draw=none, fill = none, right] {yes} (2stage);
	\draw[-] (nonparametric) to node[draw=none, fill = none, above] {no} (1stage);
	\draw[-{Latex[scale=1.25]}] (1stage) to node[draw=none, fill = none, above] {} (Arch1);
	\draw[-{Latex[scale=1.25]}] (2stage) to node[draw=none, fill = none, above] {} (2stage1);
	\draw[-{Latex[scale=1.25]}] (2stage1) to node[draw=none, fill = none, above] {} (Arch2);
	\draw[-] (Arch1) to node[draw=none, fill = none, left] {yes} (1_Arch);
	\draw[-] (Arch1) to node[draw=none, fill = none, right] {no} (1_vine);
	\draw[-] (Arch2) to node[draw=none, fill = none, left] {yes} (2_Arch);
	\draw[-] (Arch2) to node[draw=none, fill = none, right] {no} (2_vine);
	\path[-{Latex[scale=1.25]}] (1_Arch) edge [bend right = 20] (1_global);
	\path[-{Latex[scale=1.25]}] (2_Arch) edge [bend right = 20] (2_global);
	\draw[-{Latex[scale=1.25]}] (1_vine) to node[draw=none, fill = none, left] {} (1_dim);
	\draw[-{Latex[scale=1.25]}] (2_vine) to node[draw=none, fill = none, right] {} (2_dim);
	\draw[-{Latex[scale=1.25]}] (1_dim) to node[draw=none, fill = none, right] {yes} (1_seq);
	\draw[-{Latex[scale=1.25]}] (2_dim) to node[draw=none, fill = none, right] {yes} (2_seq);	
	\draw[-{Latex[scale=1.25]}] (1_dim) to node[draw=none, fill = none, left] {no} (1_global);
	\draw[-{Latex[scale=1.25]}] (2_dim) to node[draw=none, fill = none, left] {no} (2_global);		
	\end{tikzpicture}
	\caption{Overview and guidelines for usage of the different estimation strategies.}
	\label{fig:FlowchartEstMethods}
\end{figure}
%\end{landscape}

\section{Extensive simulation study}\label{Sec:SimStudy}
To %further explore the finite sample performance of the suggested estimation approaches and to 
demonstrate the gain in flexibility of D-vine copulas over Archimedean copulas with regard to dependence modeling, we additionally investigate simulation scenarios in which the association varies over time, either in strength or in type. %Moreover, we illustrate the effect of using an incorrect copula specification and explore the use of AIC as a model selection tool.

\begin{table}[H]
	\renewcommand{\arraystretch}{1.35}
	\centering
	\small
	\caption{Simulation results using \textbf{global one-stage parametric and two-stage semiparametric estimation} for four-dimensional data. In the top panels, the D-vine copula model captures tail-behavior for subsequent gap times changing from lower tail-dependence (Clayton (C)) over no tail-dependence (Frank (F)) to upper tail-dependence (Gumbel (G)) with same overall dependence of Kendall's $\tau_{1,2}=\tau_{2,3}=\tau_{3,4}=0.5$. In the bottom panels, the D-vine copula model captures for Clayton (C) copulas in $\mathcal{T}_1$ increasing dependence with $\tau_{1,2}=0.3$, $\tau_{2,3}=0.5$, $\tau_{3,4}=0.7$. The \textbf{empirical mean (empirical standard deviation) of the Kendall's $\boldsymbol{\tau}$ estimates} are presented based on 250 replications and samples of different sizes affected by either 15\%, 30\% or heavy tail 30\% right-censoring.}
	\label{Table:4d_CFG_tau_global}
	\hspace*{-1.25cm}\begin{tabular}{p{.15cm}p{.15cm}ccccccc}
		\midrule
		& & & \multicolumn{6}{c}{D-vine copula model}\\
		&  &  & C; $\tau_{1,2}:0.50$ & F; $\tau_{2,3}: 0.50$ & G; $\tau_{3,4}: 0.50$ & F; $\tau_{1,3;2}: 0.25$ & F; $\tau_{2,4;3}: 0.25$ & F; $\tau_{1,4;2,3}: 0.17$ \\
		\midrule \midrule
		\multirow{9}{*} {\begin{sideways} semiparametric two-stage \end{sideways}} &  \multirow{3}{*} {\begin{sideways} 15\% \end{sideways}} & $250$ & 0.498 (0.046) & 0.500 (0.038) & 0.507 (0.040) & 0.250 (0.050) & 0.250 (0.056) & 0.165 (0.060) \\
		&  & $500$ & 0.499 (0.032) & 0.500 (0.028) & 0.505 (0.030) & 0.251 (0.035) & 0.248 (0.037) & 0.164 (0.040) \\
		&  & $1000$ & 0.499 (0.024) & 0.498 (0.019) & 0.502 (0.021) & 0.248 (0.024) & 0.251 (0.026) & 0.164 (0.028) \\
		\cmidrule{2-9}
		& \multirow{3}{*} {\begin{sideways} 30\% \end{sideways}} & $250$ & 0.530 (0.092) & 0.519 (0.064) & 0.522 (0.059) & 0.249 (0.066) & 0.251 (0.080) & 0.157 (0.086) \\
		&  & $500$ & 0.507 (0.070) & 0.508 (0.050) & 0.512 (0.043) & 0.255 (0.047) & 0.246 (0.054) & 0.162 (0.058) \\
		&  & $1000$ & 0.509 (0.051) & 0.506 (0.034) & 0.507 (0.032) & 0.250 (0.038) & 0.247 (0.037) & 0.162 (0.041) \\
		\cmidrule{2-9}
		&  \multirow{3}{*} {\begin{sideways} {30\%HT} \end{sideways}} & $250$ & 0.637 (0.150) & 0.583 (0.114) & 0.548 (0.067) & 0.240 (0.087) & 0.255 (0.090) & 0.151 (0.091) \\
		&    & $500$ & 0.616 (0.114) & 0.572 (0.084) & 0.534 (0.058) & 0.246 (0.055) & 0.257 (0.068) & 0.160 (0.071) \\
		&    & $1000$ & 0.614 (0.115) & 0.564 (0.079) & 0.534 (0.042) & 0.246 (0.053) & 0.260 (0.049) & 0.161 (0.046) \\
		\midrule
		\multirow{6}{*} {\begin{sideways} parametric one-stage \end{sideways}} &  \multirow{2}{*} {\begin{sideways} 15\% \end{sideways}} & $250$ & 0.505 (0.033) & 0.504 (0.035) & 0.501 (0.036) & 0.253 (0.050) & 0.252 (0.055) & 0.169 (0.061) \\
		&    & $500$ & 0.505 (0.021) & 0.503 (0.024) & 0.500 (0.027) & 0.254 (0.034) & 0.250 (0.037) & 0.165 (0.040) \\
		\cmidrule{2-9}
		&   \multirow{2}{*} {\begin{sideways} 30\% \end{sideways}} & $250$ & 0.508 (0.043) & 0.510 (0.046) & 0.509 (0.052) & 0.255 (0.062) & 0.253 (0.079) & 0.167 (0.089) \\
		&    & $500$ & 0.504 (0.030) & 0.507 (0.034) & 0.503 (0.037) & 0.257 (0.044) & 0.250 (0.052) & 0.166 (0.057) \\
		\cmidrule{2-9}
		&   \multirow{2}{*} {\begin{sideways}{30\%HT} \end{sideways}} & $250$ & 0.507 (0.054) & 0.513 (0.051) & 0.512 (0.057) & 0.257 (0.070) & 0.249 (0.087) & 0.165 (0.096) \\
		&    & $500$ & 0.504 (0.036) & 0.510 (0.037) & 0.506 (0.043) & 0.259 (0.045) & 0.250 (0.061) & 0.165 (0.063) \\
		\midrule\midrule
		& &  & C; \ $\tau_{1,2}: 0.30$ & C; \ $\tau_{2,3}: 0.50$ & C; \ $\tau_{3,4}: 0.70$ & F; \ $\tau_{1,3;2}: 0.25$ & F; \ $\tau_{2,4;3}: 0.25$ & F; \ $\tau_{1,4;2,3}: 0.17$ \\
		\midrule\midrule
		\multirow{9}{*} {\begin{sideways} semiparametric two-stage \end{sideways}} & \multirow{3}{*} {\begin{sideways} 15\% \end{sideways}} & $250$ & 0.309 (0.054) & 0.499 (0.046) & 0.693 (0.035) & 0.247 (0.051) & 0.253 (0.055) & 0.164 (0.060) \\
		&  & $500$ & 0.308 (0.039) & 0.500 (0.034) & 0.696 (0.024) & 0.249 (0.037) & 0.252 (0.038) & 0.162 (0.041) \\
		&  & $1000$ & 0.303 (0.027) & 0.496 (0.025) & 0.696 (0.017) & 0.246 (0.024) & 0.253 (0.025) & 0.164 (0.027) \\
		\cmidrule{2-9}
		& \multirow{3}{*} {\begin{sideways} 30\% \end{sideways}} & $250$ & 0.362 (0.119) & 0.523 (0.089) & 0.697 (0.061) & 0.232 (0.070) & 0.254 (0.078) & 0.157 (0.093) \\
		&  & $500$ & 0.330 (0.082) & 0.509 (0.067) & 0.695 (0.044) & 0.244 (0.054) & 0.249 (0.055) & 0.162 (0.056) \\
		&  & $1000$ & 0.329 (0.061) & 0.508 (0.050) & 0.697 (0.031) & 0.244 (0.039) & 0.251 (0.035) & 0.159 (0.041) \\
		\cmidrule{2-9}
		& \multirow{3}{*} {\begin{sideways} {30\%HT} \end{sideways}} & $250$ & 0.519 (0.189) & 0.614 (0.142) & 0.736 (0.085) & 0.211 (0.087) & 0.250 (0.090) & 0.136 (0.100) \\
		&  & $500$ & 0.490 (0.159) & 0.594 (0.117) & 0.721 (0.075) & 0.214 (0.067) & 0.253 (0.071) & 0.137 (0.077) \\
		&  & $1000$ & 0.496 (0.140) & 0.596 (0.101) & 0.730 (0.058) & 0.221 (0.061) & 0.251 (0.050) & 0.141 (0.055) \\
		\midrule
		\multirow{6}{*} {\begin{sideways} parametric one-stage \end{sideways}} & \multirow{2}{*} {\begin{sideways} 15\% \end{sideways}} & $250$ & 0.299 (0.043) & 0.500 (0.039) & 0.701 (0.028) & 0.249 (0.049) & 0.253 (0.053) & 0.170 (0.062) \\
		&  & $500$ & 0.300 (0.028) & 0.500 (0.027) & 0.700 (0.020) & 0.251 (0.036) & 0.251 (0.037) & 0.165 (0.041) \\
		\cmidrule{2-9}
		& \multirow{2}{*} {\begin{sideways} 30\% \end{sideways}} & $250$ & 0.300 (0.063) & 0.499 (0.059) & 0.702 (0.044) & 0.248 (0.067) & 0.253 (0.079) & 0.169 (0.094) \\
		&  & $500$ & 0.299 (0.044) & 0.499 (0.043) & 0.700 (0.031) & 0.251 (0.051) & 0.251 (0.053) & 0.167 (0.056) \\
		\cmidrule{2-9}
		& \multirow{2}{*} {\begin{sideways} 30\%HT \end{sideways}} & $250$ & 0.300 (0.080) & 0.499 (0.079) & 0.699 (0.060) & 0.244 (0.079) & 0.254 (0.088) & 0.163 (0.102) \\
		&  & $500$ & 0.295 (0.052) & 0.497 (0.053) & 0.698 (0.042) & 0.251 (0.054) & 0.253 (0.062) & 0.166 (0.069) \\
		\midrule\midrule
	\end{tabular}\vspace*{-2cm}
\end{table}
\renewcommand{\arraystretch}{1.00}

\subsection{Settings}
In all %considered 
scenarios, the results are based on 250 data sets. We consider samples of 250, 500 or 1000 clusters, each with a maximum %cluster 
size of 4. The fourth gap time follows the same distribution as gap times 2 and 3. The censoring times are again generated from a Weibull survival function, with shape and scale parameters %chosen 
such that 15\% or 30\% of the data are censored. We also consider 30\% censoring with large event times being more prone to right-censoring (heavy tail) (see Table 9 of the supplementary material for details). 

The dependence between the four gap times is modeled via a D-vine. In trees $\mathcal{T}_2$ and $\mathcal{T}_3$, we consider Frank copulas with $\tau_{1,3;2} = \tau_{2,4;3} = 0.25$ and $\tau_{1,4;23} = 0.167$. In $\mathcal{T}_1$ we increase the complexity. In a first setting, we fix the dependence strength, but allow the type of association to vary over time: $\scd_{1,2}$ is Clayton, $\scd_{2,3}$ is Frank, $\scd_{3,4}$ is Gumbel. This reflects a slow change from lower to upper tail-dependence. As a second setting, we fix the association type to be Clayton, but allow the strength to increase: $\tau_{1,2} = 0.3$, $\tau_{2,3} = 0.5$, $\tau_{3,4} = 0.7$. %\autoref{table:CopulaSimSettings_4d} gives an overview using C for Clayton, G for Gumbel and F for Frank. 

\subsection{Results}
%Under the assumption of a correct copula, 
The obtained results for Kendall's $\tau$ in case of one-stage parametric and two-stage semi-parametric estimation are summarized in \autoref{Table:4d_CFG_tau_global}. Since in the illustrating simulations global and sequential estimation showed very similar performance only results for global proceeding are shown. The results in case of sequential estimation as well as the results in terms of the copula parameters and for the marginal parameters %in case of one-stage parametric estimation, 
are given in Table 11 to Table 14 of the supplementary material. As in the illustrating simulations, the new settings indicate that under a correct copula format
the one-stage parametric approaches perform well in all censoring scenarios, while the two-stage semiparametric approaches are more sensitive to the underlying censoring scheme. Clearly, the proposed estimation strategies allow to investigate a dependence pattern more complex than that of an Archimedean copula, including varying types and strengths of association. 

The effect of using an incorrect copula specification and the role of AIC as a valid model selection tool is explored in Section 6.4 of the supplementary material.

\section{Data application}\label{Sec:DataApplication}
In this section, we use the proposed modeling and estimation strategies to analyze the asthma data, which were introduced in \autoref{Sec:AsthmaMotivation}. \cite{meyer2015bayesian} analyze the association in the asthma data via Archimedean copulas, where the marginal survival functions of the gap times are assumed to be Weibull. An Archimedean copula imposes the same type and strength of dependence between all asthma attacks. However, an asthma attack further weakens the lungs and thus makes a child more prone to subsequent attacks. Therefore, the dependence between subsequent pairs is expected to change over time. As the simulations in \autoref{Sec:Methodology} and \autoref{Sec:SimStudy} have shown, D-vine copulas can be used to capture such features.

To explore the asthma data and to decide on the estimation strategy, we investigate the Nelson-Aalen estimate for the survival function of the total times. We consider the full data sample as well as the data subsamples based on treatment to accommodate a possible effect of the latter on the dependence structure. Each sample showed a high censoring rate with accumulation of censored observations at late time points, i.e. the Nelson-Aalen estimates show a heavily right-censored tail, leading to a leveling off at a survival value around $0.6$ for the full data set, around $0.7$ for the treated children and around $0.5$ for the placebo group (see Figure 2 of the supplementary material). Based on the simulation results and the guidelines given in \autoref{fig:FlowchartEstMethods}, we therefore opt to apply a one-stage parametric estimation approach. % to model the dependence structure in the asthma data. 
As in \cite{meyer2015bayesian}, we assume Weibull survival margins, but opposed to them, we allow for a flexible association pattern as modeled by diverse D-vine copulas.

The induced dependent right-censoring present in the asthma data makes model specification challenging. Common data exploration tools cannot be applied. For example, due to the heavy censoring for larger gap times, pairs plots on the time scale, resp. on the copula scale, would show an empty upper right corner, resp.\ an empty lower left corner, and thus visual inspection is obscured. To unravel the association in the asthma data, we therefore fit a large variety of different copula models. We consider the independence copula as well as the four-dimensional Clayton, Gumbel and Frank copulas, together with several four-dimensional D-vine copulas. For these D-vine models, we consider in tree $\mathcal{T}_1$ all possible permutations of Clayton, Gumbel and/or Frank copulas. In trees $\mathcal{T}_2$ and $\mathcal{T}_3$, all pair-copulas are taken to be Frank. This results in a total of $27$ D-vine copulas.

\autoref{Table:AsthmaResults_AIC} gives the results of global one-stage parametric estimation in terms of Kendall's $\tau$ for the three best D-vine copulas as selected by AIC, the independence copulas as well as for the Archimedean copulas. %Since focus in our analysis is on the dependence between gap times, only Kendall's $\tau$ estimates with standard errors in parentheses are shown. 
Results on marginal estimation are listed in Table 19 of the supplementary material. %In addition, AIC values for the Archimedean copulas and the independence copula are shown. 
For each data sample, all D-vine copulas perform better than the best Archimedean copula based on AIC. While the best D-vine copula is the same for all samples, the best Archimedean copula varies among the three data sets. %This can be explained by the estimates of the Kendall's $\tau$ values (standard errors in parentheses). 
%, which are presented in \autoref{Table:AsthmaResults} together with their standard errors (in parentheses). 
Recall that there is only one parameter in an Archimedean copula to describe the dependence between all gap times. In the asthma data, this dependence is very small and close to independence (as confirmed by AIC). D-vine copulas %, on the other hand, are able to 
focus on the dependence structure more locally and therewith %to 
capture varying dependence between %subsequent 
gap times. While the Kendall's $\tau$ values in trees $\mathcal{T}_2$ and $\mathcal{T}_3$ are quite small, the estimates in tree $\mathcal{T}_1$, i.e.\ for $\tau_{1,2}$, $\tau_{2,3}$ and $\tau_{3,4}$, increase over time. This finding supports the initial intuition that with each additional asthma attack, children are more prone to a relapse. The fact that a Gumbel copula is chosen for the pair 2-3 suggests that the smaller gap time 2 is, the faster a third asthma attack will follow. The same holds true for pair 3-4. For pair 1-2 there is no clear best copula family, which might be explained by the low Kendall's $\tau$ values of on average 0.10. For such a low value the specific features of a copula family such as lower or upper tail-dependence are less pronounced. Interestingly, the estimates for $\tau_{2,3}$ and $\tau_{3,4}$ for the treatment and control group are quite alike, while there is a significant difference for $\tau_{1,2}$. For treated children the occurrences of a first and a second asthma attack are close to independence, while for children in the placebo group the estimate for $\tau_{1,2}$ is about 0.18. This suggests that the medical treatment has a clear influence on the (time to) occurrence of a second asthma attack. However, whenever a treated child has a relapse, subsequent %asthma 
attacks are as likely as for untreated children. In general and most pronounced for the treatment group, Kendall's $\tau$ values including the first gap are smaller as compared to those not including the first gap. %Thus, the latter seems to behave different than subsequent gaps.

The standard errors of the estimates are obtained via bootstrapping. The algorithm %accounts for induced dependent right-censoring and the copula framework. 
is given in Section 7 of the supplementary material together with extra details on the bootstrap samples of the asthma data (Table 20). %Note that while in the asthma data there are no clusters of size 1, this case is possible for the bootstrap samples (as in many data settings). \autoref{Table:bootInfo} in \autoref{Sec:AddOnAsthma} contains information on average cluster sizes and the average censoring percentage among the bootstrap replications showing that the data generation within the bootstrap succeeds to mimic the original data characteristics as given in \autoref{Sec:AsthmaMotivation} quite accurately. 
In general, standard errors increase for estimates corresponding to later gap times. Due to the unbalanced data setting, fewer data are available for these gap times.

\begin{table}[h]
	\small
	\renewcommand{\arraystretch}{1.0}
	\centering	
	\caption{AIC values and Kendall's $\tau$ estimates with standard errors (in parentheses) of copula models fitted to each of the three samples of the asthma data using global one-stage parametric estimation. In case of Archimedean copulas the Frank (4dF), Gumbel (4dG), Clayton (4dC) and the Independence (4dInd) copula are considered. In case of D-vine copulas only the three best models are shown with Frank being the pair-copula family in trees $\mathcal{T}_2$ and $\mathcal{T}_3$.
	}
	\label{Table:AsthmaResults_AIC}
	\hspace*{-.8cm}\begin{tabular}{p{.15cm}cccccccc}
		\midrule
		&  & AIC & $\tau_{12}/ \tau$ & $\tau_{23}$ & $\tau_{34}$ & $\tau_{13;2}$ & $\tau_{24;3}$ & $\tau_{14;23}$ \\
		\hline\midrule
		\multirow{6}{*}{\begin{sideways} Full \end{sideways}} & FGG & 210.10 & 0.12 (0.052) & 0.26 (0.059) & 0.33 (0.062) & -0.05 (0.064) & 0.29 (0.080) & -0.09 (0.079) \\
		&	 CGG & 212.58 &0.13 (0.065) & 0.27 (0.059) & 0.34 (0.062) & -0.05 (0.063) & 0.30 (0.080) & -0.09 (0.078) \\
		& GGG & 213.02 &0.10 (0.048) & 0.25 (0.059) & 0.33 (0.062) & -0.05 (0.065) & 0.29 (0.080) & -0.09 (0.081) \\ 	
		& 4dF & 233.67 & 0.06 (0.025)&  &  &  &  &  \\
		& 4dG & 235.38 & 0.05 (0.030)&  &  &  &  &  \\
		& 4dC & 236.46 & 0.06 (0.040)&  &  &  &  &  \\ 	
		& 4dInd & 237.48 & &  &  &  &  &  \\ 			 		
		\midrule
		\multirow{6}{*}{\begin{sideways} Treatment \end{sideways}} & FGG & 147.80 &0.05 (0.078)&  0.26 (0.093) & 0.33 (0.106) & 0.02 (0.105) & 0.42 (0.117) & -0.16 (0.136) \\
		& CGG & 147.88 &0.06 (0.083) & 0.26 (0.093) & 0.34 (0.106) & 0.02 (0.102) & 0.43 (0.115) & -0.16 (0.133) \\
		& GGG & 148.36 &0.00 (0.036) & 0.26 (0.093) & 0.33 (0.106) & 0.02 (0.108) & 0.42 (0.116) & -0.16 (0.138) \\
		& 4dF & 154.80 &0.04 (0.032) &  &  &  &  &  \\
		& 4dG & 155.94 &0.00 (0.022) &  &  &  &  &  \\
		& 4dC & 155.50 & 0.04 (0.053)&  &  &  &  &  \\
		& 4dInd & 153.94 & &  &  &  &  &  \\ 
		\midrule	
		\multirow{6}{*}{\begin{sideways} Control \end{sideways}} & FGG & 67.08 &0.18 (0.070) & 0.26 (0.075) & 0.31 (0.086) & -0.11 (0.086) & 0.19 (0.102) & -0.03 (0.101) \\
		& FGF & 68.72 &0.18 (0.070) & 0.24 (0.075) & 0.29 (0.100) & -0.11 (0.086) & 0.17 (0.105) & -0.03 (0.101) \\
		& GGG & 69.62 &0.17 (0.068) & 0.25 (0.076) & 0.30 (0.087) & -0.11 (0.087) & 0.19 (0.103) & -0.03 (0.104)\\
		& 4dF & 78.34 & 0.06 (0.030)&  &  &  &  &  \\
		& 4dG & 77.86 & 0.06 (0.035) &  &  &  &  &  \\
		& 4dC & 79.95 & 0.07 (0.050)&  &  &  &  &  \\
		& 4dInd & 80.28 & &  &  &  &  &  \\ 
		\hline\hline
	\end{tabular}	
\end{table}
\renewcommand{\arraystretch}{1.00}

%\begin{figure}[H]
%	\centering
%		\caption{Pairs plots of the three gap time pairs $1-2$, $2-3$ and $3-4$ based on pseudo-observations
%		generated using Nelson-Aalen weights for the marginals (see \autoref{fig:asthmanelson-aalen}). The effect of right censoring is reflected by the empty lower left corner in the pairs plots. Observations shown as $\bullet$ are event
%		times for both gaps; $\downarrow$ is an event time only for the earlier observation.}
%	 \includegraphics[width=.33\linewidth]{Plots/Scatterplot12}\includegraphics[width=.33\linewidth]{Plots/Scatterplot23}\includegraphics[width=.33\linewidth]{Plots/Scatterplot34}
%
%	\label{fig:scatterplot12}
%\end{figure}

\section{Discussion}\label{Sec:Discussion}
In this paper, we address several challenges that arise when modeling the association between gap times, e.g.\ the presence of induced dependent right-censoring and the unbalanced nature of the data. We introduce D-vine copulas as a flexible class of models, that naturally captures the inherent serial dependence. Moreover, we allow nonparametric estimation of the survival margins by introducing a modified version of the nonparametric estimator by \cite{UnaAlvarezMachado2008nonparaCensGap}. As such, we extend previous work by \cite{prenen2017extending} and \cite{meyer2015bayesian} on recurrent event time data. Both use Archimedean copulas in combination with parametric survival margins. In total, four estimation strategies are suggested. First, a one-stage parametric approach, in which marginal and copula parameters are jointly estimated via likelihood maximization, is proposed. Second, flexibility is increased by nonparametric marginal modeling in a two-stage semiparametric estimation approach. For both global approaches alternative sequential procedures are developed to reduce the computational demand when extending the methodology to higher dimensions. Simulations in three and four dimensions provide evidence for the good finite sample performance of the estimation strategies. Further, they reveal the limits of each approach, especially when data are heavily distorted by right-censoring. Guidelines for appropriate handling of recurrent event time data are formulated. An application of the proposed methodology to real data on children suffering from asthma provides new insights on the evolution of the disease. These findings could not be detected by Archimedean copulas, which impose a too restrictive dependence structure to the data. This stresses the need for flexible copula models such as D-vines when interest is in the dependence of right-censored recurrent event time data.

%  The \backmatter command formats the subsequent headings so that they
%  are in the journal style.  Please keep this command in your document
%  in this position, right after the final section of the main part of 
%  the paper and right before the Acknowledgements, Supplementary Materials,
%  and References sections. 

%  This section is optional.  Here is where you will want to cite
%  grants, people who helped with the paper, etc.  But keep it short!
\section*{Acknowledgements}
The authors wish to thank Matthias Killiches for discussion on early drafts of this manuscript. Numerical calculations were performed on a Linux cluster supported by DFG
grant INST 95/919-1 FUGG. This work was supported by the Deutsche Forschungsgemeinschaft [DFG CZ 86/4-1], the Interuniversity Attraction Poles Programme (IAP-network P7/06), Belgian Science Policy Office and the Research Foundation Flanders (FWO), Scientific Research Community on ``Asymptotic Theory for Multidimensional Statistics" [W000817N].
\vspace*{-.8pt}

%  If your paper refers to supplementary web material, then you MUST
%  include this section!!  See Instructions for Authors at the journal
%  website http://www.biometrics.tibs.org

%\section*{Supplementary Materials}
%
%Extensive supplementary material is available at the Biometrics website on Wiley Online
%Library.\vspace*{-.8pt}

%  Here, we create the bibliographic entries manually, following the
%  journal style.  If you use this method or use natbib, PLEASE PAY
%  CAREFUL ATTENTION TO THE BIBLIOGRAPHIC STYLE IN A RECENT ISSUE OF
%  THE JOURNAL AND FOLLOW IT!  Failure to follow stylistic conventions
%  just lengthens the time spend copyediting your paper and hence its
%  position in the publication queue should it be accepted.

%  We greatly prefer that you incorporate the references for your
%  article into the body of the article as we have done here 
%  (you can use natbib or not as you choose) than use BiBTeX,
%  so that your article is self-contained in one file.
%  If you do use BiBTeX, please use the .bst file that comes with 
%  the distribution.  In this case, replace the thebibliography
%  environment below by 
%

\bibliographystyle{apalike}
\bibliography{References}

\newpage

\setcounter{page}{1}
\setcounter{table}{0}
\setcounter{figure}{0}
\setcounter{section}{0}

\begin{center}
	{\centering {\Large SUPPLEMENTARY MATERIAL}}
\end{center}
\vspace*{.05cm}
\section{Required data format} \label{Sec:DataAppendix}

\begin{table}[h]
	%\footnotesize
	\centering
	%\begin{centering}
	\caption{Data format assumed for the induced dependent right-censored gap time data $(y_{i,1},y_{i,2},\ldots,y_{i,d-1},y_{i,d},\delta_{i,1},\delta_{i,2},\ldots,\delta_{i,d-1},\delta_{i,d})$, $i=1,\ldots,n$: ordering by decreasing cluster size.}
	\label{table:DataFormat}
	\begin{tabular}{ l c c c c c} 	
		\midrule	
		$i$ & $y_{i,1}$ & $y_{i,2}$ & $\cdots$ & $y_{i,d-1}$ & $y_{i,d}$ \\  \hline
		\midrule
		1 & $g_{1,1}$ & $g_{1,2}$ & $\cdots$ & $g_{1,d-1}$ & $y_{1,d}$ \\[-0.25ex]
		$\vdots$ & $\vdots$ & $\vdots$ &  & $\vdots$ & $\vdots$ \\[-0.25ex]
		$n_d$ & $g_{n_d,1}$ & $g_{n_d,2}$ & $\cdots$ & $g_{n_d,d-1}$ & $y_{n_d,d}$ \\
		$n_d+1$ & $g_{n_d+1,1}$ & $g_{n_d+1,2}$ & $\cdots$ & $y_{n_d+1,d-1}$ & \\[-0.25ex]
		$\vdots$ & $\vdots$ & $\vdots$ & & $\vdots$ & \\[-0.25ex]
		$n_d + n_{d-1}$ & $g_{n_d + n_{d-1},1}$ & $g_{n_d + n_{d-1},2}$ & $\cdots$ & $y_{n_d+n_{d-1},d-1}$ & \\[-0.25ex]
		$\vdots$ & $\vdots$ & $\vdots$ &  &  & \\[-0.25ex]
		%$n_d+\ldots+n_2$ & $g_{n_d+\ldots+n_2,1}$ & $y_{n_d+\ldots+n_2,2}$ &  & &  \\
		$n_d+\ldots+n_2+1$ & $y_{n_d+\ldots+n_2+1,1}$ & &  &  &\\[-0.25ex]
		$\vdots$ & $\vdots$ &  &  &  & \\[-0.25ex]
		$n_d+\ldots+n_2+n_1=n$ & $y_{n,1}$ &  &  &  & \\
		\midrule
		$i$ & $\delta_{i,1}$ & $\delta_{i,2}$ & $\cdots$ & $\delta_{i,d-1}$ & $\delta_{i,d}$\\  \hline
		\midrule
		1 &  1 & 1 & $\cdots$ & 1 & $\delta_{1,d}$\\[-0.25ex]
		$\vdots$  & $\vdots$ & $\vdots$ &  & $\vdots$ & $\vdots$\\[-0.25ex]
		$n_d$  & 1 & 1 & $\cdots$ & 1 & $\delta_{n_d,d}$\\
		$n_d+1$ & 1 & 1 & $\cdots$ & $\delta_{n_d+1,d-1}$ &\\[-0.25ex]
		$\vdots$ & $\vdots$ & $\vdots$ &  & $\vdots$ & \\[-0.25ex]
		$n_d + n_{d-1}$ & 1 & 1 & $\cdots$ & $\delta_{n_d+n_{d-1},d-1}$ & \\[-0.25ex]
		$\vdots$ & $\vdots$ & $\vdots$ &  &  & \\[-0.25ex]
		%$n_d+\ldots+n_2$ & 1 & $\delta_{n_d+\ldots+n_2,2}$  & & & \\
		$n_d+\ldots+n_2+1$ & $\delta_{n_d+\ldots+n_2+1,1}$ & & & & \\[-0.25ex]
		$\vdots$& $\vdots$ &  &  &  & \\[-0.25ex]
		$n_d+\ldots+n_2+n_1=n$& $\delta_{n,1}$ & & & & \\
		\hline
		\hline
	\end{tabular}
	%\end{centering}
	\\
	%{\footnotesize where $D_{1}(\theta) = \frac{1}{\theta}  \int_{0}^{\theta} \frac{t}{e^{t}-1} dt $ is the Debye function.}
\end{table}

\section{Details on popular Archimedean copulas}
\renewcommand{\arraystretch}{1.1}
\begin{table}[H]
	%\small
	\centering
	\caption{Popular bivariate Archimedean copulas with the range of their dependence parameter $\theta$, the formula of $\phi$ and corresponding Kendall's $\tau$ value.}
	\label{table:Copulas}
	\begin{tabular}{ p{1.25cm} c c c} 			\midrule
		&  Clayton & Gumbel & Frank\\
		\hline\midrule
		\centering \multirow{2}{*}{$\theta \in $} & \multirow{2}{*}{$(0,\infty)$}  & \multirow{2}{*}{$[1,\infty) $} & \multirow{2}{*}{$(-\infty,\infty) \setminus \{0\}$}\\
		& & & \\
		\centering\multirow{2}{*}{$\phi(s)$} & \multirow{2}{*}{$(1+\theta s)^{-1/ \theta}$}  & \multirow{2}{*}{$e^{-s^{1/\theta}}$} & \multirow{2}{*}{$-\frac{1}{\theta} \ln\{1-(1-e^{-\theta})e^{-s}\}$} \\
		& & & \\
		\multirow{2}{*}{$\sC(u_{1},u_{2})$} &  \multirow{2}{*}{$(u_{1}^{-\theta}+u_{2}^{-\theta}-1)^{-\frac{1}{\theta}}$}  & \multirow{2}{*}{$e^{\Big\{ - [(-\ln u_{1})^{\theta} + (- \ln u_{2})^{\theta} ]^{\frac{1}{\theta}} \Big\}}$} & \multirow{2}{*}{$- \frac{1}{\theta}  \ln \Big\{  1+ \frac{(e^{-\theta u_{1}}-1)    (e^{-\theta u_{2}}-1)}{   e^{-\theta }-1} \Big\}   $}\\
		& & & \\
		\centering\multirow{2}{*}{$\tau$} & \multirow{2}{*}{$\tau = \frac{\theta}{\theta+2}$}  &\multirow{2}{*}{ $\tau = 1-\frac{1}{\theta}$} & $\tau =
		1-\frac{4}{\theta}+4\frac{D_1(\theta)}{\theta} $\\
		& & & {\scriptsize with} $D_1(\theta) =
		\int_0^{\theta} \frac{t/\theta}{e^t-1}dt$ \\
		\midrule			\midrule
	\end{tabular}
\end{table}
\renewcommand{\arraystretch}{1.0}

\section{Derivation of the likelihood contributions for D-vine copulas}\label{Sec:VinesAppendix}

In case of ordered D-vine copulas, there are explicit expressions for the loglikelihood contributions used in the four likelihood based estimation strategies. These expressions are analytically tractable and easy to apply in arbitrary dimensions.

For ease of notation, we subsequently consider data on the copula level. Further, we assume clusters of maximum size $d=4$ and derive the two possible loglikelihood contributions -- depending on whether the last observed gap time corresponds to a true event or to a right-censoring value. Let $\sC_{1:4}$ with density $\scd_{1:4}$ be the copula describing the vector $(U_1,U_2,U_3,U_4)$, which corresponds to the vector of observed gap times $(Y_1, Y_2, Y_3, Y_4)$.  Assuming that $\scd_{1:4}$ arises from a four-dimensional ordered D-vine copula, we have

\begin{align}
\scd_{1:4}\left(u_1,\right.& \left.\hspace*{-.1cm} u_2, u_3, u_4\right)\label{eq:deriv_density}\\
\quad = & \ \scd_{1,2}\left(u_1, u_2\right)\scd_{2,3}\left(u_2, u_3\right)\scd_{3,4}\left(u_3, u_4\right)\notag\\
& \times \scd_{1,3;2}\{\sC_{1|2}\left(u_1|u_2\right), \sC_{3|2}\left(u_3|u_2\right)\}\scd_{2,4;3}\{\sC_{2|3}\left(u_2|u_3\right), \sC_{4|3}\left(u_4|u_3\right)\}\notag\\
& \times \scd_{1,4;2:3}\{\sC_{1|2:3}\left(u_1|u_2,u_3\right),\sC_{4|2:3}\left(u_4|u_2,u_3\right)\}\notag\\
\quad = & \ \scd_{1,2}\left(u_1, u_2\right)\scd_{2,3}\left(u_2, u_3\right)\scd_{3,4}\left(u_3, u_4\right)\notag\\
& \times \scd_{1,3;2}\{h_{1|2}\left(u_1|u_2\right), h_{3|2}\left(u_3|u_2\right)\}\scd_{2,4;3}\{h_{2|3}\left(u_2|u_3\right), h_{4|3}\left(u_4|u_3\right)\}\notag\\
& \times \scd_{1,4;2:3}\left[h_{1|3;2}\{h_{1|2}\left(u_1|u_2\right)\big\vert h_{3|2}\left(u_3|u_2\right)\},h_{4|2;3}\{h_{4|3}\left(u_4|u_3\right)\big\vert h_{2|3}\left(u_2|u_3\right)\}\right].\notag
\end{align}
%According to \eqref{eq:loglik} and \eqref{eq:2stage_contribution} 
The loglikelihood contribution for a cluster, of which the last observed gap time corresponds to a true event, corresponds to the copula density evaluated at the observed gap times. In case of a four-dimensional ordered D-vine copula the loglikelihood contribution is thus given by \eqref{eq:deriv_density}. For a cluster, of which the last observed gap time corresponds to a right-censoring value, the loglikelihood contribution equals the partial derivative of $\sC_{1:4}$ with respect to the variables $U_1$, $U_2$ and $U_3$ evaluated at the observed copula data. In case of an underlying four-dimensional ordered D-vine copula, the following holds:
\allowdisplaybreaks
\begin{align}
\frac{\partial^3 \sC_{1:4}\left(u_1,u_2,u_3,u_4\right)}{\partial u_1 \partial u_2 \partial u_3} \hspace{-3cm}\label{eq:deriv_partDeriv}\\
= & \  \int_{0}^{u_4} \scd_{1:4}\left(u_1, u_2, u_3, v_4\right)dv_4\notag\\
\ \stackrel{\text{\eqref{eq:deriv_density}}}{=} & \ \int_{0}^{u_4} \scd_{1,2}\left(u_1, u_2\right)\scd_{2,3}\left(u_2, u_3\right)\scd_{3,4}\left(u_3, v_4\right)\scd_{1,3;2}\{\sC_{1|2}\left(u_1|u_2\right), \sC_{3|2}\left(u_3|u_2\right)\}\notag\\
& \times \scd_{2,4;3}\{\sC_{2|3}\left(u_2|u_3\right), \sC_{4|3}\left(v_4|u_3\right)\}\scd_{1,4;2,3}\{\sC_{1|2:3}\left(u_1|u_2, u_3\right), \sC_{4|2:3}\left(v_4|u_2, u_3\right)\} dv_4\notag\\
= & \  \scd_{1,2}\left(u_1, u_2\right)\scd_{2,3}\left(u_2, u_3\right)\scd_{1,3;2}\{\sC_{1|2}\left(u_1|u_2\right), \sC_{3|2}\left(u_3|u_2\right)\}\notag\\
& \times \int_{0}^{u_4} \frac{\partial^2}{\partial u_3 \partial v_4} \sC_{3,4}\left(u_3, v_4\right)\frac{\partial^2}{\partial u \partial v} \sC_{2,4;3}\left(u,v\right)\bigg \vert_{\myrel{u = \sC_{2|3}\left(u_2|u_3\right)}{v = \sC_{4|3}\left(v_4|u_3\right)}}\notag\\
& \times \frac{\partial^2}{\tilde{u} \partial \tilde{v}} \sC_{1,4;2,3}\left(\tilde{u}, \tilde{v}\right)\bigg \vert_{\myrel{\tilde{u} = \sC_{1|2:3}\left(u_1|u_2, u_3\right)}{\tilde{v} = \sC_{4|2:3}\left(v_4|u_2, u_3\right)}}dv_4\notag\\
= & \  \scd_{1:3}\left(u_1, u_2, u_3\right)\notag\\
& \times \int_{0}^{u_4} \frac{\partial}{\partial v_4}\underbrace{\left[\frac{\partial}{\partial u_3}\sC_{3,4}\left(u_3, v_4\right)\right]}_{=\sC_{4|3}\left(v_4|u_3\right)=v}\frac{\partial}{\partial v}\underbrace{\left[\frac{\partial}{\partial u}\sC_{2,4;3}\left(u, v\right)\right]}_{=\sC_{4|2:3}\left(v_4|u_2,u_3\right)=\tilde{v}}\bigg\vert_{\myrel{u = \sC_{2|3}\left(u_2|u_3\right)}{v = \sC_{4|3}\left(v_4|u_3\right)}}\notag\\
& \times \frac{\partial^2}{\partial \tilde{u} \partial \tilde{v}} \sC_{1,4;2,3}\left(\tilde{u}, \tilde{v}\right)\bigg \vert_{\myrel{\tilde{u} = \sC_{1|2:3}\left(u_1|u_2, u_3\right)}{\tilde{v} = \sC_{4|2:3}\left(u_4|u_2, u_3\right)}}dv_4\notag\\
= & \  \scd_{1:3}\left(u_1, u_2, u_3\right)\int_{0}^{u_4} \frac{\partial v}{\partial v_4} \frac{\partial \tilde{v}}{\partial v}\frac{\partial^2}{\partial \tilde{u} \partial \tilde{v}} \sC_{1,4;2,3}\left(\tilde{u}, \tilde{v}\right)\bigg \vert_{\myrel{v = \sC_{4|3}\left(v_4|u_3\right)\hspace{0.65cm}}{\myrel{\tilde{u} = \sC_{1|2:3}\left(u_1|u_2, u_3\right)}{\tilde{v} = \sC_{4|2:3}\left(v_4|u_2, u_3\right)}}}dv_4\notag\\
= & \  \scd_{1:3}\left(u_1, u_2, u_3\right)\int_{0}^{u_4} \frac{\partial}{\partial v_4} \left[\frac{\partial}{\partial \tilde{u}} \sC_{1,4;2,3}\{\tilde{u}, \sC_{4|2:3}\left(v_4|u_2, u_3\right)\}\bigg \vert_{\tilde{u} = \sC_{1|2:3}\left(u_1|u_2, u_3\right)}\right]dv_4\notag\\
= & \  \scd_{1:3}\left(u_1, u_2, u_3\right)\frac{\partial}{\partial \tilde{u}} \sC_{1,4;2,3}\{\tilde{u}, \sC_{4|2:3}\left(u_4|u_2, u_3\right)\}\bigg \vert_{\tilde{u} = \sC_{1|2:3}\left(u_1|u_2, u_3\right)}\notag\\
= & \  \scd_{1:3}\left(u_1, u_2, u_3\right)\sC_{4|1:3}\left(u_4|u_1, u_2, u_3\right)\notag\\
= & \  \scd_{1,2}\left(u_1, u_2\right)\scd_{2,3}\left(u_2, u_3\right)\scd_{1,3;2}\{h_{1|2}\left(u_1|u_2\right),h_{3|2}\left(u_3|u_2\right)\}\notag\\
& \times h_{4|1;2:3}\left[h_{4|2;3}\{h_{4|3}\left(u_4|u_3\right)\big\vert h_{2|3}\left(u_2|u_3\right)\}\bigg\vert h_{1|3;2}\{h_{1|2}\left(u_1|u_2\right)\big\vert h_{3|2}\left(u_3|u_2\right)\}\right].\notag
\end{align}
The first part of the final expression equals the three-dimensional copula density $\scd_{1:3}$, which arises from a three-dimensional ordered D-vine copula. The second part is a univariate conditional distribution, which according to \cite{Joe1997} can be recursively evaluated in terms of the pair-copulas in $\mathcal{T}_1$ to $\mathcal{T}_{2}$. Similarly, for arbitrary dimension $d$, one can show that
\begin{align}\label{eq:condCDF}
\frac{\partial^{d-1} \sC_{1:d}(u_1,\ldots,u_{d})}{\partial u_1 \cdots \partial u_{d-1}}=\scd_{1:d-1}(u_1,\ldots,u_{d-1})\sC_{d|1:d-1}(u_{d}|u_1,\ldots,u_{d-1}).
\end{align}
\noindent
To conclude, both loglikelihood contributions only depend on the bivariate building blocks, namely the pair-copulas, of the ordered D-vine copula.

\section{Data sampling procedure}\label{Sec:AppendixDataSampling}
Throughout this paper, we support our findings via simulations. Here, we briefly outline the procedure to generate unbalanced induced dependent right-censored data for $d=4$.

First, we sample data from the underlying $4$-dimensional copula. Next, we apply -- using appropriate assumptions for the survival margins -- the inverse probability transform to create gap times $G_{i,j}$ ($i=1,\ldots,n$ and $j = 1, \dots, 4$). The corresponding event times $T_{i,j}$ are defined as $T_{i,1}=G_{i,1}$ and $T_{i,j} = \sum_{\ell=1}^{j} G_{i,\ell}$. Based on sampled censoring times, the observed data are obtained as follows: if $T_{i,1}>C_{i}$ we set $d_i = 1$ and retain $C_{i}$, if $T_{i,2}>C_{i}$ we set $d_i = 2$ and retain $(T_{i,1},C_{i})$, etc. For $d_i = 4$ we distinguish between three events, i.e.\ $(T_{i,1},T_{i,2},T_{i,3},C_{i})$ and four events, i.e.\ $(T_{i,1},T_{i,2},T_{i,3},T_{i,4})$. Finally, the observed gap times for cluster $i$ with $d_i > 1$ are given by $(Y_{i,1},\ldots,Y_{i,d_{i}})$ with $Y_{i,j} = G_{i,j}$ for $j = 1, \ldots, d_i - 1$ and $Y_{i,d_i}=\min(G_{i,d_i}, C_i - \sum_{\ell=1}^{d_i-1} G_{i,\ell})$ together with the right-censoring indicator $\delta_{i,d_i}=I(Y_{i,d_i} = G_{i,d_i})$. Note that with this procedure the last event/gap time in a cluster of size $d_i < 4$ is always right-censored. Given that many studies have a limited follow-up period, the latter most often holds true in practice, see e.g.\ the asthma data.

\newpage
\section{Additional material for illustrating simulations}\label{Sec:AddOnsIlluSims1}

\subsection{Simulation settings}
\begin{table}[h]
	\centering
	%\small
	\captionof{table}{Simulation settings for the marginal survival functions of the three gap times and for the survival function of the censoring times leading to 15\%, 30\% or 30\% HT (heavy tail) censoring.}
	\label{table:marginalSimSettings_3d}
	\begin{tabular}{ r  c c  c c c} \midrule
		\multirow{2}{*}{\centering Weibull parameters} & \multirow{2}{*}{Gap time 1} & \multirow{2}{*}{Gap time 2 - 3} & \multicolumn{3}{c}{Censoring} \\
		& & & 15\%  & 30\% & 30\% HT\\
		\hline
		\midrule
		scale $\lambda$ & 0.5 & 1 & 0.1 & 0.25 & 0.1 \\
		shape $\rho$ & 1.5 & 1.5 & 1.5 & 1.5 & 3\\
		\hline\hline	
	\end{tabular}
\end{table}

\begin{figure}[H]
	\centering
	\includegraphics[width=.9\linewidth]{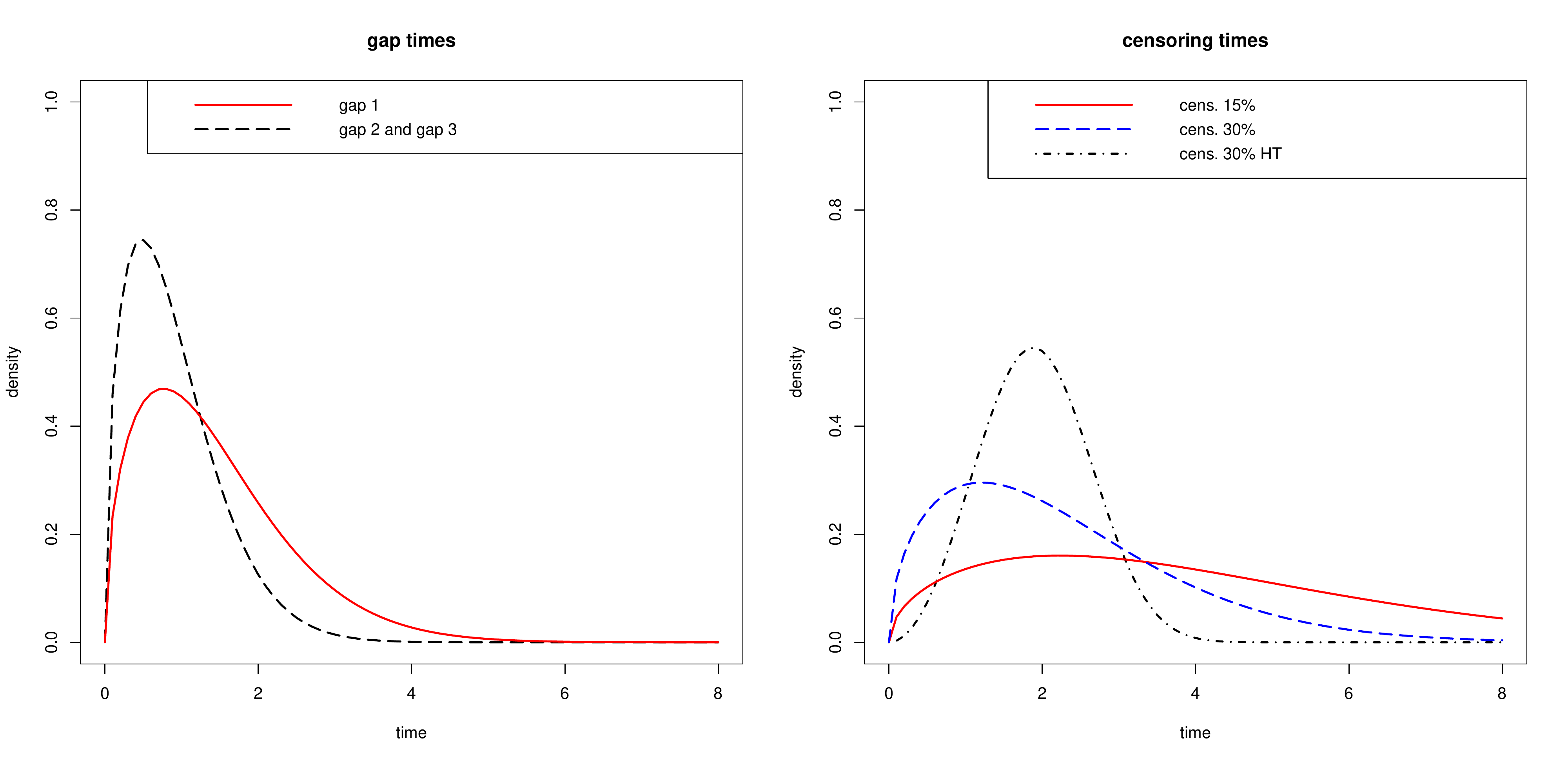}
	\caption{Density functions for the Weibull specifications in \autoref{table:marginalSimSettings_3d}.}
	\label{fig:choices_weibull}
\end{figure}

\begin{table}[H]
	\centering
	%\small
	\captionof{table}{Simulation settings for Archimedean copulas and D-vine copulas.}
	\label{table:CopulaSimSettings_3d}
	\begin{tabular}{ccc} \hline\midrule
		\multicolumn{3}{c}{3d Archimedean copula (copula family; Kendall's $\tau$; parameter)}\\
		\multicolumn{3}{c}{$\scd; \tau; \theta$}\\
		\midrule
		\multicolumn{3}{c}{C; 0.5; 2.00}\\
		\multicolumn{3}{c}{G; 0.5; 2.00}\\
		\hline
		\midrule
		\multicolumn{3}{c}{D-vine copula (pair-copula family; Kendall's $\tau$; parameter)}\\
		$\scd_{1,2}; \tau_{1,2}; \theta_{1,2}$ & $\scd_{2,3}; \tau_{2,3};\theta_{2,3}$  &  $\scd_{1,3;2}; \tau_{1,3;2}; \theta_{1,3;2}$\\
		\midrule
		C; 0.5; 2.00 & C; 0.5; 2.00 & F; 0.25; 2.37 \\
		G; 0.5; 2.00 & G; 0.5; 2.00 & F; 0.25; 2.37 \\
		\hline
		\hline	
	\end{tabular}
\end{table}

\subsection{One-stage parametric estimation}
\subsubsection*{Copula parameter estimates}
\begin{table}[H]
	\centering
	\small
	\caption{Simulation results using a \textbf{one-stage parametric estimation approach for three-dimensional data}. A \textbf{Clayton (3dC) copula} (top panel right) and a \textbf{Gumbel (3dG) copula} (bottom panel right) each with Kendall's $\tau = 0.5$ are considered. A \textbf{D-vine copula including Clayton copulas} (top panel left), \textbf{resp.\ Gumbel copulas} (bottom panel left), with $\tau_{1,2} = \tau_{2,3} = 0.5$ in $\mathcal{T}_1$ and a Frank (F) copula with $\tau_{1,3;2} = 0.25$ in $\mathcal{T}_2$ is considered. For the D-vine copulas \textbf{global and sequential likelihood estimation} is reported. The \textbf{empirical mean (empirical standard deviation) for the copula parameter estimates} are presented based on 250 replications and samples of size 250 and 500 affected by either 15\%, 30\% or heavy tail 30\% right-censoring.}
	\label{Table:3d_ClaytonGumbel_1stage_theta}
	\begin{tabular}{cccccccc}
		\midrule
		& & & & \multicolumn{3}{c}{D-vine copula model} & Archimedean copula\\
		&  &  &  & C; \ $\theta_{1,2}: 2.00$ & C; \ $\theta_{2,3}: 2.00$ & F; \ $\theta_{1,3;2}: 2.37$ & 3dC; \ $\theta: 2.00$ \\
		\midrule\midrule
		\multirow{12}{*} {\begin{sideways} parametric one-stage \end{sideways}} & \multirow{6}{*} {\begin{sideways} global \end{sideways}} & \multirow{2}{*} {15\%} & $250$ & 2.036 (0.292) & 2.071 (0.353) & 2.396 (0.554) & 2.039 (0.266)\\
		&  &  & $500$ & 2.019 (0.218) & 2.039 (0.242) & 2.389 (0.375) & 2.020 (0.198) \\
		\cmidrule{3-8}
		&  & \multirow{2}{*} {30\%} & $250$ & 2.044 (0.412) & 2.091 (0.503) & 2.407 (0.722) & 2.069 (0.345) \\
		&  &  & $500$ & 2.030 (0.269) & 2.068 (0.341) & 2.398 (0.489) & 2.018 (0.233) \\
		\cmidrule{3-8}
		&  & \multirow{2}{*} {30\% HT} & $250$ & 2.082 (0.494) & 2.121 (0.670) & 2.388 (0.870) & 2.078 (0.424) \\
		&  &  & $500$ & 2.054 (0.339) & 2.061 (0.471) & 2.380 (0.563) & 2.019 (0.307) \\
		\cmidrule{2-8}
		& \multirow{6}{*} {\begin{sideways} sequential \end{sideways}} & \multirow{2}{*} {15\%} & $250$ & 2.033 (0.293) & 2.067 (0.359) & 2.391 (0.551) & \\
		&  &  & $500$ & 2.020 (0.224) & 2.039 (0.247) & 2.387 (0.375) &\\
		\cmidrule{3-7}
		&  & \multirow{2}{*} {30\%} & $250$ & 2.042 (0.415) & 2.088 (0.506) & 2.401 (0.721) &\\
		&  &  & $500$ & 2.028 (0.273) & 2.066 (0.343) & 2.395 (0.488) &\\
		\cmidrule{3-7}
		&  & \multirow{2}{*} {30\% HT} & $250$ & 2.084 (0.499) & 2.122 (0.665) & 2.387 (0.867) &\\
		&  &  & $500$ & 2.055 (0.340) & 2.060 (0.472) & 2.377 (0.563) &\\
		\midrule\midrule
		&  &  &  & G; \ $\theta_{1,2}: 2.00$ & G; \ $\theta_{2,3}: 2.00$ & F; \ $\theta_{1,3;2}: 2.37$ & 3dG; \ $\theta: 2.00$ \\
		\midrule\midrule
		\multirow{12}{*} {\begin{sideways} parametric one-stage \end{sideways}} & \multirow{6}{*} {\begin{sideways} global \end{sideways}} & \multirow{2}{*} {15\%} & $250$ & 2.001 (0.131) & 2.014 (0.145) & 2.397 (0.536) & 2.006 (0.119) \\
		&  &  & $500$ & 2.003 (0.108) & 2.009 (0.109) & 2.381 (0.374) & 2.005 (0.084) \\
		\cmidrule{3-8}
		&  & \multirow{2}{*} {30\%} & $250$ & 2.015 (0.159) & 2.030 (0.179) & 2.387 (0.705) & 2.024 (0.143) \\
		&  &  & $500$ & 2.016 (0.120) & 2.024 (0.134) & 2.401 (0.485) & 2.013 (0.102) \\
		\cmidrule{3-8}
		&  & \multirow{2}{*} {30\% HT} & $250$ & 2.041 (0.173) & 2.047 (0.191) & 2.357 (0.840) & 2.029 (0.165) \\
		&  &  & $500$ & 2.033 (0.128) & 2.035 (0.149) & 2.369 (0.509) & 2.019 (0.117)\\
		\cmidrule{2-8}
		& \multirow{6}{*} {\begin{sideways} sequential \end{sideways}} & \multirow{2}{*} {15\%} & $250$ & 2.001 (0.133) & 2.016 (0.145) & 2.393 (0.535) & \\
		&  &  & $500$ & 2.003 (0.110) & 2.012 (0.109) & 2.381 (0.373) &\\
		\cmidrule{3-7}
		&  & \multirow{2}{*} {30\%} & $250$ & 2.015 (0.159) & 2.030 (0.179) & 2.387 (0.705)& \\
		&  &  & $500$ & 2.008 (0.122) & 2.019 (0.135) & 2.406 (0.484) &\\
		\cmidrule{3-7}
		&  & \multirow{2}{*} {30\% HT} & $250$ & 2.012 (0.181) & 2.027 (0.192) & 2.383 (0.851) &\\
		&  &  & $500$ & 2.011 (0.132) & 2.020 (0.150) & 2.382 (0.513) &\\
		\midrule\midrule				
	\end{tabular}
\end{table}

\subsubsection*{Marginal estimates}
\renewcommand{\arraystretch}{1.5}
\begin{table}[H]
	\centering
	\small
	\caption{Simulation results using a \textbf{one-stage parametric estimation approach for three-dimensional data}. A \textbf{Clayton copula} (top panel) with $\tau=0.5$ and a \textbf{D-vine copula including Clayton copulas} (bottom panel) with $\tau_{1,2} = \tau_{2,3} = 0.5$ in $\mathcal{T}_1$ and a Frank (F) copula with $\tau_{1,3;2} = 0.25$ in $\mathcal{T}_2$ is considered. For the D-vine copula \textbf{global and sequential likelihood estimation} is reported. The \textbf{empirical mean (empirical standard deviation) for the marginal parameter estimates} are presented based on 250 replications and samples of size 250 and 500 affected by either 15\%, 30\% or heavy tail 30\% right-censoring.}
	\label{Table:3d_Clayton_1stage_margins}
	\hspace*{-.75cm}\begin{tabular}{p{.15cm}cccccccc}
		\midrule
		& & & \multicolumn{6}{c}{3d Clayton copula}\\
		&  &  & $\lambda_{1}: 0.50$ & $\rho_{1}: 1.50$ & $\lambda_{2}: 1.00$ & $\rho_{2}: 1.50$ & $\lambda_{3}: 1.00$ & $\rho_{3}: 1.50$ \\
		\midrule\midrule
		\multirow{6}{*} {\begin{sideways} global \end{sideways}} & \multirow{2}{*} {\begin{sideways} 15\% \end{sideways}} & $250$ & 0.495 (0.042) & 1.510 (0.072) & 0.994 (0.077) & 1.512 (0.083) & 0.997 (0.079) & 1.517 (0.088) \\
		&  & $500$ & 0.500 (0.032) & 1.504 (0.052) & 1.001 (0.055) & 1.505 (0.055) & 1.001 (0.055) & 1.501 (0.058) \\
		\cmidrule{2-9}
		& \multirow{2}{*} {\begin{sideways} 30\% \end{sideways}} & $250$ & 0.493 (0.041) & 1.512 (0.083) & 0.995 (0.094) & 1.518 (0.114) & 1.001 (0.101) & 1.525 (0.119) \\
		&  & $500$ & 0.498 (0.033) & 1.509 (0.064) & 1.006 (0.066) & 1.507 (0.069) & 1.006 (0.072) & 1.498 (0.076) \\
		\cmidrule{2-9}
		& \multirow{2}{*} {\begin{sideways} 30\% HT \end{sideways}} & $250$ & 0.494 (0.041) & 1.516 (0.092) & 0.995 (0.119) & 1.516 (0.130) & 1.014 (0.154) & 1.526 (0.146) \\
		& & $500$ & 0.499 (0.033) & 1.508 (0.070) & 1.010 (0.086) & 1.510 (0.079) & 1.013 (0.109) & 1.503 (0.103) \\
		\midrule\midrule
		& & & \multicolumn{6}{c}{Clayton based D-vine model}\\
		&  &  & $\lambda_{1}: 0.50$ & $\rho_{1}: 1.50$ & $\lambda_{2}: 1.00$ & $\rho_{2}: 1.50$ & $\lambda_{3}: 1.00$ & $\rho_{3}: 1.50$ \\
		\midrule\midrule
		\multirow{6}{*} {\begin{sideways} global \end{sideways}} & \multirow{2}{*} {\begin{sideways} 15\% \end{sideways}} & $250$ & 0.496 (0.044) & 1.516 (0.076) & 0.998 (0.081) & 1.512 (0.079) & 1.004 (0.082) & 1.512 (0.087) \\
		&  & $500$ & 0.498 (0.029) & 1.508 (0.049) & 0.999 (0.055) & 1.503 (0.058) & 1.000 (0.056) & 1.505 (0.062) \\
		\cmidrule{2-9}
		& \multirow{2}{*} {\begin{sideways} 30\% \end{sideways}} & $250$ & 0.497 (0.048) & 1.518 (0.084) & 1.007 (0.102) & 1.521 (0.102) & 1.017 (0.132) & 1.519 (0.123) \\
		&  & $500$ & 0.498 (0.030) & 1.507 (0.059) & 0.999 (0.065) & 1.502 (0.069) & 1.001 (0.080) & 1.509 (0.084) \\
		\cmidrule{2-9}
		& \multirow{2}{*} {\begin{sideways} 30\% HT \end{sideways}} & $250$ & 0.497 (0.046) & 1.517 (0.091) & 1.001 (0.122) & 1.509 (0.128) & 1.035 (0.218) & 1.519 (0.159) \\
		&  & $500$ & 0.498 (0.030) & 1.505 (0.063) & 0.994 (0.091) & 1.496 (0.085) & 1.018 (0.139) & 1.510 (0.106) \\
		\midrule
		\multirow{6}{*} {\begin{sideways} sequential \end{sideways}} & \multirow{2}{*} {\begin{sideways} 15\% \end{sideways}} & $250$ & 0.497 (0.046) & 1.517 (0.085) & 0.999 (0.082) & 1.513 (0.080) & 1.005 (0.083) & 1.512 (0.088) \\
		&  & $500$ & 0.498 (0.030) & 1.508 (0.054) & 0.999 (0.055) & 1.503 (0.061) & 1.000 (0.057) & 1.505 (0.062) \\
		\cmidrule{2-9}
		& \multirow{2}{*} {\begin{sideways} 30\% \end{sideways}} & $250$ & 0.497 (0.048) & 1.518 (0.090) & 1.008 (0.103) & 1.522 (0.102) & 1.017 (0.133) & 1.519 (0.124) \\
		&  & $500$ & 0.498 (0.031) & 1.508 (0.061) & 0.999 (0.066) & 1.502 (0.070) & 1.001 (0.080) & 1.510 (0.085) \\
		\cmidrule{2-9}
		& \multirow{2}{*} {\begin{sideways} 30\% HT \end{sideways}} & $250$ & 0.497 (0.047) & 1.513 (0.091) & 1.002 (0.124) & 1.508 (0.128) & 1.033 (0.211) & 1.517 (0.157) \\
		&  & $500$ & 0.498 (0.030) & 1.504 (0.064) & 0.994 (0.092) & 1.496 (0.084) & 1.019 (0.139) & 1.510 (0.107) \\
		\midrule\midrule
	\end{tabular}
	%\vspace*{-1cm}
\end{table}
%\newpage

\begin{table}[H]
	\centering
	\small
	\caption{Simulation results using a \textbf{one-stage parametric estimation approach for three-dimensional data}. A  \textbf{Gumbel copula} (top panel) with $\tau=0.5$ and a \textbf{D-vine copula including Gumbel copulas} (bottom panel) with $\tau_{1,2} = \tau_{2,3} = 0.5$ in $\mathcal{T}_1$ and a Frank (F) copula with $\tau_{1,3;2} = 0.25$ in $\mathcal{T}_2$ is considered. For the D-vine copula \textbf{global and sequential likelihood estimation} is reported. The \textbf{empirical mean (empirical standard deviation) for the marginal parameter estimates} are presented based on 250 replications and samples of size 250 and 500 affected by either 15\%, 30\% or heavy tail 30\% right-censoring.}
	\label{Table:3d_Gumbel_1stage_margins}
	\hspace*{-.75cm}\begin{tabular}{p{.15cm}cccccccc}
		\midrule
		& & & \multicolumn{6}{c}{3d Gumbel copula}\\
		&  &  & $\lambda_{1}: 0.50$ & $\rho_{1}: 1.50$ & $\lambda_{2}: 1.00$ & $\rho_{2}: 1.50$ & $\lambda_{3}: 1.00$ & $\rho_{3}: 1.50$ \\
		\midrule\midrule
		\multirow{6}{*} {\begin{sideways} global \end{sideways}} & \multirow{2}{*} {\begin{sideways} 15\% \end{sideways}} & $250$ & 0.500 (0.043) & 1.508 (0.084) & 1.004 (0.084) & 1.513 (0.079) & 1.004 (0.080) & 1.501 (0.085) \\
		&  & $500$ & 0.498 (0.031) & 1.505 (0.054) & 1.002 (0.057) & 1.503 (0.055) & 0.989 (0.059) & 1.496 (0.062) \\
		\cmidrule{2-9}
		& \multirow{2}{*} {\begin{sideways} 30\% \end{sideways}} & $250$ & 0.501 (0.043) & 1.506 (0.090) & 1.002 (0.097) & 1.509 (0.099) & 1.003 (0.112) & 1.498 (0.103) \\
		&  & $500$ & 0.500 (0.032) & 1.505 (0.060) & 1.006 (0.068) & 1.505 (0.070) & 0.991 (0.084) & 1.497 (0.076) \\
		\cmidrule{2-9}
		& \multirow{2}{*} {\begin{sideways} 30\% HT \end{sideways}} & $250$ & 0.501 (0.043) & 1.507 (0.095) & 0.997 (0.111) & 1.507 (0.116) & 1.016 (0.161) & 1.503 (0.132) \\
		& & $500$ & 0.500 (0.030) & 1.503 (0.067) & 1.006 (0.082) & 1.502 (0.082) & 0.995 (0.124) & 1.495 (0.098) \\
		\midrule\midrule
		& & & \multicolumn{6}{c}{Gumbel based D-vine model}\\
		&  &  & $\lambda_{1}: 0.50$ & $\rho_{1}: 1.50$ & $\lambda_{2}: 1.00$ & $\rho_{2}: 1.50$ & $\lambda_{3}: 1.00$ & $\rho_{3}: 1.50$ \\
		\midrule\midrule
		\multirow{6}{*} {\begin{sideways} global \end{sideways}} & \multirow{2}{*} {\begin{sideways} 15\% \end{sideways}} & $250$ & 0.485 (0.038) & 1.525 (0.081) & 0.961 (0.038) & 1.517 (0.085) & 0.987 (0.078) & 1.526 (0.094) \\
		&  & $500$ & 0.489 (0.025) & 1.515 (0.053) & 0.969 (0.027) & 1.507 (0.058) & 0.985 (0.049) & 1.514 (0.061) \\
		\cmidrule{2-9}
		& \multirow{2}{*} {\begin{sideways} 30\% \end{sideways}} & $250$ & 0.486 (0.041) & 1.521 (0.089) & 0.954 (0.048) & 1.510 (0.097) & 0.991 (0.133) & 1.522 (0.130) \\
		&  & $500$ & 0.491 (0.028) & 1.511 (0.060) & 0.963 (0.035) & 1.496 (0.069) & 0.984 (0.085) & 1.508 (0.082) \\
		\cmidrule{2-9}
		& \multirow{2}{*} {\begin{sideways} 30\% HT \end{sideways}} & $250$ & 0.490 (0.042) & 1.510 (0.091) & 0.948 (0.056) & 1.488 (0.103) & 1.014 (0.197) & 1.517 (0.152) \\
		&  & $500$ & 0.493 (0.028) & 1.503 (0.063) & 0.955 (0.048) & 1.481 (0.071) & 0.993 (0.128) & 1.501 (0.098) \\
		\midrule
		\multirow{6}{*} {\begin{sideways} sequential \end{sideways}} & \multirow{2}{*} {\begin{sideways} 15\% \end{sideways}} & $250$ & 0.497 (0.046) & 1.517 (0.085) & 0.999 (0.079) & 1.516 (0.086) & 1.009 (0.091) & 1.518 (0.095) \\
		&  & $500$ & 0.489 (0.025) & 1.515 (0.053) & 0.969 (0.027) & 1.507 (0.058) & 0.985 (0.049) & 1.514 (0.061) \\
		\cmidrule{2-9}
		& \multirow{2}{*} {\begin{sideways} 30\% \end{sideways}} & $250$ & 0.498 (0.047) & 1.518 (0.090) & 1.007 (0.102) & 1.521 (0.099) & 1.015 (0.142) & 1.518 (0.123) \\
		&  & $500$ & 0.499 (0.031) & 1.507 (0.061) & 0.999 (0.070) & 1.503 (0.071) & 1.002 (0.093) & 1.505 (0.080) \\
		\cmidrule{2-9}
		& \multirow{2}{*} {\begin{sideways} 30\% HT \end{sideways}} & $250$ & 0.497 (0.047) & 1.515 (0.093) & 1.006 (0.117) & 1.513 (0.112) & 1.046 (0.211) & 1.528 (0.153) \\
		&  & $500$ & 0.498 (0.030) & 1.504 (0.064) & 0.996 (0.088) & 1.499 (0.080) & 1.016 (0.137) & 1.509 (0.100) \\
		\midrule\midrule
	\end{tabular}
\end{table}
\renewcommand{\arraystretch}{1}	
\vspace*{-.00cm}
\subsection{Two-stage semiparametric estimation}
\subsubsection*{Copula parameter estimates}

\begin{table}[H]
	\centering
	\small
	\caption{Simulation results using a \textbf{two-stage semiparametric estimation approach for three-dimensional data}. A \textbf{Clayton (3dC) copula} (top panel right) and a \textbf{Gumbel (3dG) copula} (bottom panel right) with Kendall's $\tau = 0.5$ is considered. A \textbf{D-vine copula including Clayton copulas} (top panel left), \textbf{resp. Gumbel copulas} (bottom panel left), with $\tau_{1,2} = \tau_{2,3} = 0.5$ in $\mathcal{T}_1$ and a Frank (F) copula with $\tau_{1,3;2} = 0.25$ in $\mathcal{T}_2$ is considered. For the D-vine copulas \textbf{global and sequential likelihood estimation} is reported. The \textbf{empirical mean (empirical standard deviation) for the copula parameter estimates} are presented based on 250 replications and samples of size 250, 500 and 1000 affected by either 15\%, 30\% or heavy tail 30\% right-censoring.}
	\label{Table:3d_ClaytonGumbel_2stage_theta}
	\begin{tabular}{cccccccc}
		\midrule
		& & & & \multicolumn{3}{c}{D-vine copula model} & Archimedean copula \\
		&  &  &  & C; \ $\theta_{1,2}: 2.00$ & C; \ $\theta_{2,3}: 2.00$ & F; \ $\theta_{1,3;2}: 2.37$ & 3dC; \ $\theta: 2.00$ \\
		\midrule\midrule
		\multirow{18}{*} {\begin{sideways} semiparametric two-stage \end{sideways}} & \multirow{9}{*} {\begin{sideways} global \end{sideways}} & \multirow{3}{*} {15\%} & $250$ & 1.985 (0.329) & 2.011 (0.370) & 2.422 (0.567) & 1.994 (0.315)\\
		&  &  & $500$ & 1.990 (0.274) & 2.003 (0.288) & 2.416 (0.388) & 1.984 (0.225)\\
		&  &  & $1000$ & 1.991 (0.180) & 1.994 (0.188) & 2.384 (0.265) & 1.993 (0.148)\\
		\cmidrule{3-8}
		&  & \multirow{3}{*} {30\%} & $250$ & 2.114 (0.601) & 2.131 (0.617) & 2.392 (0.747) & 2.090 (0.511) \\
		&  &  & $500$ & 2.041 (0.359) & 2.075 (0.414) & 2.420 (0.494) & 2.012 (0.338)\\
		&  &  & $1000$ & 2.041 (0.288) & 2.035 (0.299) & 2.393 (0.375) & 1.996 (0.245)\\
		\cmidrule{3-8}
		&  & \multirow{3}{*} {30\% HT} & $250$ & 2.785 (1.094) & 2.736 (1.062) & 2.377 (0.959) & 2.590 (0.923) \\
		&  &  & $500$ & 2.701 (0.881) & 2.599 (0.836) & 2.352 (0.653) & 2.505 (0.752)\\
		&  &  & $1000$ & 2.554 (0.662) & 2.402 (0.625) & 2.302 (0.439) & 2.409 (0.601) \\
		\cmidrule{2-8}
		& \multirow{9}{*} {\begin{sideways} sequential  \end{sideways}} & \multirow{3}{*} {15\%} & $250$ & 1.990 (0.327) & 2.009 (0.368) & 2.414 (0.564) &\\
		&  &  & $500$ & 1.995 (0.272) & 2.003 (0.284) & 2.411 (0.387) &\\
		&  &  & $1000$ & 1.993 (0.178) & 1.992 (0.186) & 2.382 (0.265) &\\
		\cmidrule{3-7}
		&  & \multirow{3}{*} {30\%} & $250$ & 2.116 (0.596) & 2.176 (0.607) & 2.365 (0.737) &\\
		&  &  & $500$ & 2.046 (0.356) & 2.123 (0.415) & 2.400 (0.487) &\\
		&  &  & $1000$ & 2.041 (0.285) & 2.079 (0.289) & 2.379 (0.372) &\\
		\cmidrule{3-7}
		&  & \multirow{3}{*} {30\% HT} & $250$ & 2.769 (1.072) & 2.773 (0.956) & 2.336 (0.939) &\\
		&  &  & $500$ & 2.682 (0.861) & 2.641 (0.767) & 2.326 (0.640) &\\
		&  &  & $1000$ & 2.541 (0.646) & 2.463 (0.568) & 2.279 (0.428) &\\
		\midrule\midrule
		&  &  &  & G; \ $\theta_{1,2}: 2.00$ & G; \ $\theta_{2,3}: 2.00$ & F; \ $\theta_{1,3;2}: 2.37$ & 3dG; \ $\theta: 2.00$ \\
		\midrule\midrule
		\multirow{18}{*} {\begin{sideways} semiparametric two-stage \end{sideways}} & \multirow{9}{*} {\begin{sideways} global \end{sideways}} & \multirow{3}{*} {15\%} & $250$ & 2.014 (0.152) & 2.036 (0.162) & 2.397 (0.552) & 2.026 (0.135)\\
		&  &  & $500$ & 2.018 (0.114) & 2.028 (0.114) & 2.389 (0.381) &2.018 (0.092)\\
		&  &  & $1000$ & 2.009 (0.079) & 2.012 (0.084) & 2.371 (0.263) & 2.016 (0.060)\\
		\cmidrule{3-8}
		&  & \multirow{3}{*} {30\%} & $250$ & 2.034 (0.205) & 2.067 (0.206) & 2.395 (0.731) & 2.061 (0.177) \\
		&  &  & $500$ & 2.035 (0.135) & 2.051 (0.156) & 2.429 (0.496) &2.037 (0.116)\\
		&  &  & $1000$ & 2.025 (0.098) & 2.024 (0.103) & 2.375 (0.388) &2.029 (0.080)\\
		\cmidrule{3-8}
		&  & \multirow{3}{*} {30\% HT} & $250$ & 2.191 (0.299) & 2.164 (0.261) & 2.442 (0.963) & 2.184 (0.281) \\
		&  &  & $500$ & 2.183 (0.232) & 2.139 (0.224) & 2.437 (0.679) & 2.154 (0.206)\\
		&  &  & $1000$ & 2.157 (0.177) & 2.105 (0.152) & 2.372 (0.468) & 2.122 (0.167)\\
		\cmidrule{2-8}
		& \multirow{9}{*} {\begin{sideways} sequential  \end{sideways}} & \multirow{3}{*} {15\%} & $250$ & 2.014 (0.152) & 2.033 (0.162) & 2.392 (0.548) &\\
		&  &  & $500$ & 2.018 (0.115) & 2.022 (0.116) & 2.389 (0.381) &\\
		&  &  & $1000$ & 2.008 (0.079) & 2.007 (0.085) & 2.373 (0.264) &\\
		\cmidrule{3-7}
		&  & \multirow{3}{*} {30\%} & $250$ & 2.034 (0.205) & 2.084 (0.211) & 2.373 (0.718) &\\
		&  &  & $500$ & 2.036 (0.134) & 2.065 (0.159) & 2.416 (0.491) &\\
		&  &  & $1000$ & 2.024 (0.098) & 2.035 (0.106) & 2.366 (0.385) &\\
		\cmidrule{3-7}
		&  & \multirow{3}{*} {30\% HT} & $250$ & 2.191 (0.299) & 2.164 (0.261) & 2.442 (0.963) &\\
		&  &  & $500$ & 2.186 (0.233) & 2.165 (0.221) & 2.409 (0.653) &\\
		&  &  & $1000$ & 2.158 (0.177) & 2.131 (0.149) & 2.350 (0.457) &\\
		\midrule
		\midrule	
	\end{tabular}
\end{table}

\section{Additional material for extensive simulations}\label{Sec:AddOnsIlluSims2}

\subsection{Simulation settings}

\begin{table}[ht]
	\centering
	%	\small
	\captionof{table}{Simulation settings for the marginal survival functions of the four gap times and for the survival function of the censoring times leading to 15\%, 30\% or 30\% HT (heavy tail) censoring.}
	\label{table:marginalSimSettings_4d}
	\begin{tabular}{ r c c c c c} \midrule
		\multirow{2}{*}{\centering Weibull parameters} & \multirow{2}{*}{Gap time 1} & \multirow{2}{*}{Gap time 2 - 4} & \multicolumn{3}{c}{Censoring} \\
		& & & 15\%  & 30\% & 30\% HT\\
		\hline
		\midrule
		scale $\lambda$ & 0.5 & 1 & 0.085 & 0.25 & 0.085 \\
		shape $\rho$ & 1.5 & 1.5 & 1.5 & 1.5 & 3\\
		\hline\hline	
	\end{tabular}
	
\end{table}

\begin{table}[ht]
	\centering
	%\small
	\captionof{table}{Simulation settings for D-vine copulas.}
	\label{table:CopulaSimSettings_4d}
	\hspace*{-.75cm}\begin{tabular}{c c c c}\midrule
		& \multicolumn{3}{c}{D-vine copula (pair-copula families; Kendall's $\tau$; parameter)}\\
		&	$\scd_{1,2}; \tau_{1,2}; \theta_{1,2}$ & $\scd_{2,3}; \tau_{2,3};\theta_{2,3}$  & $\scd_{3,4}; \tau_{3,4};\theta_{3,4}$\\
		\midrule
		Setting 1&	C; 0.5; 2.00 & F; 0.5; 5.76 & G; 0.5; 2.00\\
		Setting 2 &	C; 0.3; 0.86 & C; 0.5; 2.00 & C; 0.7; 4.67\\
		\hline\midrule		
		&	 $\scd_{1,3;2}; \tau_{1,3;2}; \theta_{1,3;2}$ & $\scd_{2,4;3}; \tau_{2,4;3}; \theta_{2,4;3}$  & $\scd_{1,4;2:3}; \tau_{1,4;2:3}; \theta_{1,4;2:3}$\\
		\midrule
		Setting 1&	 F; 0.25; 2.37 & F; 0.25; 2.37 & F; 0.167; 1.53\\
		Setting 2 &	 F; 0.25; 2.37 & F; 0.25; 2.37 & F; 0.167; 1.53\\
		\hline
		\hline	
	\end{tabular}	
\end{table}

\subsection{One-stage parametric estimation: Marginal estimates}

\begin{landscape}
	\begin{table}[H]
		\centering
		\vspace*{-1.2cm}
		\caption{Simulation results for the \textbf{marginal parameter estimates} in case of \textbf{one-stage parametric (global and sequential) estimation}. In the top panels, the underlying D-vine copula model captures tail-behavior for subsequent gap times changing from lower tail-dependence (Clayton (C)) over no tail-dependence (Frank (F)) to upper tail-dependence (Gumbel (G)) with same overall dependence of Kendall's $\tau_{1,2}=\tau_{2,3}=\tau_{3,4}=0.5$. In the bottom panels, the underlying D-vine copula model captures for Clayton (C) copulas in $\mathcal{T}_1$ increasing dependence with $\tau_{1,2}=0.3$, $\tau_{2,3}=0.5$, $\tau_{3,4}=0.7$. The \textbf{empirical mean (empirical standard deviation) of the marginal parameter estimates} are presented based on 250 replications and samples of different sizes affected by either 15\%, 30\% or heavy tail 30\% right-censoring.}
		\label{Table:4d_CFG_margins}
		\begin{tabular}{cccccccccccc}
			\midrule
			&  &  &  & $\lambda_{1}: 0.50$ & $\rho_{1}: 1.50$ & $\lambda_{2}: 1.00$ & $\rho_{2}: 1.50$ & $\lambda_{3}: 1.00$ & $\rho_{3}: 1.50$ & $\lambda_{4}: 1.00$ & $\rho_{4}: 1.50$ \\
			\midrule\midrule
			\multirow{12}{*} {\begin{sideways} Setting 1 (\autoref{table:CopulaSimSettings_4d}) \end{sideways}} & \multirow{6}{*} {\begin{sideways} global \end{sideways}} & \multirow{2}{*} {15\%} & $250$ & 0.491 (0.036) & 1.507 (0.074) & 0.980 (0.061) & 1.517 (0.068) & 0.965 (0.034) & 1.513 (0.087) & 0.984 (0.074) & 1.530 (0.101) \\
			&  &  & $500$ & 0.492 (0.026) & 1.508 (0.053) & 0.986 (0.045) & 1.513 (0.052) & 0.972 (0.023) & 1.509 (0.063) & 0.986 (0.058) & 1.518 (0.066) \\
			\cmidrule{3-12}
			&  & \multirow{2}{*} {30\%} & $250$ & 0.493 (0.040) & 1.508 (0.093) & 0.981 (0.079) & 1.513 (0.093) & 0.951 (0.053) & 1.499 (0.113) & 1.004 (0.158) & 1.528 (0.143) \\
			&  &  & $500$ & 0.494 (0.029) & 1.505 (0.064) & 0.989 (0.061) & 1.515 (0.070) & 0.964 (0.037) & 1.500 (0.085) & 0.987 (0.111) & 1.512 (0.101) \\
			\cmidrule{3-12}
			&  & \multirow{2}{*} {30\% HT} & $250$ & 0.494 (0.040) & 1.507 (0.095) & 0.984 (0.106) & 1.509 (0.112) & 0.939 (0.075) & 1.485 (0.122) & 1.025 (0.259) & 1.526 (0.188) \\
			&  &  & $500$ & 0.495 (0.029) & 1.505 (0.064) & 0.988 (0.074) & 1.510 (0.082) & 0.949 (0.054) & 1.483 (0.087) & 0.986 (0.165) & 1.500 (0.122) \\
			\cmidrule{2-12}
			& \multirow{6}{*} {\begin{sideways} sequential \end{sideways}} & \multirow{2}{*} {15\%} & $250$ & 0.500 (0.042) & 1.506 (0.082) & 1.001 (0.073) & 1.510 (0.073) & 1.005 (0.074) & 1.518 (0.091) & 1.005 (0.083) & 1.525 (0.105) \\
			&  &  & $500$ & 0.501 (0.031) & 1.504 (0.058) & 1.004 (0.053) & 1.507 (0.056) & 1.006 (0.055) & 1.511 (0.065) & 1.004 (0.065) & 1.511 (0.068) \\
			\cmidrule{3-12}
			&  & \multirow{2}{*} {30\%} & $250$ & 0.501 (0.044) & 1.508 (0.095) & 1.003 (0.091) & 1.509 (0.093) & 1.016 (0.119) & 1.522 (0.121) & 1.031 (0.169) & 1.533 (0.142) \\
			&  &  & $500$ & 0.501 (0.033) & 1.502 (0.067) & 1.007 (0.068) & 1.510 (0.072) & 1.013 (0.088) & 1.516 (0.093) & 1.010 (0.120) & 1.515 (0.102) \\
			\cmidrule{3-12}
			&  & \multirow{2}{*} {30\% HT} & $250$ & 0.502 (0.042) & 1.502 (0.093) & 1.007 (0.113) & 1.508 (0.111) & 1.019 (0.163) & 1.515 (0.140) & 1.046 (0.252) & 1.527 (0.180) \\
			&  &  & $500$ & 0.501 (0.031) & 1.502 (0.067) & 1.008 (0.081) & 1.509 (0.084) & 1.020 (0.128) & 1.514 (0.108) & 1.023 (0.179) & 1.514 (0.125) \\
			\midrule\midrule
			%		&  &  &  & $\lambda_{1}: 0.50$ & $\rho_{1}: 1.50$ & $\lambda_{2}: 1.00$ & $\rho_{2}: 1.50$ & $\lambda_{3}: 1.00$ & $\rho_{3}: 1.50$ & $\lambda_{4}: 1.00$ & $\rho_{4}: 1.50$ \\
			%		\midrule\midrule
			\multirow{12}{*} {\begin{sideways} Setting 2 (\autoref{table:CopulaSimSettings_4d}) \end{sideways}} & \multirow{6}{*} {\begin{sideways} global \end{sideways}} & \multirow{2}{*} {15\%} & $250$ & 0.500 (0.042) & 1.507 (0.080) & 1.000 (0.077) & 1.511 (0.074) & 1.003 (0.072) & 1.521 (0.078) & 1.521 (0.078) & 1.524 (0.084) \\
			&  &  & $500$ & 0.501 (0.030) & 1.505 (0.055) & 1.004 (0.058) & 1.509 (0.053) & 1.005 (0.056) & 1.512 (0.053) & 1.512 (0.053) & 1.511 (0.058) \\
			\cmidrule{3-12}
			&  & \multirow{2}{*} {30\%} & $250$ & 0.501 (0.043) & 1.508 (0.095) & 1.005 (0.110) & 1.512 (0.100) & 1.017 (0.122) & 1.534 (0.111) & 1.534 (0.111) & 1.531 (0.131) \\
			&  &  & $500$ & 0.501 (0.033) & 1.503 (0.065) & 1.006 (0.079) & 1.512 (0.070) & 1.013 (0.082) & 1.514 (0.077) & 1.514 (0.077) & 1.517 (0.088) \\
			\cmidrule{3-12}
			&  & \multirow{2}{*} {30\% HT} & $250$ & 0.501 (0.042) & 1.505 (0.096) & 1.008 (0.139) & 1.509 (0.113) & 1.027 (0.172) & 1.527 (0.143) & 1.527 (0.143) & 1.559 (0.186) \\
			&  &  & $500$ & 0.501 (0.031) & 1.502 (0.066) & 1.011 (0.091) & 1.512 (0.080) & 1.016 (0.122) & 1.514 (0.096) & 1.514 (0.096) & 1.521 (0.121) \\
			\cmidrule{2-12}
			& \multirow{6}{*} {\begin{sideways} sequential \end{sideways}} & \multirow{2}{*} {15\%} & $250$ & 0.500 (0.042) & 1.506 (0.082) & 1.001 (0.077) & 1.508 (0.077) & 1.003 (0.072) & 1.515 (0.084) & 1.515 (0.084) & 1.519 (0.088) \\
			&  &  & $500$ & 0.499 (0.029) & 1.503 (0.058) & 0.999 (0.056) & 1.506 (0.057) & 0.998 (0.053) & 1.511 (0.059) & 1.511 (0.059) & 1.509 (0.063) \\
			\cmidrule{3-12}
			&  & \multirow{2}{*} {30\%} & $250$ & 0.501 (0.044) & 1.507 (0.095) & 1.004 (0.110) & 1.511 (0.101) & 1.015 (0.121) & 1.527 (0.116) & 1.527 (0.116) & 1.527 (0.134) \\
			&  &  & $500$ & 0.502 (0.033) & 1.500 (0.066) & 1.008 (0.079) & 1.510 (0.073) & 1.016 (0.084) & 1.514 (0.082) & 1.514 (0.082) & 1.513 (0.088) \\
			\cmidrule{3-12}
			&  & \multirow{2}{*} {30\% HT} & $250$ & 0.502 (0.043) & 1.501 (0.094) & 1.005 (0.138) & 1.502 (0.112) & 1.009 (0.169) & 1.509 (0.134) & 1.509 (0.134) & 1.531 (0.185) \\
			&  &  & $500$ & 0.501 (0.031) & 1.502 (0.067) & 1.010 (0.092) & 1.507 (0.079) & 1.018 (0.123) & 1.515 (0.096) & 1.515 (0.096) & 1.517 (0.119) \\
			\midrule\midrule
		\end{tabular}\vspace*{-5cm}
	\end{table}
\end{landscape}

\subsection{One-stage parametric and two-stage semiparametric estimation:\\ Kendall's $\tau$ and/or copula parameter estimates}

\renewcommand{\arraystretch}{1.25}
\begin{table}[H]
	\centering
	\small
	\vspace*{-.35cm}
	\caption{Simulation results using \textbf{sequential one-stage parametric and sequential two-stage semiparametric estimation} for four-dimensional data. In the top panels, the D-vine copula model captures tail-behavior for subsequent gap times changing from lower tail-dependence (Clayton (C)) over no tail-dependence (Frank (F)) to upper tail-dependence (Gumbel (G)) with same overall dependence of Kendall's $\tau_{1,2}=\tau_{2,3}=\tau_{3,4}=0.5$. In the bottom panels, the D-vine copula model captures for Clayton (C) copulas in $\mathcal{T}_1$ increasing dependence with $\tau_{1,2}=0.3$, $\tau_{2,3}=0.5$, $\tau_{3,4}=0.7$. The \textbf{empirical mean (empirical standard deviation) of the Kendall's $\boldsymbol{\tau}$ estimates} are presented based on 250 replications and samples of different sizes affected by either 15\%, 30\% or heavy tail 30\% right-censoring.}
	\label{Table:4d_CFG_tau}
	\hspace*{-1.1cm}\begin{tabular}{p{.15cm}p{.15cm}ccccccc}
		\midrule
		& & & \multicolumn{6}{c}{D-vine copula model}\\
		&  &  & C; $\tau_{1,2}:0.50$ & F; $\tau_{2,3}: 0.50$ & G; $\tau_{3,4}: 0.50$ & F; $\tau_{1,3;2}: 0.25$ & F; $\tau_{2,4;3}: 0.25$ & F; $\tau_{1,4;2,3}: 0.17$ \\
		\midrule \midrule
		\multirow{9}{*} {\begin{sideways} semiparametric two-stage \end{sideways}} &   \multirow{3}{*} {\begin{sideways} 15\% \end{sideways}} & $250$ & 0.497 (0.047) & 0.500 (0.039) & 0.507 (0.041) & 0.249 (0.049) & 0.251 (0.057) & 0.163 (0.059) \\
		&  &   $500$ & 0.499 (0.033) & 0.500 (0.028) & 0.504 (0.030) & 0.251 (0.035) & 0.249 (0.037) & 0.163 (0.040) \\
		&  &   $1000$ & 0.499 (0.024) & 0.498 (0.019) & 0.501 (0.022) & 0.248 (0.024) & 0.252 (0.026) & 0.164 (0.028) \\
		\cmidrule{2-9}
		&   \multirow{3}{*} {\begin{sideways} 30\% \end{sideways}} & $250$ & 0.529 (0.092) & 0.523 (0.062) & 0.530 (0.057) & 0.247 (0.065) & 0.248 (0.079) & 0.152 (0.083) \\
		&    & $500$ & 0.506 (0.071) & 0.515 (0.047) & 0.520 (0.041) & 0.253 (0.047) & 0.249 (0.051) & 0.159 (0.056) \\
		&    & $1000$ & 0.509 (0.052) & 0.511 (0.033) & 0.515 (0.031) & 0.248 (0.038) & 0.247 (0.036) & 0.159 (0.041) \\
		\cmidrule{2-9}
		& \multirow{3}{*} {\begin{sideways} 30\%HT \end{sideways}} & $250$ & 0.638 (0.147) & 0.583 (0.108) & 0.559 (0.060) & 0.239 (0.086) & 0.242 (0.089) & 0.143 (0.085) \\
		&  & $500$ & 0.616 (0.113) & 0.573 (0.077) & 0.546 (0.051) & 0.245 (0.054) & 0.247 (0.066) & 0.154 (0.068) \\
		&  & $1000$ & 0.614 (0.114) & 0.565 (0.074) & 0.546 (0.036) & 0.245 (0.053) & 0.250 (0.047) & 0.155 (0.044) \\
		\midrule
		\multirow{6}{*} {\begin{sideways} parametric one-stage \end{sideways}} &  \multirow{2}{*} {\begin{sideways} 15\% \end{sideways}} & $250$ & 0.499 (0.037) & 0.499 (0.036) & 0.500 (0.037) & 0.251 (0.049) & 0.252 (0.055) & 0.168 (0.060) \\
		&  & $500$ & 0.500 (0.023) & 0.499 (0.026) & 0.500 (0.027) & 0.252 (0.035) & 0.250 (0.037) & 0.165 (0.040) \\
		\cmidrule{2-9}
		& \multirow{2}{*} {\begin{sideways} 30\% \end{sideways}} & $250$ & 0.500 (0.046) & 0.500 (0.052) & 0.504 (0.053) & 0.249 (0.061) & 0.255 (0.077) & 0.164 (0.087) \\
		&  & $500$ & 0.499 (0.032) & 0.499 (0.037) & 0.500 (0.038) & 0.253 (0.046) & 0.250 (0.052) & 0.165 (0.056) \\
		\cmidrule{2-9}
		& \multirow{2}{*} {\begin{sideways} 30\%HT \end{sideways}} & $250$ & 0.500 (0.057) & 0.501 (0.059) & 0.506 (0.058) & 0.252 (0.071) & 0.248 (0.086) & 0.164 (0.095) \\
		&  & $500$ & 0.498 (0.038) & 0.498 (0.043) & 0.500 (0.045) & 0.251 (0.048) & 0.252 (0.062) & 0.164 (0.063) \\
		\midrule\midrule
		&   &  & C; \ $\tau_{1,2}: 0.30$ & C; \ $\tau_{2,3}: 0.50$ & C; \ $\tau_{3,4}: 0.70$ & F; \ $\tau_{1,3;2}: 0.25$ & F; \ $\tau_{2,4;3}: 0.25$ & F; \ $\tau_{1,4;2,3}: 0.17$ \\
		\midrule\midrule
		\multirow{9}{*} {\begin{sideways} semiparametric two-stage \end{sideways}} & \multirow{3}{*} {\begin{sideways} 15\% \end{sideways}} & $250$ & 0.311 (0.055) & 0.498 (0.047) & 0.693 (0.035) & 0.248 (0.050) & 0.251 (0.055) & 0.162 (0.059) \\
		&    & $500$ & 0.309 (0.040) & 0.499 (0.034) & 0.696 (0.024) & 0.250 (0.037) & 0.251 (0.039) & 0.161 (0.040) \\
		&    & $1000$ & 0.304 (0.028) & 0.495 (0.025) & 0.696 (0.017) & 0.247 (0.025) & 0.252 (0.026) & 0.164 (0.027) \\
		\cmidrule{2-9}
		&   \multirow{3}{*} {\begin{sideways} 30\% \end{sideways}} & $250$ & 0.364 (0.116) & 0.526 (0.085) & 0.700 (0.058) & 0.233 (0.069) & 0.251 (0.077) & 0.152 (0.090) \\
		&    & $500$ & 0.331 (0.082) & 0.513 (0.063) & 0.698 (0.042) & 0.244 (0.054) & 0.249 (0.055) & 0.160 (0.055) \\
		&    & $1000$ & 0.330 (0.060) & 0.512 (0.047) & 0.700 (0.029) & 0.244 (0.040) & 0.250 (0.036) & 0.158 (0.041) \\
		\cmidrule{2-9}
		&  \multirow{3}{*} {\begin{sideways} 30\%HT \end{sideways}} & $250$ & 0.300 (0.080) & 0.499 (0.079) & 0.699 (0.060) & 0.244 (0.079) & 0.254 (0.088) & 0.163 (0.102) \\
		&    & $500$ & 0.489 (0.155) & 0.596 (0.107) & 0.726 (0.064) & 0.212 (0.067) & 0.246 (0.072) & 0.133 (0.075) \\
		&    & $1000$ & 0.495 (0.134) & 0.595 (0.095) & 0.733 (0.050) & 0.220 (0.061) & 0.242 (0.050) & 0.138 (0.053) \\
		\midrule\midrule
		\multirow{6}{*} {\begin{sideways} parametric one-stage \end{sideways}} 
		& \multirow{2}{*} {\begin{sideways} 15\% \end{sideways}} & $250$ & 0.300 (0.045) & 0.501 (0.039) & 0.701 (0.028) & 0.249 (0.048) & 0.252 (0.054) & 0.168 (0.061) \\
		&  & $500$ & 0.302 (0.030) & 0.503 (0.026) & 0.701 (0.020) & 0.252 (0.035) & 0.249 (0.038) & 0.164 (0.041) \\
		\cmidrule{2-9}
		& \multirow{2}{*} {\begin{sideways} 30\% \end{sideways}} & $250$ & 0.300 (0.064) & 0.499 (0.060) & 0.702 (0.044) & 0.248 (0.066) & 0.252 (0.079) & 0.167 (0.092) \\
		& & $500$ & 0.298 (0.045) & 0.498 (0.044) & 0.699 (0.031) & 0.253 (0.050) & 0.250 (0.052) & 0.166 (0.056) \\
		\cmidrule{2-9}
		& \multirow{2}{*} {\begin{sideways} 30\%HT \end{sideways}} & $250$ & 0.305 (0.082) & 0.504 (0.080) & 0.702 (0.060) & 0.250 (0.081) & 0.251 (0.087) & 0.162 (0.098) \\
		& & $500$ & 0.296 (0.054) & 0.498 (0.054) & 0.698 (0.042) & 0.251 (0.053) & 0.254 (0.061) & 0.164 (0.068) \\
		\midrule\midrule
	\end{tabular}\vspace*{-2.5cm}
\end{table}
\renewcommand{\arraystretch}{1.25}

\begin{table}[H]
	\centering
	\small
	\caption{Simulation results using \textbf{global one-stage parametric and global two-stage semiparametric estimation} for four-dimensional data. In the top panels, the D-vine copula model captures tail-behavior for subsequent gap times changing from lower tail-dependence (Clayton (C)) over no tail-dependence (Frank (F)) to upper tail-dependence (Gumbel (G)) with same overall dependence of Kendall's $\tau_{1,2}=\tau_{2,3}=\tau_{3,4}=0.5$. In the bottom panels, the D-vine copula model captures for Clayton (C) copulas in $\mathcal{T}_1$ increasing dependence with $\tau_{1,2}=0.3$, $\tau_{2,3}=0.5$, $\tau_{3,4}=0.7$. The \textbf{empirical mean (empirical standard deviation) of the copula parameter estimates} are presented based on 250 replications and samples of different sizes affected by either 15\%, 30\% or heavy tail 30\% right-censoring.}
	\label{Table:4d_CFG_theta_global}
	\hspace*{-1.1cm}\begin{tabular}{p{.15cm}p{.15cm}ccccccc}
		\midrule
		&  &  & C; $\theta_{1,2}: 2.00$ & F; $\theta_{2,3}: 5.74$ & G; $\theta_{3,4}: 2.00$ & F; $\theta_{1,3;2}: 2.37$ & F; $\theta_{2,4;3}: 2.37$ & F; $\theta_{1,4;2,3}: 1.54$ \\
		\midrule \midrule
		\multirow{9}{*} {\begin{sideways} semiparametric two-stage \end{sideways}}  & \multirow{3}{*} {\begin{sideways} 15\% \end{sideways}} & $250$ & 2.015 (0.369) & 5.776 (0.689) & 2.042 (0.163) & 2.388 (0.537) & 2.395 (0.602) & 1.529 (0.584) \\
		&  & $500$ & 2.009 (0.258) & 5.757 (0.513) & 2.027 (0.124) & 2.392 (0.373) & 2.358 (0.385) & 1.512 (0.384) \\
		&  & $1000$ & 1.999 (0.192) & 5.708 (0.353) & 2.012 (0.087) & 2.353 (0.251) & 2.386 (0.273) & 1.511 (0.268) \\
		\cmidrule{2-9}
		& \multirow{3}{*} {\begin{sideways} 30\% \end{sideways}} & $250$ & 2.417 (0.872) & 6.236 (1.312) & 2.122 (0.260) & 2.393 (0.711) & 2.422 (0.856) & 1.475 (0.840) \\
		&  & $500$ & 2.138 (0.593) & 5.971 (0.956) & 2.065 (0.184) & 2.446 (0.511) & 2.352 (0.564) & 1.508 (0.563) \\
		&  & $1000$ & 2.122 (0.434) & 5.874 (0.634) & 2.036 (0.136) & 2.379 (0.405) & 2.351 (0.387) & 1.493 (0.399) \\
		\cmidrule{2-9}
		& \multirow{3}{*} {\begin{sideways} 30\%HT \end{sideways}} & $250$ & 4.303 (2.138) & 8.131 (2.795) & 2.261 (0.328) & 2.322 (0.929) & 2.479 (0.981) & 1.420 (0.889) \\
		&  & $500$ & 3.643 (1.576) & 7.545 (1.964) & 2.178 (0.249) & 2.350 (0.578) & 2.482 (0.748) & 1.495 (0.697) \\
		&  & $1000$ & 3.549 (1.347) & 7.303 (1.678) & 2.164 (0.189) & 2.347 (0.575) & 2.496 (0.519) & 1.487 (0.451) \\
		\midrule 
		\multirow{6}{*} {\begin{sideways} parametric one-stage \end{sideways}}  & \multirow{2}{*} {\begin{sideways} 15\% \end{sideways}} & $250$ & 2.062 (0.273) & 5.837 (0.631) & 2.013 (0.143) & 2.425 (0.531) & 2.414 (0.588) & 1.574 (0.594) \\
		&  & $500$ & 2.049 (0.173) & 5.805 (0.455) & 2.005 (0.109) & 2.417 (0.362) & 2.377 (0.386) & 1.530 (0.387) \\
		\cmidrule{2-9}
		& \multirow{2}{*} {\begin{sideways} 30\% \end{sideways}} & $250$ & 2.096 (0.363) & 5.983 (0.871) & 2.061 (0.224) & 2.456 (0.675) & 2.447 (0.850) & 1.571 (0.871) \\
		&  & $500$ & 2.049 (0.247) & 5.902 (0.637) & 2.021 (0.154) & 2.465 (0.476) & 2.386 (0.549) & 1.539 (0.552) \\
		\cmidrule{2-9}
		& \multirow{2}{*} {\begin{sideways} 30\%HT \end{sideways}} & $250$ & 2.106 (0.455) & 6.065 (0.981) & 2.078 (0.247) & 2.484 (0.766) & 2.414 (0.938) & 1.559 (0.953) \\
		&  & $500$ & 2.055 (0.295) & 5.969 (0.692) & 2.039 (0.175) & 2.479 (0.485) & 2.395 (0.643) & 1.536 (0.615) \\
		\midrule \midrule
		&  &  & C; $\theta_{1,2}: 0.86$ & C; $\theta_{2,3}: 2.00$ & C; $\theta_{3,4}: 4.67$ & F; $\theta_{1,3;2}: 2.37$ & F; $\theta_{2,4;3}: 2.37$ & F; $\theta_{1,4;2,3}: 1.54$ \\
		\midrule\midrule
		\multirow{9}{*} {\begin{sideways} semiparametric two-stage \end{sideways}}  & \multirow{3}{*} {\begin{sideways} 15\% \end{sideways}} & $250$ & 0.914 (0.227) & 2.025 (0.372) & 4.596 (0.739) & 2.353 (0.539) & 2.427 (0.591) & 1.527 (0.588) \\
		&  & $500$ & 0.902 (0.167) & 2.021 (0.273) & 4.616 (0.514) & 2.369 (0.397) & 2.406 (0.400) & 1.500 (0.393) \\
		&  & $1000$ & 0.875 (0.112) & 1.982 (0.200) & 4.608 (0.374) & 2.339 (0.256) & 2.403 (0.266) & 1.515 (0.262) \\
		\cmidrule{2-9}
		& \multirow{3}{*} {\begin{sideways} 30\% \end{sideways}} & $250$ & 1.255 (0.674) & 2.351 (0.882) & 4.868 (1.418) & 2.215 (0.737) & 2.461 (0.842) & 1.476 (0.926) \\
		&  & $500$ & 1.032 (0.381) & 2.152 (0.581) & 4.702 (0.971) & 2.332 (0.575) & 2.378 (0.580) & 1.505 (0.548) \\
		&  & $1000$ & 1.006 (0.278) & 2.107 (0.408) & 4.659 (0.673) & 2.316 (0.413) & 2.389 (0.371) & 1.472 (0.400) \\
		\cmidrule{2-9}
		& \multirow{3}{*} {\begin{sideways} 30\%HT \end{sideways}} & $250$ & 2.761 (1.686) & 3.818 (1.901) & 6.297 (2.415) & 2.017 (0.894) & 2.426 (0.979) & 1.283 (0.981) \\
		&  & $500$ & 2.282 (1.245) & 3.307 (1.407) & 5.622 (1.812) & 2.030 (0.690) & 2.442 (0.773) & 1.275 (0.760) \\
		&  & $1000$ & 2.238 (1.036) & 3.211 (1.133) & 5.693 (1.405) & 2.096 (0.637) & 2.397 (0.532) & 1.299 (0.530) \\
		\midrule
		\multirow{6}{*} {\begin{sideways} parametric one-stage \end{sideways}} &  \multirow{2}{*} {\begin{sideways} 15\% \end{sideways}} & $250$ & 0.866 (0.178) & 2.024 (0.315) & 4.756 (0.616) & 2.381 (0.523) & 2.420 (0.573) & 1.579 (0.607) \\
		&  & $500$ & 0.864 (0.117) & 2.014 (0.216) & 4.702 (0.440) & 2.389 (0.378) & 2.394 (0.388) & 1.523 (0.393) \\
		\cmidrule{2-9}
		& \multirow{2}{*} {\begin{sideways} 30\% \end{sideways}} & $250$ & 0.878 (0.256) & 2.046 (0.490) & 4.861 (1.046) & 2.377 (0.717) & 2.453 (0.860) & 1.600 (0.932) \\
		&  & $500$ & 0.864 (0.178) & 2.023 (0.351) & 4.731 (0.694) & 2.405 (0.539) & 2.398 (0.561) & 1.548 (0.552) \\
		\cmidrule{2-9}
		& \multirow{2}{*} {\begin{sideways} 30\%HT \end{sideways}} & $250$ & 0.897 (0.339) & 2.090 (0.639) & 4.912 (1.381) & 2.358 (0.861) & 2.475 (0.959) & 1.549 (1.033) \\
		&  & $500$ & 0.852 (0.211) & 2.023 (0.422) & 4.754 (0.932) & 2.407 (0.577) & 2.427 (0.664) & 1.544 (0.671) \\
		\midrule \midrule
	\end{tabular}
\end{table}

\renewcommand{\arraystretch}{1.25}
\begin{table}[H]
	\centering
	\small
	\caption{Simulation results using \textbf{sequential one-stage parametric and sequential two-stage semiparametric estimation} for four-dimensional data. In the top panels, the D-vine copula model captures tail-behavior for subsequent gap times changing from lower tail-dependence (Clayton (C)) over no tail-dependence (Frank (F)) to upper tail-dependence (Gumbel (G)) with same overall dependence of Kendall's $\tau_{1,2}=\tau_{2,3}=\tau_{3,4}=0.5$. In the bottom panels, the D-vine copula model captures for Clayton (C) copulas in $\mathcal{T}_1$ increasing dependence with $\tau_{1,2}=0.3$, $\tau_{2,3}=0.5$, $\tau_{3,4}=0.7$. The \textbf{empirical mean (empirical standard deviation) of the copula parameter estimates} are presented based on 250 replications and samples of different sizes affected by either 15\%, 30\% or heavy tail 30\% right-censoring.}
	\label{Table:4d_CFG_theta}
	\hspace*{-1.1cm}\begin{tabular}{p{.15cm}p{.15cm}ccccccc}
		\midrule
		&  &  & C; $\theta_{1,2}: 2.00$ & F; $\theta_{2,3}: 5.74$ & G; $\theta_{3,4}: 2.00$ & F; $\theta_{1,3;2}: 2.37$ & F; $\theta_{2,4;3}: 2.37$ & F; $\theta_{1,4;2,3}: 1.54$ \\
		\midrule \midrule
		\multirow{9}{*} {\begin{sideways} semiparametric two-stage \end{sideways}} & \multirow{3}{*} {\begin{sideways} 15\% \end{sideways}} & $250$ & 2.012 (0.374) & 5.776 (0.701) & 2.041 (0.167) & 2.379 (0.530) & 2.401 (0.608) & 1.513 (0.577) \\
		&  & $500$ & 2.006 (0.262) & 5.753 (0.515) & 2.022 (0.125) & 2.390 (0.374) & 2.373 (0.390) & 1.507 (0.382) \\
		&  & $1000$ & 1.999 (0.196) & 5.703 (0.354) & 2.008 (0.089) & 2.351 (0.252) & 2.393 (0.277) & 1.509 (0.268) \\
		\cmidrule{2-9}
		& \multirow{3}{*} {\begin{sideways} 30\% \end{sideways}} & $250$ & 2.413 (0.871) & 6.313 (1.292) & 2.158 (0.258) & 2.367 (0.701) & 2.394 (0.839) & 1.423 (0.802) \\
		&  & $500$ & 2.131 (0.598) & 6.083 (0.921) & 2.099 (0.181) & 2.415 (0.504) & 2.374 (0.541) & 1.469 (0.543) \\
		&  & $1000$ & 2.120 (0.440) & 5.970 (0.619) & 2.071 (0.137) & 2.360 (0.405) & 2.353 (0.384) & 1.467 (0.390) \\
		\cmidrule{2-9}
		& \multirow{3}{*} {\begin{sideways} 30\%HT \end{sideways}} & $250$ & 4.300 (2.125) & 8.059 (2.647) & 2.312 (0.311) & 2.301 (0.922) & 2.338 (0.943) & 1.334 (0.826) \\
		&  & $500$ & 3.643 (1.561) & 7.523 (1.816) & 2.228 (0.232) & 2.334 (0.572) & 2.370 (0.707) & 1.433 (0.660) \\
		&  & $1000$ & 3.550 (1.339) & 7.276 (1.558) & 2.217 (0.172) & 2.340 (0.574) & 2.385 (0.496) & 1.430 (0.426) \\
		\midrule \midrule
		\multirow{6}{*} {\begin{sideways} parametric one-stage \end{sideways}} &  \multirow{2}{*} {\begin{sideways} 15\% \end{sideways}} & $250$ & 2.016 (0.295) & 5.756 (0.655) & 2.011 (0.145) & 2.398 (0.521) & 2.411 (0.589) & 1.563 (0.588) \\
		&  & $500$ & 2.012 (0.184) & 5.745 (0.476) & 2.006 (0.109) & 2.398 (0.366) & 2.381 (0.388) & 1.524 (0.384) \\
		\cmidrule{2-9}
		& \multirow{2}{*} {\begin{sideways} 30\% \end{sideways}} & $250$ & 2.038 (0.378) & 5.812 (0.954) & 2.038 (0.221) & 2.384 (0.655) & 2.462 (0.829) & 1.544 (0.854) \\
		&  & $500$ & 2.007 (0.254) & 5.767 (0.686) & 2.011 (0.157) & 2.420 (0.489) & 2.394 (0.550) & 1.535 (0.547) \\
		\cmidrule{2-9}
		& \multirow{2}{*} {\begin{sideways} 30\%HT \end{sideways}} & $250$ & 2.051 (0.467) & 5.866 (1.081) & 2.054 (0.250) & 2.427 (0.772) & 2.396 (0.917) & 1.545 (0.941) \\
		&  & $500$ & 2.010 (0.311) & 5.757 (0.777) & 2.016 (0.182) & 2.398 (0.509) & 2.416 (0.652) & 1.523 (0.610) \\
		\midrule \midrule
		&  &  & C; $\theta_{1,2}: 0.86$ & C; $\theta_{2,3}: 2.00$ & C; $\theta_{3,4}: 4.67$ & F; $\theta_{1,3;2}: 2.37$ & F; $\theta_{2,4;3}: 2.37$ & F; $\theta_{1,4;2,3}: 1.54$ \\
		\midrule\midrule
		\multirow{9}{*} {\begin{sideways} semiparametric two-stage \end{sideways}} & \multirow{3}{*} {\begin{sideways} 15\% \end{sideways}} & $250$ & 0.919 (0.231) & 2.017 (0.375) & 4.608 (0.743) & 2.366 (0.530) & 2.402 (0.592) & 1.506 (0.575) \\
		&  & $500$ & 0.904 (0.170) & 2.013 (0.269) & 4.609 (0.515) & 2.380 (0.395) & 2.397 (0.407) & 1.491 (0.390) \\
		&  & $1000$ & 0.877 (0.116) & 1.971 (0.199) & 4.602 (0.370) & 2.343 (0.260) & 2.395 (0.272) & 1.511 (0.262) \\
		\cmidrule{2-9}
		& \multirow{3}{*} {\begin{sideways} 30\% \end{sideways}} & $250$ & 1.259 (0.660) & 2.365 (0.850) & 4.923 (1.393) & 2.220 (0.727) & 2.426 (0.825) & 1.427 (0.882) \\
		&  & $500$ & 1.037 (0.378) & 2.181 (0.556) & 4.748 (0.946) & 2.330 (0.573) & 2.379 (0.575) & 1.480 (0.532) \\
		&  & $1000$ & 1.011 (0.276) & 2.133 (0.394) & 4.726 (0.645) & 2.315 (0.418) & 2.381 (0.378) & 1.455 (0.392) \\
		\cmidrule{2-9}
		& \multirow{3}{*} {\begin{sideways} 30\%HT \end{sideways}} & $250$ & 0.897 (0.339) & 2.090 (0.639) & 4.912 (1.381) & 2.358 (0.861) & 2.475 (0.959) & 1.549 (1.033) \\
		&  & $500$ & 2.256 (1.205) & 3.266 (1.284) & 5.670 (1.621) & 2.010 (0.688) & 2.362 (0.780) & 1.236 (0.735) \\
		&  & $1000$ & 2.211 (0.994) & 3.165 (1.051) & 5.722 (1.260) & 2.088 (0.638) & 2.307 (0.527) & 1.270 (0.507) \\
		\midrule\midrule
		\multirow{6}{*} {\begin{sideways} parametric one-stage \end{sideways}} & \multirow{2}{*} {\begin{sideways} 15\% \end{sideways}} & $250$ & 0.868 (0.184) & 2.030 (0.314) & 4.754 (0.623) & 2.375 (0.503) & 2.416 (0.573) & 1.567 (0.598) \\
		&  & $500$ & 0.870 (0.123) & 2.032 (0.211) & 4.716 (0.447) & 2.400 (0.371) & 2.376 (0.396) & 1.513 (0.393) \\
		\cmidrule{2-9}
		& \multirow{2}{*} {\begin{sideways} 30\% \end{sideways}} & $250$ & 0.880 (0.264) & 2.049 (0.501) & 4.859 (1.042) & 2.381 (0.711) & 2.441 (0.853) & 1.579 (0.916) \\
		&  & $500$ & 0.862 (0.183) & 2.018 (0.358) & 4.722 (0.710) & 2.423 (0.527) & 2.389 (0.553) & 1.538 (0.548) \\
		\cmidrule{2-9}
		& \multirow{2}{*} {\begin{sideways} 30\%HT \end{sideways}} & $250$ & 0.920 (0.351) & 2.134 (0.659) & 4.978 (1.386) & 2.415 (0.879) & 2.439 (0.952) & 1.527 (0.979) \\
		&  & $500$ & 0.858 (0.222) & 2.026 (0.429) & 4.752 (0.922) & 2.402 (0.569) & 2.435 (0.654) & 1.531 (0.666) \\
		\midrule \midrule
	\end{tabular}
\end{table}

\subsection{Results on copula selection by AIC}
To illustrate the effect of using an incorrect copula specification, we fitted in addition to the correct D-vine specification as given in \autoref{table:CopulaSimSettings_4d} a four-dimensional Clayton copula and an incorrect D-vine with all pair-copulas being of type Clayton (Clayton vine). In case of one-stage parametric 
estimation the format of the survival margins 
is correctly specified. 

In \autoref{Table:4d_CFG_aic} and \autoref{Table:4d_CCC_aic} we list the preference by AIC, i.e.\ the proportion of data sets, for which each of the three model specifications performed best based on AIC. It follows that AIC is able to detect the correct copula for the majority of simulated data sets, indicating that AIC is a valid tool for model selection. As expected, the AIC preference for the correct model increases as sample size grows, but decreases for a higher censoring rate. Also, AIC selects the correct model more often for one-stage parametric estimation as compared to two-stage semiparametric estimation. Finally, for the D-vine with CFG in tree level $\mathcal{T}_1$ the correct vine is selected more often as compared to the D-vine with CCC in tree level $\mathcal{T}_1$. This is to be expected. Indeed, the latter resembles a Clayton-vine and a Clayton copula more closely, making model detection more difficult.

%\begin{landscape}
\begin{table}[h]
	\centering
	\small%\vspace*{-.75cm}
	\caption{Results on copula selection by AIC, under both a one-stage parametric (global and sequential) and a two-stage semiparametric estimation approach (global and sequential) for four-dimensional data (4d). The D-vine copula model captures tail-behavior for subsequent variable pairs changing from lower tail-dependence (Clayton (C)) over no tail-dependence (Frank (F)) to upper tail-dependence (Gumbel (G)) with same overall dependence of Kendall's $\tau_{1,2}=\tau_{2,3}=\tau_{3,4}=0.5$. We consider the fit of a correctly specified D-vine copula (CFG), an incorrect Clayton D-vine and a Clayton copula. The AIC preference rate is based on 250 replications and samples of different sizes affected by either 15\%, 30\% or heavy tail 30\% right-censoring.}
	\label{Table:4d_CFG_aic}
	\begin{tabular}{cccccccccc}
		\midrule
		& & & & \multicolumn{2}{c}{D-vine global} & \multirow{2}{*}{4d Clayton} & \multicolumn{2}{c}{D-vine sequential} & \multirow{2}{*}{4d Clayton} \\
		& &  &  & correct & incorrect &  & correct & incorrect & \\
		\midrule \midrule
		\multirow{9}{*} {\begin{sideways} semiparametric \end{sideways}} & \multirow{9}{*}{\begin{sideways} two-stage \end{sideways}}  & \multirow{3}{*} {15\%} & $250$ & 1 & 0 & 0 & 1 & 0 & 0 \\
		& &  &  $500$   & 1 & 0 & 0 & 1 & 0 & 0 \\
		& &  &  $1000$  & 1 & 0 & 0 & 1 & 0 & 0 \\
		\cmidrule{3-10}
		& & \multirow{3}{*} {30\%} & $250$ & 0.956 & 0.028 & 0.016 & 0.964 & 0.016 & 0.020 \\
		& &  &  $500$   & 1 & 0 & 0 & 1 & 0 & 0 \\
		& &  &  $1000$  & 1 & 0 & 0 & 1 & 0 & 0 \\
		\cmidrule{3-10}
		& & \multirow{3}{*} {30\% HT} & $250$ & 0.844 & 0.116 & 0.040 & 0.836 & 0.108 & 0.056 \\
		& &  &  $500$   & 0.928 & 0.064 & 0.008 & 0.928 & 0.064 & 0.008 \\
		& &  &  $1000$  & 0.972 & 0.028 & 0 & 0.972 & 0.028 & 0 \\
		\midrule\midrule
		\multirow{6}{*} {\begin{sideways} parametric \end{sideways}} & \multirow{6}{*}{\begin{sideways} one-stage \end{sideways}} &\multirow{2}{*} {15\%} & $250$ & 1 & 0 & 0 & 1 & 0 & 0 \\
		& &  &  $500$  & 1 & 0 & 0 & 1 & 0 & 0 \\
		\cmidrule{3-10}
		& & \multirow{2}{*} {30\%} & $250$ & 1 & 0 & 0 & 1 & 0 & 0 \\
		& &  &  $500$   & 1 & 0 & 0 & 1 & 0 & 0 \\
		\cmidrule{3-10}
		& & \multirow{2}{*} {30\% HT} & $250$ & 0.996 & 0.004 & 0 & 0.996 & 0.004 & 0 \\
		& &  &  $500$   & 1 & 0 & 0 & 1 & 0 & 0 \\
		\midrule\midrule
	\end{tabular}
\end{table}
%\end{landscape}
\newpage

%\begin{landscape}
\begin{table}[h]
	\centering
	\small%\vspace*{-.75cm}
	\caption{Results on copula selection by AIC, under both a one-stage parametric (global and sequential) and a two-stage semiparametric estimation approach (global and sequential) for four-dimensional data (4d). The D-vine copula model captures for Clayton (C) copulas in $\mathcal{T}_1$ increasing dependence with $\tau_{1,2}=0.3$, $\tau_{2,3}=0.5$, $\tau_{3,4}=0.7$. We consider the fit of a correctly specified D-vine copula (CCC), an incorrect Clayton D-vine and a Clayton copula. The AIC preference rate is based on 250 replications and samples of different sizes affected by either 15\%, 30\% or heavy tail 30\% right-censoring.}
	\label{Table:4d_CCC_aic}
	\begin{tabular}{cccccccccc}
		\midrule
		& & & & \multicolumn{2}{c}{D-vine global} & \multirow{2}{*}{4d Clayton} & \multicolumn{2}{c}{D-vine sequential} & \multirow{2}{*}{4d Clayton} \\
		& &  &  & correct & incorrect &  & correct & incorrect & \\
		\midrule \midrule
		\multirow{9}{*} {\begin{sideways} semiparametric \end{sideways}} & \multirow{9}{*}{\begin{sideways} two-stage \end{sideways}}  & \multirow{3}{*} {15\%} & $250$ & 0.956 & 0.044 & 0 & 0.960 & 0.040 & 0 \\
		& &  &  $500$   & 1 & 0 & 0 & 1 & 0 & 0 \\
		& &  &  $1000$  & 1 & 0 & 0 & 1 & 0 & 0 \\
		\cmidrule{3-10}
		& & \multirow{3}{*} {30\%} & $250$ & 0.816 & 0.172 & 0.012 & 0.864 & 0.124 & 0.012 \\
		& &  &  $500$   & 0.908 & 0.092 & 0 & 0.936 & 0.064 & 0 \\
		& &  &  $1000$  & 0.976 & 0.024 & 0 & 0.992 & 0.008 & 0 \\
		\cmidrule{3-10}
		& & \multirow{3}{*} {30\% HT} & $250$ & 0.424 & 0.344 & 0.232 & 0.472 & 0.260 & 0.268 \\
		& &  &  $500$   & 0.552 & 0.392 & 0.056 & 0.632 & 0.292 & 0.076 \\
		& &  &  $1000$  & 0.632 & 0.364 & 0.004 & 0.740 & 0.248 & 0.012 \\
		\midrule\midrule
		\multirow{6}{*} {\begin{sideways} parametric \end{sideways}} & \multirow{6}{*}{\begin{sideways} one-stage \end{sideways}} &\multirow{2}{*} {15\%} & $250$ & 0.964 & 0.036 & 0 & 0.964 & 0.036 & 0 \\
		& &  &  $500$   & 0.996 & 0.004 & 0 & 0.996 & 0.004 & 0 \\
		\cmidrule{3-10}
		& & \multirow{2}{*} {30\%} & $250$ & 0.896 & 0.104 & 0 & 0.892 & 0.108 & 0 \\
		& &  &  $500$   & 0.960 & 0.040 & 0 & 0.960 & 0.040 & 0 \\
		\cmidrule{3-10}
		& & \multirow{2}{*} {30\% HT} & $250$ & 0.804 & 0.176 & 0.020 & 0.800 & 0.176 & 0.024 \\
		& &  &  $500$   & 0.924 & 0.076 & 0 & 0.924 & 0.076 & 0 \\
		\midrule\midrule
	\end{tabular}
\end{table}
\newpage
\section{Vine copula bootstrapping for right-censored recurrent event-time data within the one-stage estimation approach}\label{Sec:VineCopBoot}

%\subsection*{Bootstrapping algorithm}
We consider four-dimensional data.
\begin{itemize}
	\item[] \textit{Step 1:}\\ Under a prespecified parametric format for the marginal survival functions (e.g. Weibull), fit the D-vine copula model of interest to the data $\left(y_{i,j},\delta_{i,j}\right)$ as listed in \autoref{table:DataFormat} ($i=1,\ldots,n$ and $j=1,\ldots,d_i$). Obtain the marginal parameter estimates $\boldsymbol{\widehat{\alpha}}_j$ ($j=1,\ldots,4$) and the D-vine copula parameter estimates $\widehat{\boldsymbol{\theta}}_{1:4}$.
	\bigskip
	\item[] \textit{Step 2:}\\ Obtain the Nelsen-Aalen/Kaplan-Meier estimate $\widehat{G}$ of the censoring distribution $G$ based on the observations $\left(y_{i,1}+y_{i,2}+\ldots+y_{i,d_i},1-\delta_{i,d_i}\right)$  ($i=1,\ldots,n$).
	\bigskip
	\item[] \textit{Step 3:}\\ Generate $B$ bootstrap samples in the following way: For $b=1,\ldots,B$, $i=1,\ldots,n$ and $j=1,\ldots,4$,
	\begin{itemize}
		\item[] \textit{Step 3.1:}\\ Sample vine copula data $\left(u^{(b)}_{i,1},u^{(b)}_{i,2},u^{(b)}_{i,3},u^{(b)}_{i,4}\right)$ from the fitted D-vine copula model with parameter vector $\widehat{\boldsymbol{\theta}}_{1:4}$.
		\bigskip
		\item[] \textit{Step 3.2:}\\ Generate gap times $\left(g^{(b)}_{i,1},g^{(b)}_{i,2},g^{(b)}_{i,3},g_{i,4}^{(b)}\right)$ via $g^{(b)}_{i,j} = S^{-1}_j\left(u^{(b)}_{i,j};\boldsymbol{\widehat{\alpha}}_j\right)$, where $S_j\left(\cdot;\boldsymbol{\widehat{\alpha}}_j\right)$ follows the assumed marginal distribution with estimated parameters $\boldsymbol{\widehat{\alpha}}_j$.
		\bigskip
		\item[] \textit{Step 3.3:}\\ Obtain event times $t^{(b)}_{i,j}$ by setting $t^{(b)}_{i,j} = \sum_{\ell=1}^{j} g^{(b)}_{i,\ell}$.
		\bigskip
		\item[] \textit{Step 3.4:}\\ Generate $c^{(b)}_i$ from $\widehat{G}$.
		\bigskip
		\item[] \textit{Step 3.5:}\\ Obtain observed data. If $t^{(b)}_{i,1}>c^{(b)}_{i}$ set $d_i = 1$ and retain $y^{(b)}_{i,1} = c^{(b)}_{i}$, if $t^{(b)}_{i,2}>c^{(b)}_{i}$ set $d_i = 2$ and retain $(y^{(b)}_{i,1},y^{(b)}_{i,2}) = (g^{(b)}_{i,1},c^{(b)}_{i}-t^{(b)}_{i,1})$, etc. For $d_i = 4$ distinguish between three or four events, i.e.\ $(y^{(b)}_{i,1},y^{(b)}_{i,2},y^{(b)}_{i,3},y^{(b)}_{i,4}) = (g^{(b)}_{i,1},g^{(b)}_{i,2},g^{(b)}_{i,3},c^{(b)}_{i}-t^{(b)}_{i,3})$ or $(y^{(b)}_{i,1},y^{(b)}_{i,2},y^{(b)}_{i,3},y^{(b)}_{i,4}) = (g^{(b)}_{i,1},g^{(b)}_{i,2},g^{(b)}_{i,3},g^{(b)}_{i,4})$. Define $\delta^{(b)}_{i,j} = 1$ if $j<d_i$ and $\delta^{(b)}_{i,d_i} = I\left(y^{(b)}_{i,d_i} \leq c^{(b)}_i-t^{(b)}_{i,d_i-1}\right)$.
		\bigskip
		\item[] \textit{Step 3.6:}\\ Under a prespecified parametric format for the marginal survival functions (e.g. Weibull), fit the D-vine copula model of interest to the bootstrap data $(y^{(b)}_{i,j},\delta^{(b)}_{i,j})$ ($i=1,\ldots,n$ and $j=1,\ldots,d_i$). Obtain the marginal parameter estimates $\widehat{\boldsymbol{\alpha}}^{(b)}_j$ ($j=1,\ldots,4$) and the D-vine copula parameter estimates $\widehat{\boldsymbol{\theta}}^{(b)}_{1:4}$.
	\end{itemize}
	\item[]\textit{Step 4:}\\ Obtain the bootstrap standard errors using $\widehat{\boldsymbol{\theta}}^{(1)}_{1:4},\ldots,\widehat{\boldsymbol{\theta}}^{(B)}_{1:4}$.
\end{itemize}

\newpage
\section{Additional material on the analysis of the asthma data}\label{Sec:AddOnAsthma}

\renewcommand{\arraystretch}{1}
\begin{table}[ht]
	\caption{Frequency table of the number of asthma attacks per child ($d_i \text{ for } i=1,\ldots,232$) in the asthma data.}
	\label{Table:FreqAsthma}
	\centering
	\begin{tabular} {ccccccccccccccccc}
		\midrule\midrule
		$d_i$ & 2 & 3 & 4 & 5 & 6 & 7 & 8 & 9 & 10 & 11 & 12 & 13 & 14 & 15 & 20 & 39 \\ %\hline\hline
		\#Children & 88 & 47 & 26 & 16 & 10 & 8 & 9 & 8 & 2 & 7 & 2 & 3 & 3 & 1 & 1 & 1 \\
		\hline\hline
	\end{tabular}
\end{table}
\renewcommand{\arraystretch}{1}

\begin{table}[ht]		
	\caption{Modified asthma data: original data cut off at attack $4$.}
	\label{table:modAsthmaData}
	\centering
	\begin{tabular} {clccc}
		\midrule
		\multicolumn{2}{c}{\multirow{2}{*}{Sample}} & \multicolumn{3}{c}{Cluster size}\\
		&& 2 & 3 & $\geq$ 4 \\
		\hline\midrule
		\multirow{2}{*}{Full} &	\#Children with last attack event & 1 & 3 & 72  \\
		&	\#Children with last attack censored & 87 & 44 & 25  \\
		\midrule
		\multirow{2}{*}{Treatment} &	\#Children with last attack event & 0 & 3 & 27  \\
		&	\#Children with last attack censored & 50 & 25 &  8 \\
		\midrule
		\multirow{2}{*}{Control} &	\#Children with last attack event & 1 & 0 &  45 \\
		&	\#Children with last attack censored & 37 & 19 & 17 \\
		\midrule
		\midrule
	\end{tabular}
\end{table}

\begin{figure}[ht]
	\centering
	\begin{tabular}{ccc}
		{\footnotesize 	Full sample} & {\footnotesize Treatment subsample} & {\footnotesize Control subsample}\\
		\includegraphics[width=.29\linewidth]{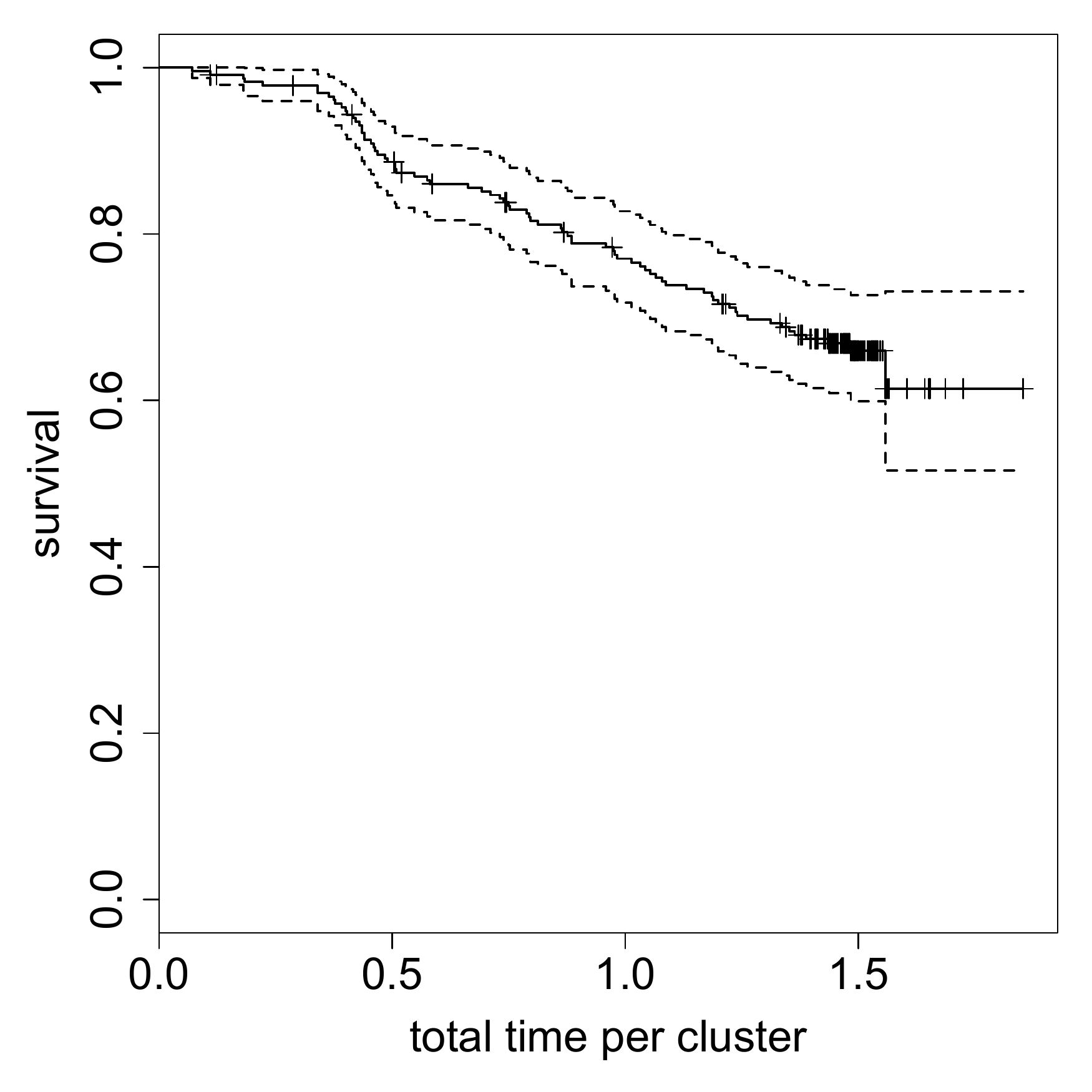}&\includegraphics[width=.29\linewidth]{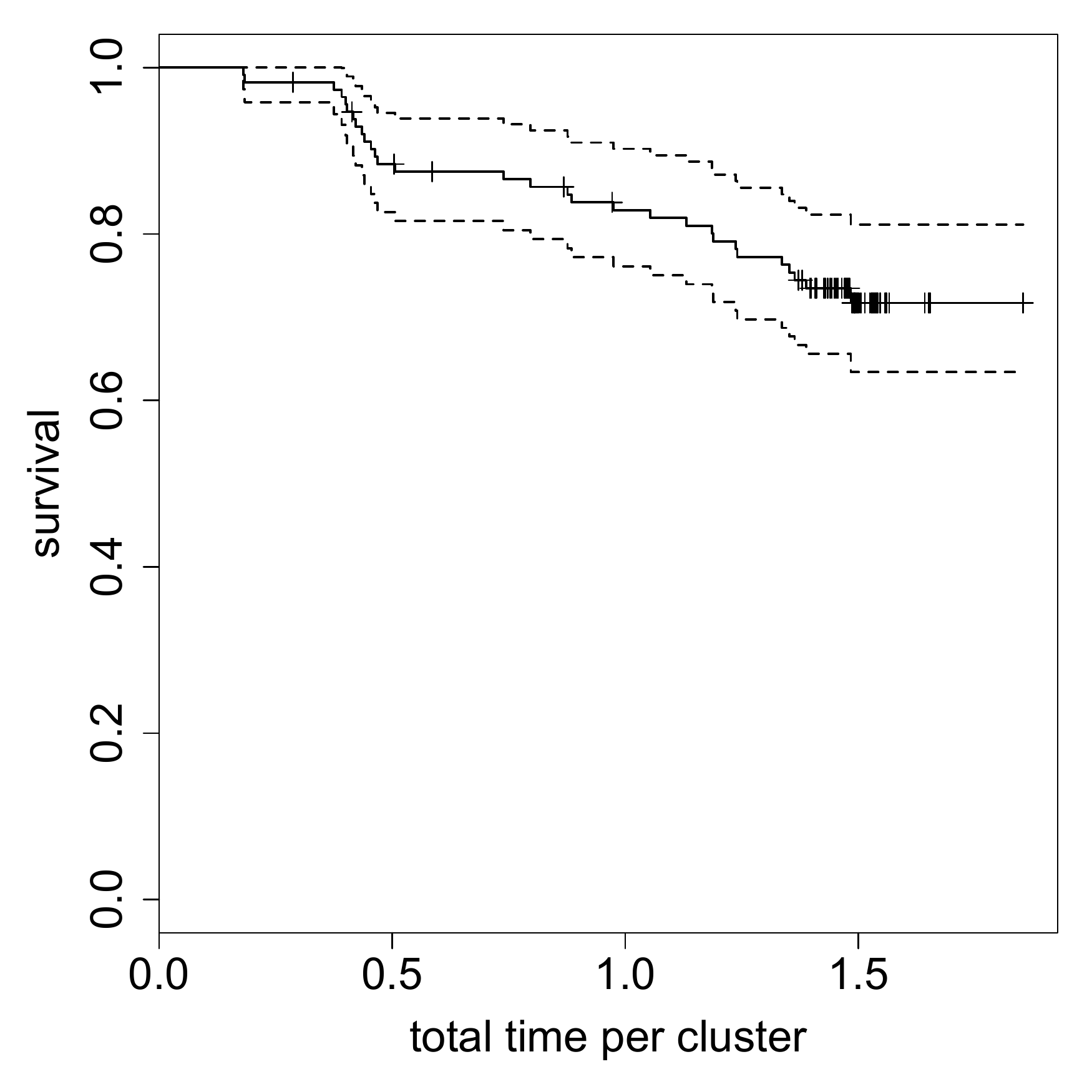}&\includegraphics[width=.29\linewidth]{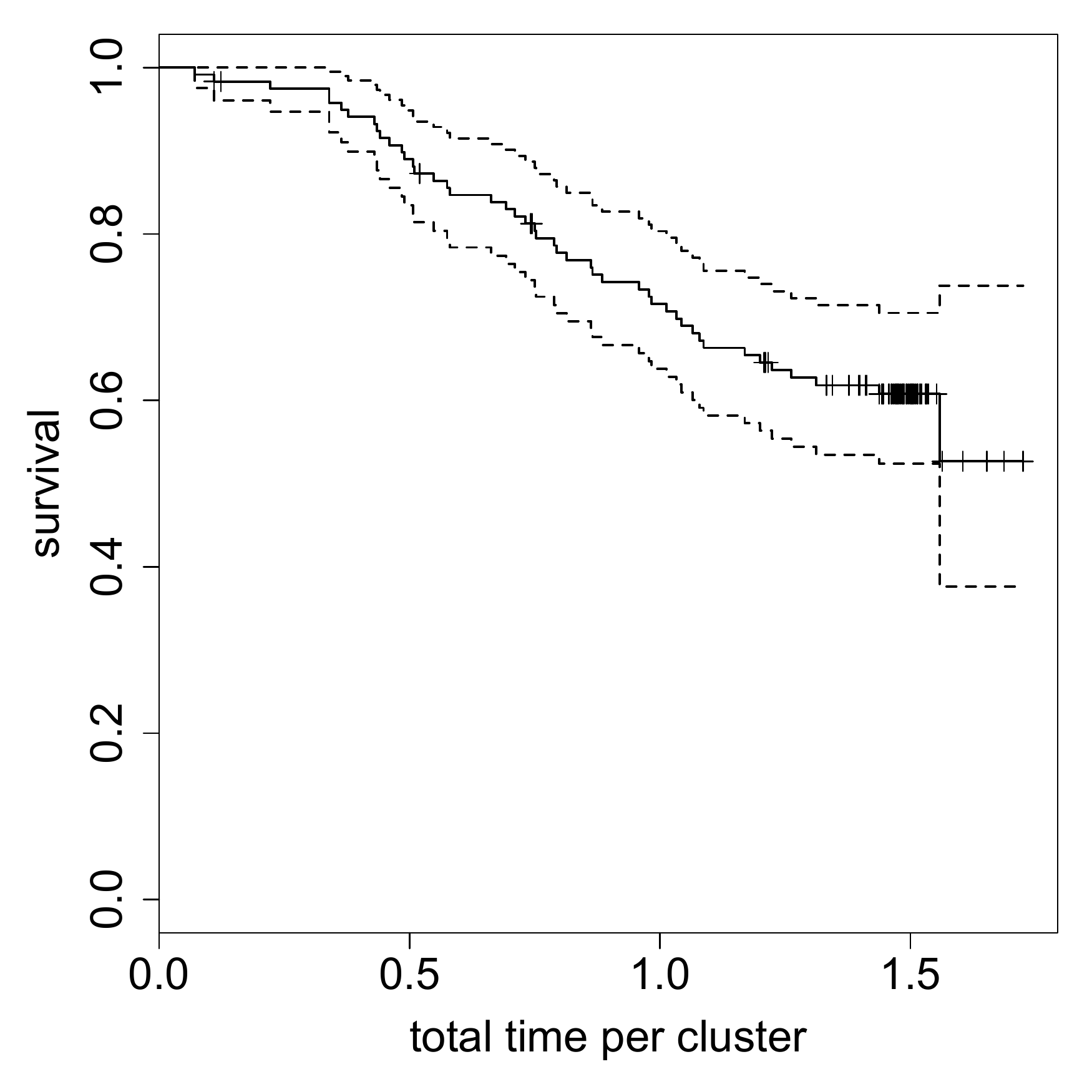}\\
	\end{tabular}
	\caption{Nelson-Aalen estimate of the survival function for the total times (in years).}	
	\label{fig:asthmanelson-aalen}
\end{figure}

\begin{landscape}
	\begin{table}[ht]
		\renewcommand{\arraystretch}{1.0}
		\small
		\centering
		\vspace*{-0.75cm}
		\caption{Estimates for the marginal parameters and standard errors (in parentheses) of copula models fitted to each of the three samples of asthma data. In case of Archimedean copulas the Frank, Gumbel, Clayton and the Independence copula are considered. In case of D-vine copulas only the three best models based on AIC are shown with Frank being the pair-copula family in trees $\mathcal{T}_2$ and $\mathcal{T}_3$. One stage-parametric estimation is considered. For D-vine copulas both sequential (top panels) and global (bottom panels) estimation are performed.
		}
		\label{Table:AsthmaResults_margins}
		\begin{tabular}{ccccccccccc}
			\midrule
			&  &  & $\lambda_{1}$ & $\rho_{1}$ & $\lambda_{2}$ & $\rho_{2}$ & $\lambda_{3}$ & $\rho_{3}$ & $\lambda_{4}$ & $\rho_{4}$\\
			\hline\midrule
			\multirow{9}{*}{\begin{sideways} Sequential estimation \end{sideways}}	& \multirow{3}{*}{Full} & FGG & 1.900 (0.135) & 1.005 (0.058) & 1.285 (0.122) & 0.612 (0.044) & 1.365 (0.181) & 0.698 (0.057) & 1.664 (0.356) & 0.726 (0.069) \\
			& & CGG & 1.900 (0.135) & 1.005 (0.058) & 1.234 (0.125) & 0.600 (0.044) & 1.347 (0.180) & 0.696 (0.057) & 1.611 (0.350) & 0.719 (0.068) \\
			&  & GGG & 1.900 (0.035) & 1.005 (0.058) & 1.273 (0.097) & 0.626 (0.044) & 1.364 (0.125) & 0.704 (0.058) & 1.655 (0.134) & 0.732 (0.069) \\  		
			\cmidrule{2-11}
			& \multirow{3}{*}{Treatment} & FGG & 1.759 (0.176) & 1.174 (0.101) & 1.043 (0.150) & 0.595 (0.064) & 1.062 (0.243) & 0.619 (0.083) & 1.554 (0.654) & 0.711 (0.114)\\
			& & CGG & 1.759 (0.176) & 1.174 (0.101) & 1.027 (0.151) & 0.591 (0.065) & 1.060 (0.241) & 0.620 (0.083) & 1.518 (0.644) & 0.706 (0.115) \\
			&  & GGG & 1.759 (0.176) & 1.174 (0.101) & 1.050 (0.146) & 0.599 (0.064) & 1.065 (0.242) & 0.621 (0.083) & 1.547 (0.661) & 0.707 (0.114) \\
			\cmidrule{2-11}
			&\multirow{3}{*}{Control} & FGG & 2.057 (0.205) & 0.898 (0.076) & 1.596 (0.207) & 0.639 (0.059) & 1.602 (0.250) & 0.756 (0.077) & 1.834 (0.460) & 0.745 (0.092) \\
			&  & FGF & 2.057 (0.205) & 0.900 (0.076) & 1.596 (0.207) & 0.639 (0.059) & 1.602 (0.250) & 0.756 (0.077) & 1.889 (0.508) & 0.763 (0.096) \\
			& &  GGG & 2.057 (0.205) & 0.900 (0.076) & 1.563 (0.199) & 0.653 (0.060) & 1.601 (0.251) & 0.763 (0.078) & 1.811 (0.494) & 0.750 (0.093) \\
			\midrule\midrule
			\multirow{21}{*}{\begin{sideways} Global estimation \end{sideways}}	& \multirow{7}{*}{Full} &  FGG & 1.891 (0.133) & 1.008 (0.058) & 1.247 (0.119) & 0.602 (0.042) & 1.330 (0.177) & 0.684 (0.055) & 1.662 (0.139) & 0.716 (0.067) \\
			&  & CGG & 1.899 (0.135) & 1.005 (0.058) & 1.200 (0.120) & 0.590 (0.042) & 1.307 (0.177) & 0.682 (0.056) & 1.568 (0.339) & 0.708 (0.067) \\	
			&  & GGG & 1.900 (0.134) & 0.989 (0.057) & 1.241 (0.114) & 0.613 (0.042) & 1.325 (0.178) & 0.689 (0.056) & 1.617 (0.359) & 0.720 (0.069) \\
			& & 4dF & 1.869 (0.137) & 1.008 (0.055) & 1.303 (0.110) & 0.620 (0.045) & 1.543 (0.178) & 0.698 (0.057) & 2.660 (0.427) & 0.761 (0.065) \\
			&  & 4dG & 1.879 (0.159) & 0.995 (0.060) & 1.302 (0.134) & 0.616 (0.049) & 1.585 (0.224) & 0.696 (0.059) & 2.827 (0.572) & 0.766 (0.080) \\	
			&  & 4dC & 1.890 (0.133) & 1.006 (0.059) & 1.273 (0.113) & 0.612 (0.044) & 1,478 (0.184) & 0.680 (0.058) & 2.525 (0.451) & 0.731 (0.065) \\
			&  & 4dInd & 1.900 (0.138) & 1.005 (0.060) & 1.293 (0.111) & 0.620 (0.043) & 1.657 (0.174) & 0.690 (0.058) & 3.068 (0.446) & 0.742 (0.066) \\ 		
			\cmidrule{2-11}
			& \multirow{7}{*}{Treatment} &  FGG & 1.757 (0.177) & 1.175 (0.102) & 1.018 (0.147) & 0.590 (0.061) & 1.025 (0.239) & 0.608 (0.083) & 1.503 (0.632) & 0.700 (0.112) \\
			& & CGG & 1.760 (0.177) & 1.173 (0.101) & 1.001 (0.149) & 0.586 (0.061) & 1.021 (0.237) & 0.609 (0.081) & 1.463 (0.608) & 0.695 (0.110) \\
			&  & GGG & 1.763 (0.177) & 1.172 (0.101) & 1.026 (0.145) & 0.595 (0.062) & 1.028 (0.238) & 0.610 (0.083) & 1.497 (0.640) & 0.696 (0.113) \\
			& & 4dF & 1.730 (0.174) & 1.175 (0.101) & 1.055 (0.138) & 0.596 (0.065) & 1.312 (0.233) & 0.609 (0.085) & 3.584 (1.104) & 0.776 (0.118) \\
			&  & 4dG & 1.759 (0.196) & 1.174 (0.104) & 1.050 (0.151) & 0.599 (0.071) & 1.379 (0.254) & 0.601 (0.084) & 3.989 (1.352) & 0.753 (0.127) \\	
			&  & 4dC & 1.753 (0.177) & 1.173 (0.095) & 1.038 (0.130) & 0.594 (0.070) & 1.290 (0.254) & 0.597 (0.084) & 3.541 (1.230) & 0.750 (0.112) \\
			&  & 4dInd & 1.759 (0.181) & 1.174 (0.095) & 1.051 (0.131) & 0.559 (0.073) & 1.378 (0.251) & 0.601 (0.084) & 3.989 (1.162) & 0.753 (0.111) \\ 	
			\cmidrule{2-11}
			&\multirow{7}{*}{Control} & FGG & 2.039 (0.205) & 0.906 (0.077) & 1.548 (0.200) & 0.626 (0.057) & 1.566 (0.244) & 0.740 (0.074) & 1.803 (0.467) & 0.736 (0.091) \\
			&  & FGF & 2.039 (0.205) & 0.906 (0.077) & 1.561 (0.203) & 0.632 (0.058) & 1.601 (0.249) & 0.754 (0.077) & 1.881 (0.513) & 0.762 (0.096) \\
			&  & GGG & 2.038 (0.205) & 0.881 (0.073) & 1.526 (0.193) & 0.637 (0.057) & 1.563 (0.248) & 0.736 (0.076) & 1.787 (0.494) & 0.740 (0.091) \\
			& & 4dF & 2.011 (0.199) & 0.902 (0.073) & 1.604 (0.187) & 0.647 (0.059) & 1.751 (0.279) & 0.767 (0.082) & 2.405 (0.469) & 0.763 (0.088) \\
			&  & 4dG & 2.028 (0.227) & 0.891 (0.075) & 1.600 (0.208) & 0.643 (0.067) & 1.786 (0.324) & 0.761 (0.084) & 2.525 (0.628) & 0.773 (0.101) \\	
			&  & 4dC & 2.047 (0.204) & 0.900 (0.068) & 1.565 (0.187) & 0.639 (0.062) & 1.655 (0.262) & 0.744 (0.080) & 2.252 (0.515) & 0.739 (0.089) \\
			&  & 4dInd & 2.057 (0.212) & 0.898 (0.071) & 1.583 (0.170) & 0.648 (0.062) & 1.883 (0.264) & 0.759 (0.081) & 2.766 (0.541) & 0.756 (0.089) \\ 		
			\hline\hline
		\end{tabular}	
	\end{table}
\end{landscape}

To obtain standard errors of the parameter estimates found for the samples of the asthma data the bootstrap algorithm given in \autoref{Sec:VineCopBoot} is used. While in the asthma data there are no clusters of size 1, this case is possible for the bootstrap samples (as in many data settings). \autoref{Table:bootInfo} contains information on average cluster sizes and the average censoring percentage among the bootstrap replications showing that the data generation within the bootstrap succeeds to mimic the features of the asthma data characteristics quite accurately. 

\begin{table}[H]
	\renewcommand{\arraystretch}{1.0}
	\small
	\centering	
	\caption{Average cluster sizes and average censoring percentage among the 250 bootstrap replications used for standard error calculation of the copula and marginal parameter estimates in the asthma data. Results for all three subsamples are shown. In case of Archimedean copulas the Frank, Gumbel, Clayton and the Independence copula are considered. In case of D-vine copulas only the three best models are shown with Frank being the pair-copula family in $\mathcal{T}_2$ and $\mathcal{T}_3$. One stage-parametric estimation is considered. For D-vine copulas both sequential (top panels) and global (bottom panels) estimation are performed.}
	\label{Table:bootInfo}
	\begin{tabular}{ccccccccc}
		\midrule
		&  &  & \multirow{2}{*}{$\#$size 1} & \multirow{2}{*}{$\#$size 2} & \multirow{2}{*}{$\#$size 3} & $\#$size 4 & $\#$size 4 & \multirow{2}{*}{$\%$censoring}\\
		&  &  & & & & (event) & (censored) & \\
		\midrule\midrule
		\multirow{9}{*}{\begin{sideways} Sequential estimation \end{sideways}}	& \multirow{3}{*}{Full} & FGG & 17.93 & 63.77 & 46.17 & 23.88 & 80.25 & 21.69 \\
		& & CGG & 17.93 & 63.48 & 46.36 & 23.96 & 80.28 & 21.67  \\
		&  & GGG & 17.93 & 64.65 & 47.23 & 24.07 & 78.13 & 22.08 \\  	
		\cmidrule{2-9}
		& \multirow{3}{*}{Treatment} & FGG & 8.98 & 39.79 & 24.51 & 7.83 & 31.89 & 25.33 \\
		& & CGG & 8.98 & 39.62 & 24.61 & 7.89 & 31.90 & 25.31 \\
		&  & GGG & 8.98 & 39.62 & 24.94 & 7.93 & 31.53 & 25.45 \\
		\cmidrule{2-9}
		&\multirow{3}{*}{Control} & FGG & 8.51 & 24.41 & 22.57 & 14.91 & 48.60 & 18.62 \\
		&  & FGF & 8.51 & 24.41 & 22.57 & 14.41 & 49.10 & 18.48 \\
		& &  GGG & 8.51 & 24.98 & 23.23 & 15.22 & 47.06 & 19.11 \\
		\midrule\midrule
		\multirow{21}{*}{\begin{sideways} Global estimation \end{sideways}}	& \multirow{7}{*}{Full} &  FGG & 18.06 & 65.77 & 45.72 & 22.96 & 79.49 & 21.92 \\
		&  & CGG & 17.94 & 65.27 & 45.91 & 23.25 & 79.64 & 21.86  \\
		&  & GGG & 18.15 & 66.07 & 46.67 & 23.27 & 77.83 & 22.22\\
		& & 4dF & 18.75 & 62.46 & 45.39 & 22.17 & 83.23 & 21.24\\
		&  & 4dG & 18.54 & 62.57 & 45.07 & 21.45 & 84.37 & 21.05 \\	
		&  & 4dC & 18.07 & 62.73 & 44.39 & 20.43 & 86.38 & 20.72 \\
		&  & 4dInd & 17.64 & 63.06 & 44.32 & 20.20 & 86.77 & 20.64 \\ 		
		\cmidrule{2-9}
		& \multirow{7}{*}{Treatment} &  FGG & 9.00 & 40.58 & 24.14 & 7.67 & 31.62 & 25.51 \\
		& & CGG & 8.98 & 40.46 & 24.17 & 7.69 & 31.70 & 25.47 \\
		&  & GGG & 8.96 & 40.43 & 24.59 & 7.76 & 31.25 & 25.63\\
		& & 4dF & 9.38 & 39.34 & 23.29 & 6.61 & 34.38 & 24.49 \\
		&  & 4dG & 8.98 & 39.79 & 23.00 & 6.31 & 34.92 & 24.28\\	
		&  & 4dC & 9.03 & 39.86 & 22.95 & 6.12 & 35.04 & 24.25 \\
		&  & 4dInd & 8.83 & 39.98 & 23.14 & 6.26 & 34.79 & 24.31\\ 	
		\cmidrule{2-9}
		&\multirow{7}{*}{Control} & FGG & 8.62 & 25.43 & 22.43 & 14.44 & 48.09 & 18.86 \\
		&  & FGF & 8.62 & 25.21 & 22.28 & 14.26 & 48.63 & 18.69 \\
		&  & GGG & 8.75 & 25.69 & 22.89 & 14.73 & 46.94 & 19.24\\
		& & 4dF & 9.15 & 23.94 & 22.26 & 14.03 & 49.62 & 18.38 \\
		&  & 4dG & 8.66 & 23.93 & 21.92 & 13.89 & 50.59 & 18.03 \\	
		&  & 4dC & 8.42 & 23.73 & 21.85 & 13.45 & 51.55 & 17.73 \\
		&  & 4dInd & 8.32 & 24.08 & 21.27 & 13.78 & 51.55 & 17.71 \\ 		
		\hline\hline
	\end{tabular}	\vspace*{-3cm}
\end{table}

\end{document}